\documentclass[fleqn,usenatbib]{mnras}

\usepackage{newtxtext,newtxmath}

\usepackage[T1]{fontenc}

\DeclareRobustCommand{\VAN}[3]{#2}
\let\VANthebibliography\thebibliography
\def\thebibliography{\DeclareRobustCommand{\VAN}[3]{##3}\VANthebibliography}

\usepackage{graphicx}	
\usepackage{amsmath}	
\usepackage{xcolor}
\usepackage{longtable}
\usepackage{multirow}

\newcommand{\Teff}{\mbox{$T_\mathrm{eff}$}}
\newcommand{\Msun}{\mbox{$\mathrm{M}_\odot$}}

\newcommand{\Ion}[2]{#1\,\textsc{#2}}
\newcommand{\kms}{\mbox{km\,s$^{-1}$}}
\newcommand{\dd}{\mathrm{d}}
\newcommand{\WD}[3]{WD\,J#1$#2$#3}
\newcommand{\DWD}[3]{\WD{#1}{#2}{#3}AB}
\newcommand{\dtauWD}{\mbox{$\Delta\tau_\mathrm{WD}$}}
\newcommand{\dtauMS}{\mbox{$\Delta\tau_\mathrm{preWD}$}}

\title[Measuring the IFMR using Gaia DWDs]{Measuring the Initial-Final Mass-Relation using
wide double white dwarf binaries from Gaia DR3}

\author[M. A. Hollands et al.]{
M. A. Hollands,$^{1}$\thanks{E-mail: m.hollands@sheffield.ac.uk}
S. P. Littlefair,$^{1}$
and S. G. Parsons$^{1}$
\\
$^{1}$ Department of Physics and Astronomy, University of Sheffield, Sheffield, S3 7RH, UK \\
}

\date{Accepted 2023 November 23. Received 2023 November 23; in original form 2023 August 7}

\pubyear{2023}

\begin{document}
\label{firstpage}
\pagerange{\pageref{firstpage}--\pageref{lastpage}}
\maketitle

\begin{abstract}
The Initial-Final Mass-Relation (IFMR) maps the masses of main sequence stars to their
white dwarf descendants. The most common approach to measure the IFMR has been to use
white dwarfs in clusters.
However, it has been shown that wide double white dwarfs can also be used to measure the IFMR
using a Bayesian approach.
We have observed a large sample of 90 \emph{Gaia} double white dwarfs using FORS2 on the VLT.
Considering 52 DA+DA, DA+DC, and DC+DC pairs, 
we applied our extended Bayesian framework to probe the IFMR in exquisite detail.
Our monotonic IFMR is well constrained by our observations for initial masses of $1$--$5$\,\Msun,
with the range $1$--$4$\,\Msun\ mostly constrained to a precision of $0.03$\,\Msun\ or better.
We add an important extension to the framework, using a Bayesian mixture-model to determine the
IFMR robustly in the presence of systems departing from single star evolution.
We find a large but uncertain outlier fraction of $59\pm21$\,per\,cent,
with outlier systems requiring an additional $0.70_{-0.22}^{+0.40}$\,Gyr uncertainty in their cooling age differences.
However, we find that this fraction is dominated by a few systems with massive components near 0.9\,\Msun,
where we are most sensitive to outliers, but are also able to establish four systems as merger candidates.
\end{abstract}




\section{Introduction}

After depleting the hydrogen fuel in their cores, almost all ($\simeq 97$~per\,cent) main sequences stars
end their lives by transitioning to the giant phase and shedding their outer layers in an intense stellar wind.
This mass loss that takes place during stellar evolution stops short of the compact central core, which
goes on to become a white dwarf. For most white dwarfs, the interior is composed of carbon and oxygen, enveloped
by a layer of helium, and finally by a thin but opaque layer of hydrogen at the surface. The most massive
white dwarfs, i.e. those with masses $>1.1$\,\Msun, will instead have interiors dominated by oxygen and neon \citep{camisassaetal19-1}.
Main sequence stars with masses above about 8\,\Msun \citep{weidemann+koester83-1},
undergo further nuclear burning during the giant phases,
leading to the formation of more exotic stellar remnants via core-collapse supernovae.

It is intuitive to consider that the mass of a main sequence star ought to be correlated with
the mass of the resulting white dwarf. More precisely the (final) white dwarf mass can be considered a function
of the (initial) main sequence star mass. This function is known as the Initial-Final Mass-Relation (IFMR).
The IFMR essentially encodes the amount of mass-loss occurring during stellar
evolution for stars of different masses \citep{bloecker95-1}, and so is an important component for understanding a wide range
of astrophysical topics, from stellar evolution itself, to the chemical enrichment of the Galaxy. At a more
basic level, the IFMR is also required simply to establish the progenitor masses for studies on
individual white dwarfs.

Despite its universal importance, the precise functional form of the IFMR cannot be simply determined. 
From a computational perspective, determining the IFMR is limited by the predicted mass-loss
occurring during the giant branches of stellar evolution, which is extremely sensitive to input physics,
such convective and rotational mixing \citep{marigo+girardi07-1,cummingsetal19-1}, and nuclear
reaction rates \citep{fieldsetal16-1}.

From the observational side, the IFMR can be measured empirically\footnotemark, taking advantage of white dwarf
cooling as an accurate clock \citep{mestel52-1,fontaineetal01-1}. Essentially, by accurately measuring the effective
temperature (\Teff) and mass of a white dwarf, comparison with white dwarf cooling models yields
the length of time since the white dwarf left the tip of the Asymptotic Giant Branch (AGB),
i.e. the white dwarf cooling age. As long as an independent clock can be sourced to constrain the pre-white dwarf
(pre-WD) lifetime, then the initial mass can be determined from stellar evolution models.
Thus for a sufficiently large sample of white dwarfs, the relation between
initial mass, and final mass can be determined.

\footnotetext{Strictly speaking, all observational approaches yield \emph{semi-empirical} IFMRs,
since they are fundamentally dependent on evolutionary models for both white dwarf cooling
and pre-WD lifetimes.
They are therefore subject to uncertainties in these models \citep{heintzetal22-1}.}

The most widely used observational method to determine the IFMR is using white dwarfs in clusters
\citep{weidemann87-1,ferrarioetal05-1,salarisetal09-1,williamsetal09-1,casewelletal09-1,dobbieetal09-1},
relying on the fact that
the age of a cluster can be measured from the main sequence turn-off. Therefore all white dwarfs within a
given cluster (which will have the same total age as the cluster itself), can have their pre-white dwarf
lifetimes and hence their initial masses determined empirically. It then becomes a simple matter of fitting some
function to the initial-mass/final-mass pairs. The most recent, and most extensive IFMR based on
cluster white dwarfs was presented by \citet{cummingsetal18-1}, having examined 79 white dwarfs
among 13 different clusters, and covering the entire range of initial masses from 0.85\,\Msun\ to
7.5\,\Msun. Even so, such cluster studies are limited by the precision with which initial masses can
be determined, and the low number of objects typically found per cluster.

However, cluster white dwarfs are not the only avenue for exploring the IFMR.
As an alternative technique, \citet{el-badryetal18-1} used the population statistics of the
\emph{Gaia} DR2 white dwarf sample, to constrain the IFMR, finding reasonable agreement
with results from cluster studies. \citet{catalanetal08-1} established a different technique to infer the IFMR, examining white dwarfs in common-proper motion pairs (i.e. wide binaries) with main sequence stars. Both components of the system can be assumed to have formed at the same time, but evolving independently
due to their wide separation. Specifically when the more massive primary undergoes stellar
evolution, no common envelope occurs, and therefore there is no mass transfer onto the secondary. 
As with clusters the final masses are determined from spectral modelling of the white dwarf,
with initial masses estimated by age-dating the main sequence companions using high-resolution
spectroscopy. More recently \citet{barrientosetal21-1} applied a similar technique to 11 white dwarfs with
turnoff/subgiant companions whose ages can be determined more precisely than for main
sequence companions. While accurate initial masses can be obtained this way, very few
white dwarfs are found in binaries with companions at this specific evolutionary stage.

Also employing wide binaries, \citet{andrewsetal15-1} prototyped a new technique
for constraining the IFMR, instead using double white dwarfs (DWDs). Once again,
both components of the binary are assumed to have evolved separately.
However, unlike in the previous examples, one cannot directly infer the pre-WD lifetimes or total ages.
Nevertheless, the cooling ages of the two white dwarfs still provide two independent clocks, where their difference (\dtauWD)
should have the same magnitude but opposite sign as the difference in pre-white dwarf lifetimes (\dtauMS),
i.e. $\dtauWD = -\dtauMS$. This can be used to limit the combination of initial masses
that can have produced that binary. By modelling DWDs within a Bayesian framework,
\citet{andrewsetal15-1} were able to construct a posterior distribution for the IFMR given
a set of 19 DWD pairs. This work predates the release of \emph{Gaia} DR2, with the authors
acknowledging the impending deluge of \emph{Gaia} data would open the opportunity for a much
higher fidelity IFMR to be determined.

In this work we adopt and extend the DWD Bayesian approach, first introduced by \citet{andrewsetal15-1}
for fitting the IFMR. In Section~\ref{sec:obs}, we introduce our spectroscopically observed DWD sample.
In Section~\ref{sec:spec_mod}, we discuss our approach to spectral modelling combined with precise
\emph{Gaia} photometry and astrometry. In Section~\ref{sec:model}, we explain the Bayesian framework
we have adopted to model the IFMR, as well as our extensions to the framework first introduced by
\citet{andrewsetal15-1}. We show and discuss our resulting IFMR fits in Section~\ref{sec:results}, testing
a variety of assumptions and comparing with other established results from different methodologies.
Finally we give our conclusions and areas for future work in Section~\ref{sec:conc}.

\section{Double white dwarf sample}
\label{sec:obs}

\subsection{Sample Selection}
Our goal was to create as large a possible sample of DWDs. To achieve this, we selected double white dwarfs from \emph{Gaia} DR2 \citep{gaiaDR2-collab-2} by combining the catalogue of \cite{el-badryetal18-2} with our own selection. We selected all pairs from the WD catalogue of \cite{gentilefusilloetal19-1} with parallaxes and proper motions consistent within 3$\sigma$. We limited our sample to those DWDs with accurate parallaxes ($\pi / \sigma_{\pi} > 10$) and projected separations of less than 10\,000\,au. We also required small errors on proper motion, $\sqrt{\sigma^2_{\rm pm\_ra} + \sigma^2_{\rm pm\_dec}} < 6$\,mas. Our chosen selection was designed to supplement \cite{el-badryetal18-2} with DWDs at larger distances, and those that have poor quality photometry in \emph{Gaia} DR2.

Our selection includes 424 DWDs, whilst the catalogue of \cite{el-badryetal18-2} contains 375. There is only moderate overlap between the two samples, with 214 DWDs being common to both: 161 objects are found in \cite{el-badryetal18-2}, but not in our sample. The majority of these (120) are close systems which pass the looser parallax and proper motion cuts of \cite{el-badryetal18-2}, that are designed to accommodate orbital motion of the DWD. A further 27 are explained by the relaxed separation constraint (50\,000 vs 10\,000 AU) of \cite{el-badryetal18-2}, and 14 are missing from our selection for unknown reasons. Our selection includes 210 DWDs which are not reported by \cite{el-badryetal18-2}. The vast majority of these are either beyond 200\,pc, or do not pass the cuts placed on the quality of photometry in \cite{el-badryetal18-2}. Of those
210, 23 systems are missing from \cite{el-badryetal18-2} for unknown reasons.

Our final sample consists of 585 high-confidence DWDs, albeit with a complex set of selection criteria. With the improved astrometry of \emph{Gaia} DR3 \citep{gaiaEDR3-collab-1}, 99.8\,percent remain in our sample, and none of the rejected objects were observed.

\subsection{Observations}
From our sample of 585 DWDs, we observed 90 systems with the European Southern Observatory (ESO) Very Large Telescope (VLT) Focal Reducer and Low Dispersion Spectrograph (FORS2) under programs 103.D-0718 and 109.213B. A summary journal of observations is provided in Table~\ref{tab:night_log}. Runs A, B and C were observed under program 103.D-0718 on the nights of June 1 2019, July 2--3 2019, Sep 24 2019 respectively. Runs D and E were observed under program 109.231B on the nights of June 2--3 2022 and 21 Sep 2022 respectively. 

\begin{figure}
    \includegraphics[width=\columnwidth]{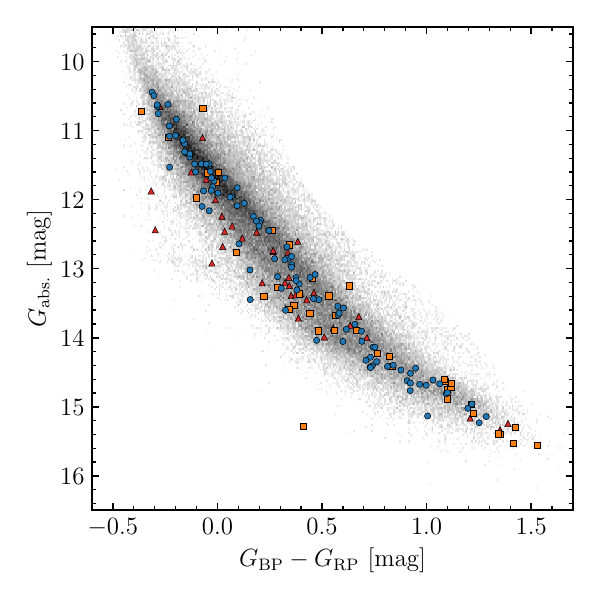}
    \caption{
    \emph{Gaia} DR3 Hertzsprung-Russel diagram of our DWD sample observed with FORS2.
    Red triangles are white dwarfs that were deemed unsuitable for spectroscopic fitting
    with DA models (e.g. due to that component or its companion having exhibiting
    strong magnetism or an unambiguous helium atmosphere). Orange squares are white dwarfs
    with spectroscopic fits, but were not used to fit the IFMR (e.g. due to that
    component or its companion having an extremely low mass or otherwise poor spectroscopic fit;
    see Section~\ref{sec:excl_spec}). The remaining white dwarfs (blue circles)
    were deemed suitable to constrain the IFMR (Section~\ref{sec:results}).
    The grey-scale background uses data from \citet{gentilefusilloetal21-2}.
    }
    \label{fig:hrd}
\end{figure}

For runs A, B and C targets were selected based on brightness and visibility. Objects with existing Sloan Digital Sky Survey (SDSS) spectra were not observed and we preferentially chose systems with large cooling age and mass differences\footnotemark,
based on the values provided in \cite{gentilefusilloetal19-1}. Prompted by the large fraction of magnetic white dwarfs amongst the systems observed in these runs (see Section~\ref{sec:rejected} for details), for runs D and E we simply selected targets at random from the visible DWD systems without SDSS spectra. A Hertzsprung-Russel diagram of the observed systems is shown in Figure~\ref{fig:hrd}, with
\emph{Gaia} astrometry/photometry and spectral classifications given in Table~\ref{tab:dwds_obs}.

\footnotetext{Binaries with near identical masses and cooling ages for both components provide little constraint on the IFMR,
as any IFMR can simultaneously explain both objects. Therefore systems with large differences in the masses and cooling ages
provide the greatest IFMR constraints.}

All observations were taken using the 1200B+97 grating with a 0.7\,\arcsec\ slit, giving a wavelength range of 3660--5010\,\AA\ at a resolution of $R=2000$. The slit was oriented to place both white dwarfs on the slit simultaneously. The data were reduced using the {\sc esoreflex} FORS2 pipeline v5.6.2 and standard procedures. Spectra were extracted using optimal extraction \citep{horne86-1} and a modification of the pipeline was made to prevent the extraction window of faint targets being reduced if a brighter target is nearby on the CCD. Wavelength calibration was performed using a combination of arc lamp images and the location of bright sky lines.

\begin{table}
    \centering
    \caption{\label{tab:night_log}
    Observing log for the different nights of observation. $N_\mathrm{obs}$ is the number of
    systems observed in each run.
    }
    \begin{tabular}{lll}
        \hline
        Run & MJD & $N_\mathrm{obs}$ \\
        \hline
        A & 58\,634--58\,635 & 24 \\
        B & 58\,666--58\,667 & 13 \\
        C & 58\,750 & 23 \\
        D & 59\,732--59\,733 & 18 \\
        E & 59\,843 & 14 \\
        \hline
    \end{tabular}
\end{table}

\subsection{Systems rejected from fitting}
\label{sec:rejected}

Out of our 90 observed systems, there were 18 systems we chose not to fit at all
as we did not believe we could derive meaningful atmospheric parameters
(red triangles in Figure~\ref{fig:hrd}).
For $12$ of these, this was because at least one component
exhibited a strong magnetic field.
These systems are \DWD{0059}{-}{2417}, \DWD{0220}{-}{1532}, \DWD{0224}{-}{4611},
\DWD{0344}{+}{1509}, \DWD{0902}{-}{3540}, \DWD{1159}{-}{4630}, \DWD{1834}{-}{6108}, \DWD{2018}{+}{2129}, \DWD{2023}{-}{1446}
\DWD{2047}{-}{8206}, \DWD{2304}{-}{0701}, and \DWD{2353}{-}{3620}.
Note that while \DWD{2023}{-}{1446} contains a strongly
magnetic DA, it is primarily excluded from fitting due to its baffling
A component (Section~\ref{sec:individual}).
Furthermore, four spectra exhibited an unambiguous helium-dominated atmosphere
(either due to helium lines or pressure broadened metal lines). These systems
are \DWD{0225}{-}{1756}, \DWD{1813}{+}{0604}, \DWD{1836}{-}{5114},
and \DWD{2355}{+}{1708}. Finally two systems,
\DWD{1310}{-}{3930} and \DWD{2115}{+}{2534},
contained white dwarfs with blue DC spectra which could plausibly
be explained by either strong magnetism or helium dominated atmospheres.
In total this left 72 systems used to perform spectroscopic fits.

\subsection{Notes on individual systems}
\label{sec:individual}

\noindent\textit{\DWD{1953}{-}{1019}} \\
This system is part of a resolved triple white dwarf, first reported by \citet{perpinyaetal19-1}.
Our observations refer to the A and C components from the discovery paper. In hindsight, it would
have been ideal to observe all three components of this system to provide further constraints
on the IFMR. That said, \citet{perpinyaetal19-1} found that the B and C components (the inner binary)
have almost identical atmospheric parameters. The potential for this system to provide a strong constraint
on the IFMR is therefore limited compared to an idealised case where all three components
have substantially different masses and cooling ages.
\medskip

\begin{figure}
    \includegraphics[width=\columnwidth]{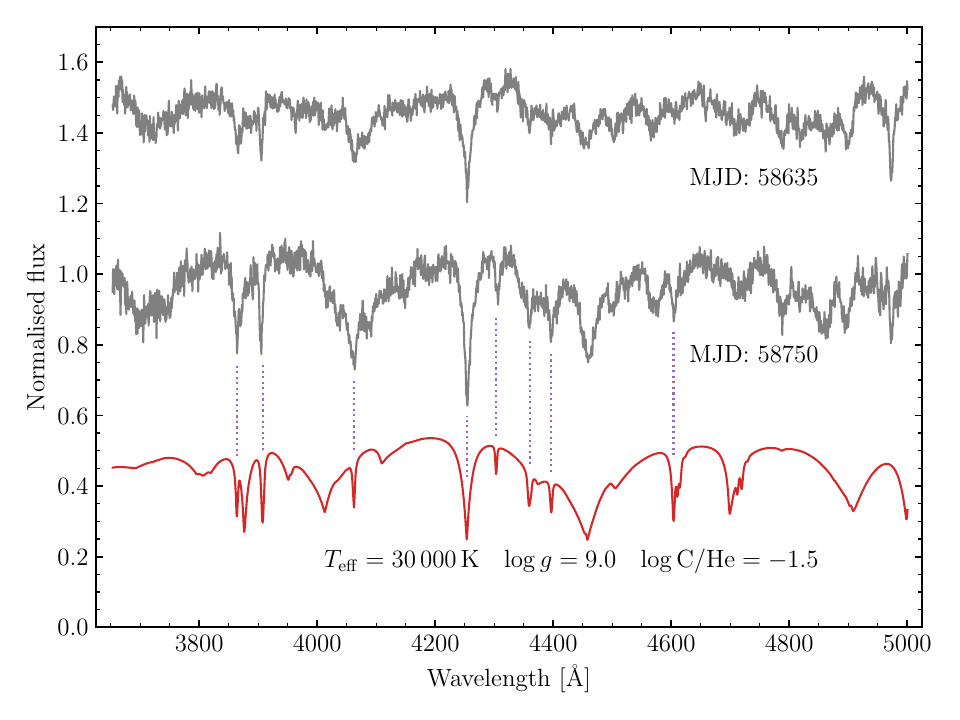}
    \caption{
    Normalised spectra of \WD{2023}{-}{1446}A (grey) with a DQB model shown in red. The spectra and model
    are offset by 0.5 from one another. The model has been blue shifted by 1050\,\kms\
    with identified \Ion{C}{ii} lines indicated by the dotted purple lines.
    }
    \label{fig:DQB}
\end{figure}

\noindent\textit{\DWD{2023}{-}{1446}} \\
While we are unable to use this system for fitting the IFMR, both components of this wide binary
are noteworthy. The B component has a spectral type DAH. The H$\beta$ absorption line shows Zeeman splitting
indicative of a $14$\,MG magnetic field. The A component, however, is extremely unusual and while
resembling a hot DQB, an assignment of DX \citep{sionetal83-1} is more accurate:
Its position in the \emph{Gaia} HR diagram indicates a mass of $\simeq 1.2$\,\Msun\ \citep{gentilefusilloetal21-2},
and a plethora of lines are apparent in the spectrum (Figure~\ref{fig:DQB}),
none of which we have been conclusively able to assign.
While a strong magnetic field with a non-hydrogen atmospheric composition provides
a plausible explanation for these observations, if we allow for a large blue-shift of 1050\,\kms\
(not accounting for the presumably substantial gravitational redshift),
some of the strongest lines can be assigned to \Ion{C}{ii}, and a few to \Ion{He}{i}
(Figure~\ref{fig:DQB}),
with no evidence of Zeeman splitting observed. Nevertheless, around one third of the spectral
features remain unassigned, and some predicted spectral lines are absent. While the proposed blue-shift
could be explained as a wavelength calibration issue, we found no evidence for this, and the A component shows no similar shift, despite being observed simultaneously. 
Furthermore, we obtained a second set of observations of this system in our third FORS2 run (run~C),
finding a similar spectrum with features occurring at the same wavelengths, though many with
different strengths.

Both components have similar proper-motions of about $17$\,mas\,yr$^{-1}$, implying
a $v_\perp$ of 15\,\kms\ for the system. Crucially, the difference in $v_\perp$
is $<1.7$\,\kms\ at the 99th percentile, which at an on-sky separation of 2000\,au
rules out the possibility of a chance alignment.
Given the somewhat convincing velocity shift, it is tempting suggest
\WD{2023}{-}{1446}A may itself be part of a close binary with a hidden companion,
where both epochs were by chance observed at the same extremum in orbital velocity.
While a much cooler white dwarf could remain hidden by the flux of \WD{2023}{-}{1446}A,
the short orbital period (required for an orbital velocity $>1050$\,\kms)
would lead to orbital smearing of the spectra.
An orbital period of a few 10\,hr could instead be achieved with a stellar-mass black hole
companion, but such a scenario is obviously contrived.
Light curves we have obtained of this object show no variability above the 1\,percent level.
As it has no bearing on our investigation of the IFMR,
we refrain from speculating further on the nature of this peculiar system.
\medskip

\noindent\textit{\DWD{1336}{-}{1620}} \\
This system was first observed during observing run~A, and subsequently re-observed during run~D.
While unintentional, it provided the opportunity to determine the level of
systematic uncertainty in our spectroscopic fits. 
With essentially four observations (two per binary component), we were able to
constrain the $\Teff$ relative precision to $0.9_{-0.4}^{+1.0}$\,percent, and the
$\log g$ precision to $0.043_{-0.017}^{+0.045}$\,dex.
These precisions can be used as priors in Section~\ref{sec:teff_logg_errs}.
\medskip

\noindent\textit{\DWD{2018}{+}{2129}} \\
This system appears to be a rare wide binary where both components are magnetic.
In the A component, Zeeman splitting of the Balmer lines is observed,
indicative of a $\simeq 1$\,MG field. At a glance, the B component spectrum resembles
a DC spectral type, though closer inspection reveals broad wavy absorption bands
characteristic of a strongly magnetic atmosphere with a field strength in the 100s
of MG (see \WD{1159}{-}{4630}B, \WD{1834}{-}{6108}A \WD{2047}{-}{8206}A,
\WD{2304}{-}{0701}A, or \WD{2353}{-}{3620}B for more obvious examples).
To our knowledge this is the only known wide DWD where both components
exhibit magnetism.

\section{Spectral modelling}
\label{sec:spec_mod}

\subsection{DA model grid}

To obtain atmospheric parameters for our double white dwarf sample, we calculated a
two-dimensional grid of pure hydrogen model spectra using the Koester white
dwarf model atmosphere code \citep{koester10-1,koester13-1}.
Effective temperatures (\Teff) were calculated from 3500\,K
to 10\,000\,K in steps of 250\,K, then up to 20\,000\,K in steps of 500\,K, and finally
up to 35\,000K in 1000\,K steps. The grid was evaluated for surface gravities ($\log g$)
from 7.00 to 9.50 in 0.25\,dex steps (cgs units). All models were calculated with
a convective mixing length parameter $\mathrm{ML2}/\alpha$ \citep{tassouletal90-1} of $0.8$. This choice of 0.8 allows
easy application of 3D corrections to our spectroscopic parameters (see below).

Each model spectrum was convolved to an instrumental resolution of 1.8\,\AA,
as appropriate for our FORS2 grating and slit-width combination.
For each combination of \Teff\ and $\log g$, we determined the stellar radius,
$R_\mathrm{WD}$,
using the mass-radius relation from \citet{bedardetal20-1} appropriate for
white dwarfs with thick hydrogen layers. Since the Koester model fluxes
are in units of $4\times$ Eddington-flux, we scaled each model in the grid
by a factor $\pi (R_\mathrm{WD}/1\,\mathrm{kpc})^2$.
As an additional step we applied gravitational redshifts to each model in our grid,
using same mass-radius relation as before.
In principle this allows our fits to be sensitive to the difference
in gravitational redshift for the two components of a binary (assuming negligible
orbital reflex motion), with the systemic radial velocity necessarily
included as an additional free parameter.

To generate fluxes for arbitrary \Teff\ and $\log g$ and wavelength,
we used tri-linear interpolation of the logarithm of the model grid fluxes.
The use of the log-fluxes is necessary as wavelengths bluer than the peak have
a strongly non-linear dependence on \Teff.

Since these models are 1-dimensional, they require correction to the corresponding
3D parameters for the most physically accurate results. We adopted the corrections
given by \citet{tremblayetal13-1}, applying their Equations (9) and (10), 
appropriate for the ML2/$\alpha=0.8$ used in our model atmosphere calculations.

\subsection{Spectrophotometric fitting}

To fit each DWD, we used not only the FORS2 spectra, but also the \emph{Gaia}
DR3 photometry to constrain the atmospheric parameters. We first fit each binary
using a least squares approach, following this up with a Markov-Chain
Monte-Carlo analysis for more informative parameter distributions.

\begin{figure*}
    \includegraphics[width=\textwidth]{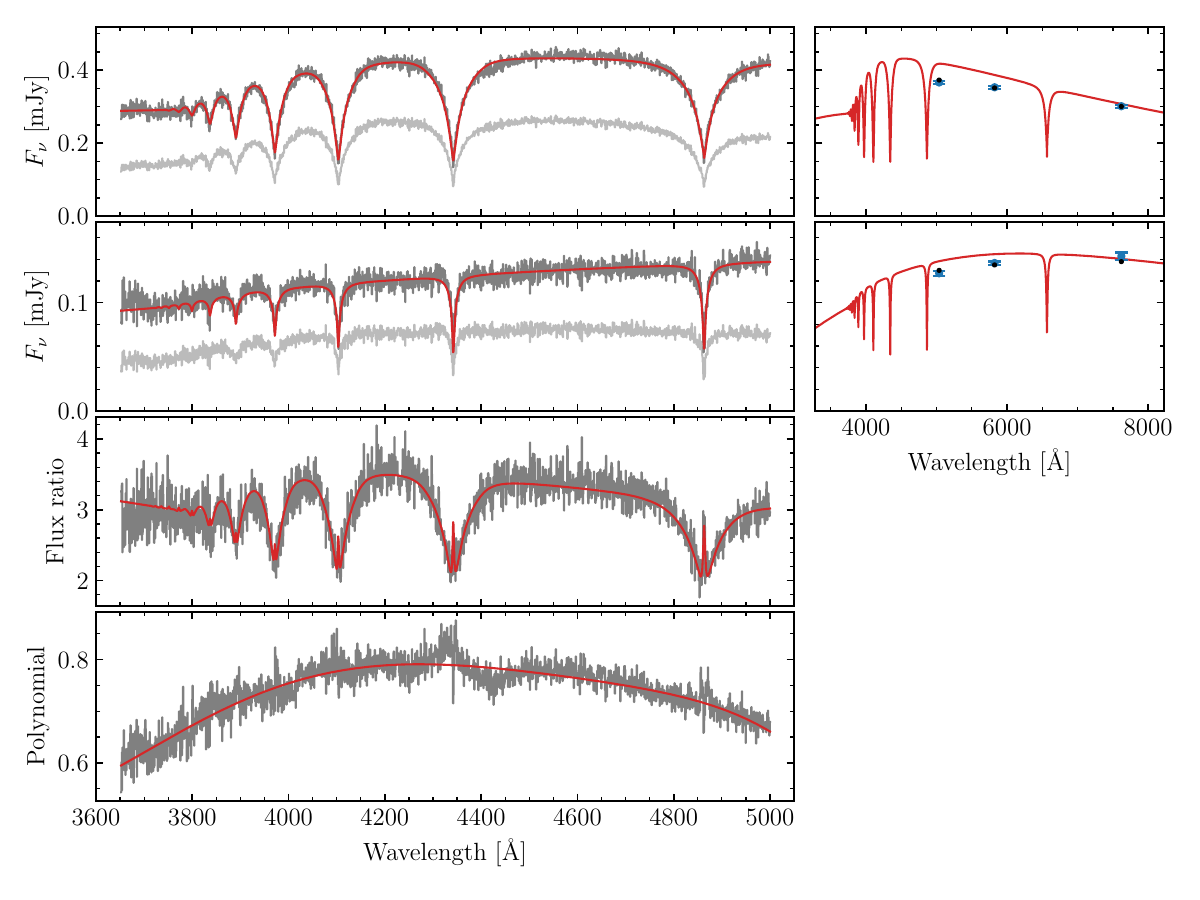}
    \caption{
    Best fit to \DWD{0023}{+}{0643}. The A and B components are shown in the top two rows.
    The left columns show the spectra with their original flux calibrations
    (light grey), compared with re-calibrated spectra (medium grey),
    and the best fitting models (red).
    The right columns show the \emph{Gaia} DR3 fluxes (blue points with error bars), against
    the model synthetic photometry (black points) -- note that the spectral models are shown
    for illustrative purposes only, and do not overlap the synthetic points due to the broadness
    of the \emph{Gaia} bandpasses.
    The second bottom panel shows the ratio of fluxes compared with the ratio of model spectra, demonstrating
    excellent agreement in the wings and line cores.
    The bottom panel shows the ratio of original spectra to their best fitting models
    (averaged over the A and B components) against the flux-correcting 5-th order polynomial.
    }
    \label{fig:fit_ex}
\end{figure*}

Inspecting the FORS2 spectra, it was clear that the flux calibrations were generally poor,
though in a consistent manner for each DWD, i.e. unphysical curvature in the spectral continuum
of the A component would also be observed in the spectrum of the B component
(as a result of both stars being observed on the slit simultaneously).
We therefore took advantage of these shared systematic uncertainties by
fitting both components of the binary simultaneously. Specifically, we modelled
the wavelength dependent flux calibration correction as a 5th-order polynomial.
While we found this approach to be generally successful, we found that the relative
fluxes can differ by a constant factor, e.g. if one component was slightly off-centre on
the slit compared to the other. We therefore introduced a further free-parameter
to re-scale the fluxes of the B-component by a constant amount.
A demonstration of our approach to flux calibrations is shown in Figure~\ref{fig:fit_ex}
for the system \DWD{0023}{+}{0643}.

Initially, we performed a least squares fit to the spectra and \emph{Gaia} photometry\footnote{
While some systems have additional photometry from surveys such as SDSS \citep{alametal15-1},
we opted not to use these to avoid introducing non-uniform systematics into our sample.}
to constrain the flux correcting polynomial, the flux scaling factor on the B-component,
the \Teff\ and $\log g$ for each component,
and the systemic radial velocity. Furthermore, we included the parallax
as a free-parameter in order to marginalise over the uncertainty of the \emph{Gaia} measurement.
We then used the best fitting parameters and covariance matrix to initialise a Markov-Chain
Monte-Carlo fit via the \textsc{python} package \texttt{emcee} \citep{foremanmackeyetal13-1}.

As part of the MCMC we introduced two new free parameters.
In a few cases, where one white dwarf is significantly brighter than its companion,
the bright component may have an extremely high S/N ratio (in order to get sufficient
counts for the fainter component). This may cause that spectrum to dominate
the fit entirely. To avoid this, we introduced a relative noise floor
parameter, $\eta$ to our fits. Simply put, we increased the flux uncertainties
according to $\sigma_i'^2 = \sigma_i^2 + (\eta\times f_i)^2$, where $\sigma_i'$ and
$\sigma_i$ are the modified and modified flux uncertainties, and the $f_i$ are
the spectroscopic fluxes at wavelength $i$. Compared with the usual $-\tfrac{1}{2}\chi^2$,
the addition of this noise-floor requires a slightly modified log-likelihood for each spectrum,
\begin{equation}
    \ln L = -\tfrac{1}{2}\chi^2 - \sum_i^N \ln \sigma_i',
    \label{eq:like_spec}
\end{equation}
where $\chi^2$ has the usual meaning and implicitly depends on the $\sigma_i'$.
A uniform prior for $\eta>0$ was used on this parameter.
Similarly we folded in a 1\,percent additional flux to the \emph{Gaia} fluxes
according to the \emph{Gaia} documentation on the uncertainty on the absolute calibration scales.

Our sample of DWDs have \emph{Gaia} parallaxes locating them at distances up to 250\,pc,
with a median of 90\,pc. Therefore it is reasonable to assume that the \emph{Gaia} fluxes
for some of the more distant objects will be affected by interstellar reddening.
Indeed in our initial attempts at fits (which did not include reddening),
we found some cases where
a good fit to the spectra resulted in models that were bluer than the \emph{Gaia} photometry,
affecting primarily the more distant objects such as \DWD{1336}{-}{1620}.
To account for this we included interstellar reddening, specifically $E(B-V)$,
as a an additional free parameter. For the MCMC fits, we used the Jeffreys prior,
$P(E(B-V)) \propto E(B-V)^{-1/2}$,
which is naturally weighted towards lower values on a linear scale,
but is also a proper prior as long as an upper-bound is provided.
For each DWD, we queried the Bayestar19 3D
extinction maps \citep{greenetal19-1} to determine the 99th percentile of $E(B-V)$
at the specified distance of the system.
This was then used as upper-bound on our reddening prior. 
For systems outside of the Bayestar footprint, we adopted $E(B-V)=10^{-4}$ as the
upper limit if located within 100\,pc of the Sun, or $0.15$ otherwise. 
For the most nearby systems, where the queried
99th percentile may return exactly zero, we also set the upper-limit at $10^{-4}$
for numerical stability.

For the priors on the remaining free parameters we used normal distributions on both $\log g$s
of $\mathcal{N}(\log g; 8.0, 0.25)$ (units of dex) to approximate the white dwarf mass distribution
in the absence of any other strongly constraining data, though in practice, even in the worst case
we found the data constrained the $\log g$ to within $\pm0.11$\,dex. For the radial velocity
$r$, we again used a normal distribution $\mathcal{N}(r; 0, 200)$ (units of \kms) specifically
to avoid $r$ becoming unbounded when fitting DC+DC pairs. Finally, for the two \Teff, we used flat priors on
a logarithmic scale to reflect the greater abundance of cool white dwarfs in volume limited samples
compared to the hottest objects.

\subsection{Spectroscopic results}

For the MCMC fits themselves, for each system, we used an ensemble of 200 walkers. We found that most of these
had fully burnt-in within about 2000 steps. Therefore we chose to run for 5000 steps,
using the final 1000 steps (thinned to every 5 steps) to report the final results.
The best parameters for all 72 fitted binaries are shown in Table~\ref{tab:dwds_fits}, including
derived parameters (masses, cooling ages, and difference in cooling ages).
Spectra, photometry and best fitting hydrogen atmosphere models for all systems are shown in
Figures~\ref{fig:fits01}--\ref{fig:fits31}.
These figures include systems that we did not fit (Section~\ref{sec:excl_spec}), but which are shown without models. 

\begin{figure}
    \includegraphics[width=\columnwidth]{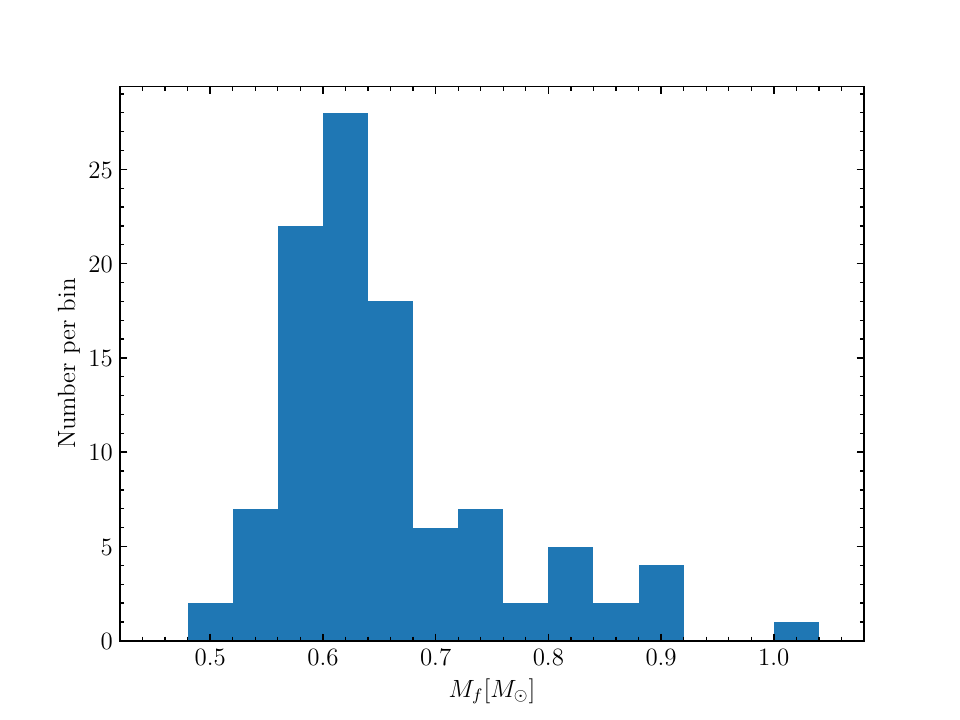}
    \caption{
    Mass distribution of our 104 white dwarfs (52 binary systems) used to fit the IFMR.
    }
    \label{fig:mdist}
\end{figure}

The white dwarf mass distribution of all objects used to perform IFMR fitting is shown in
Figure~\ref{fig:mdist} (details of additional excluded systems are given in Section~\ref{sec:excl_spec}). 
These white dwarfs have a median mass of 0.626\,\Msun, with the distribution appearing broadly
similar to white dwarf mass distributions from other works
\citep{genestetal14-1,rebassa-mansergasetal15-1,hollandsetal18-2}.

\subsubsection{Systems excluded for fitting the IFMR}
\label{sec:excl_spec}

The vast majority of our fits are consistent with both spectroscopy and
photometry (Figures\,\ref{fig:fits01}--\ref{fig:fits24}),
therefore giving us confidence that they are suitable to fit the IFMR.
Even so, there are some systems where, for one reason or another,
we do not trust our atmospheric parameters, and so exclude them from
the following parts of this work.

Four systems contain magnetic white dwarfs with field strengths of a few MG\footnotemark.
Due to Zeeman splitting, the resulting fits had synthetic photometry that were entirely inconsistent with the \emph{Gaia} data.
These systems are \DWD{0002}{+}{0733}, \DWD{0240}{-}{3248}, \DWD{1314}{+}{1732},
\DWD{1636}{+}{0927}, and \DWD{2259}{+}{1404}. Note that \WD{1535}{+}{2125}B exhibits a weak $<1MG$ magnetic
field, which does not appear to have affected the quality of its spectroscopic fit.
Therefore we see no reason to reject this system.

\footnotetext{Of course, for the most magnetic systems, where the Balmer lines are not
even approximately in the same positions as in the non-magnetic cases, we did not attempt
any spectroscopic or photometric fits (See Section~\ref{sec:rejected}).}

Many systems contain one or more DC white dwarfs. Some of these remain ambiguous as to
whether they could have hydrogen or helium dominated atmospheres. However a few systems,
show severe disagreement with their best fitting hydrogen-dominated model spectra.
The white dwarfs in question are \WD{1124}{-}{1234}A, \WD{1350}{-}{5025}A,
\WD{1729}{+}{2916}B, \WD{1929}{-}{3000}B, and \WD{2122}{+}{3005}A. In most of these cases
the fit attempts to reduce the strength of the hydrogen lines by reducing the $\Teff$,
resulting in extremely poor agreement with the photometry. Regardless, these systems
have been fitted with an incorrect atmospheric model, and so must be rejected.

As stated above, some systems containing one or more DCs remain ambiguous.
These are generally DCs with $\Teff < 5000$\,K,
which would not show strong hydrogen lines regardless of their atmospheric composition.
Therefore we rely on the photometry to assess the quality of the fits.
Indeed, some DA+DC systems such as \DWD{0109}{-}{1042} show excellent agreement with the
photometry of the DC component, suggesting that object does indeed have a hydrogen dominated
atmosphere. However, \WD{0007}{-}{1605}A, \WD{1827}{+}{0403}A,
and \WD{1859}{-}{5529}B\footnotemark\ show disagreement suggesting they instead
likely have helium dominated atmospheres. We therefore exclude these three systems.

\footnotetext{The other, brighter, component of this system,
only just shows a weak H$\beta$ line in our spectrum -- enough, however, to confirm its
hydrogen dominated nature.}

Our sample also contains DC+DC pairs introducing further ambiguity, as the atmospheric
parameters for both components are essentially determined only by the photometry.
All of these DC+DC systems, e.g. \DWD{0104}{+}{2120}, show good agreement with their photometry.
However, we found that several of these DC+DC pairs contained one or
more white dwarfs with low derived masses implying they cannot have formed via standard
evolution, as the universe is not old enough to have produced such low mass objects assuming
single star evolution for both components.
Therefore such systems must be rejected on the basis that they cannot constrain the
IFMR. These systems are \DWD{1014}{+}{0305}, \DWD{1211}{-}{4551}, \DWD{1557}{-}{3832},
\DWD{1804}{-}{6617}, \DWD{1827}{+}{0403}, \DWD{1929}{-}{4313}, \DWD{2122}{+}{3005},
\DWD{2230}{-}{7513}, and \DWD{2248}{-}{5830}.

After rejecting these 20 problematic systems (\DWD{1827}{+}{0403} and \DWD{2122}{+}{3005} are rejected on account
of multiple reasons), our final sample we used to fit the IFMR contained 52 systems
(mass distribution shown in Figure~\ref{fig:mdist}).

\subsubsection{Comparison of results}

\begin{figure*}
    \includegraphics[width=\textwidth]{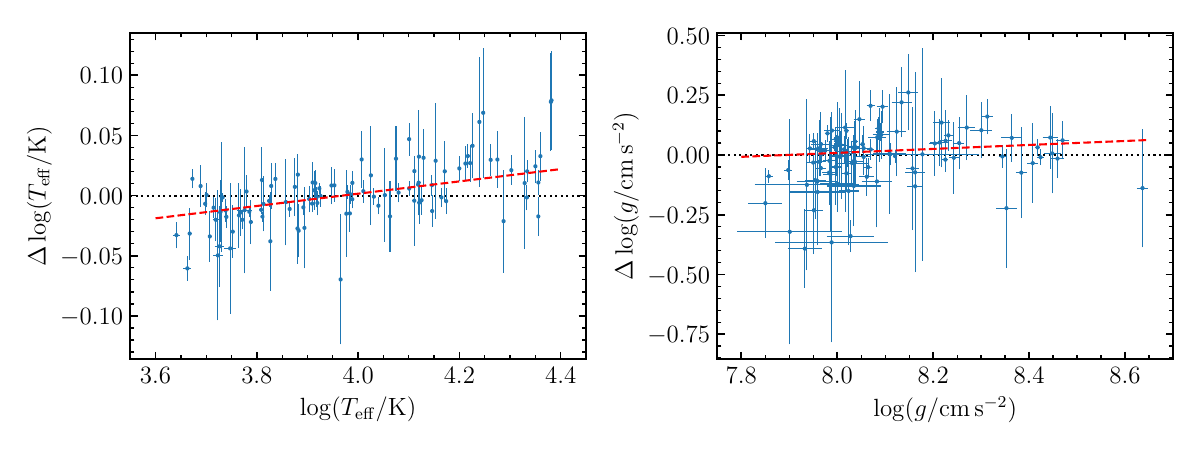}
    \caption{
    Comparison between our combined spectroscopic+photometric fits, and the
    photometric fits of \citet{gentilefusilloetal21-2}. This includes only
    the 52 systems (104 white dwarfs) used to constrain the IFMR, i.e.
    systems with poor fits due to magnetism or helium dominated atmospheres
    or extremely low masses are not shown. The dashed red lines indicate
    linear fits to the $\log\Teff$ and $\log g$ differences.
    }
    \label{fig:Tg_comp}
\end{figure*}

Since all objects in our sample were drawn from the \emph{DR2} catalogue of \citet{gentilefusilloetal19-1}
and subsequently are present in the EDR3 catalogue of \citet{gentilefusilloetal21-2}, we sought to compare
our combined spectroscopic+photometric fits with the photometric results of \citet{gentilefusilloetal21-2}.
In Figure~\ref{fig:Tg_comp}, we show the difference between these sets of results for both \Teff\
and $\log g$, though specifically for the 52 binaries (104 white dwarfs) selected to constrain the IFMR.
In both panels, the uncertainties in \Teff/$\log g$ differences are generally dominated
by the contribution from the \citet{gentilefusilloetal21-2} results, since we use the same photometric
data for our fits, but have the benefit of spectroscopy.

In the left panel, a small but statistically significant linear trend is seen
in the difference in results as a function of temperature. Specifically, we find
a gradient of $0.051\pm0.007$, or equivalently $11.8\pm1.5$\,percent\,dex$^{-1}$.
Across the range of data, this means we find 4\,percent lower \Teff\ at
$\log(\Teff/\mathrm{K})=3.65$ ($\simeq 4500$\,K), consistent results at 9300\,K,
and 5\,percent higher \Teff\ at $\log(\Teff/\mathrm{K})=4.4$ ($\simeq 25\,000$\,K).
The results for the surface gravities are much more consistent,
with the best fitting linear trend only varying by about 0.005\,dex across
the observed range of data (Figure~\ref{fig:Tg_comp}).

The \Teff\ differences, while generally small are still worth investigating
as small changes in \Teff\ will result in larger differences in cooling ages. 
The source of this discrepancy could result from differences in the model atmospheres
used between our work (Koester models) and the work of \citet{gentilefusilloetal21-2}
(Tremblay models),
the fact that we used spectroscopy \emph{and} photometry rather than just photometry,
or fitting methodology.
As a test we tried fitting the \emph{Gaia} photometry alone. In this case, the linear
trend in \Teff\ differences disappeared, suggesting the addition of spectroscopy
is responsible. We note that despite finding no linear trends, our photometric
fits were on average 3\,percent lower in \Teff\ compensated by results 0.04\,dex lower in $\log g$,
indicating some remaining difference between atmosphere models or fitting methodology. 

Differences in spectroscopic and photometric stellar parameters are well documented \citep{bergeronetal19-1}.
Despite these differences we note that given the size of the error-bars, we are
still generally within 1--2$\sigma$ of the results from \citet{gentilefusilloetal21-2}.
Given that our spectroscopy provides significantly more data with which to work
with, it is likely that our results are more accurate. Even so, in Section~\ref{sec:outliers},
we demonstrate the ability of our IFMR model to account for systematic uncertainty in
\Teff\ and $\log g$ parameters, and so this is not expected to affect our results on
fitting the IFMR. 

\section{Modelling the IFMR}

\label{sec:model}

In this section, we explain our Bayesian model for investigating the IFMR using
wide DWDs. This is essentially an adaptation of the framework first introduced
by \citet{andrewsetal15-1}, though presented in a mathematically simplified way.
Furthermore, we have expanded the framework, accounting for
correlated and underestimated \Teff\ and $\log g$ uncertainties, as well as
allowing for the possibility that some fraction of systems are outliers that
violate the assumption of coeval stellar evolution.
Our implementation of this model will be made available at \url{https://github.com/mahollands/IFMR_DWDs}.

\subsection{Bayesian framework}

Given a set of observed wide DWDs with measured parameters,
we ultimately wish to determine the distribution of IFMRs consistent with our data.
In Bayesian notation this means our desired posterior distribution can be written as
\begin{equation}
    P(\mathrm{IFMR} | \mathrm{DWDs}) \propto P(\mathrm{IFMR}) \times P(\mathrm{DWDs} | \mathrm{IFMR}),
    \label{eq:posterior}
\end{equation}
where the terms on the right are the prior probability distribution on the IFMR,
and the likelihood of obtaining our data given a specific IFMR, respectively.
Since each of the DWDs represent independent observations,
the likelihood in Equation~\eqref{eq:posterior} 
can be written as a product of likelihoods for each double white dwarf (DWD$_k$),
\begin{equation}
    P(\mathrm{DWDs} | \mathrm{IFMR}) = \prod_k^N P(\mathrm{DWD}_k | \mathrm{IFMR}).
    \label{eq:likelihood_prod}
\end{equation}

More explicitly, the relevant measured observables for each DWD are the two white dwarf
(final) masses ($M_{f1}$ and $M_{f2}$) and the difference in their white dwarf cooling ages,
\dtauWD. Furthermore these parameters have measurement uncertainties, which
we encapsulate in a covariance matrix, $\Sigma$. Written out explicitly, the individual
DWD likelihood in Equation~\eqref{eq:likelihood_prod} can be written as
\begin{equation}
    P(\mathrm{DWD}|\mathrm{IFMR}) = P(M_{f1}, M_{f2}, \dtauWD | \mathrm{IFMR}, \Sigma),
    \label{eq:likelihood_explicit}
\end{equation}
where we have dropped the index, as the DWD in question is no longer explicitly from a larger set of DWDs.
All that remains is to determine the functional form of this likelihood,
though how to do so is not immediately obvious.

If instead we had prior knowledge of the initial masses, $M_{i1}$ and $M_{i2}$,
things become much clearer.
The predicted final masses could then be determined directly from the initial masses
and IFMR. The difference in cooling ages could also be predicted from the main
sequence lifetimes corresponding to these initial masses,
recalling that $\dtauWD = -\dtauMS$ for coeval systems.
Mathematically the likelihood for a DWD with a given IFMR \emph{and} given initial masses
can be written as
\begin{equation}
    P(\mathrm{DWD} | \mathrm{IFMR}, M_{i1}, M_{i2}) = \frac{1}{\sqrt{(2\pi)^3 |\Sigma|}}
    \exp\left(-\tfrac{1}{2}\mathbf{X}^T \Sigma^{-1} \mathbf{X}\right),
    \label{eq:likelihood_exp}
\end{equation}
where
$\mathbf{X}$ is defined as
\begin{equation}
    \mathbf{X} = \left(\begin{array}{c}
         M_{f1} - \mathrm{IFMR}(M_{i1}) \\
         M_{f2} - \mathrm{IFMR}(M_{i2}) \\
         \dtauWD + \dtauMS(M_{i1}, M_{i2}) \\
    \end{array}\right),
\end{equation}
In principle this likelihood \emph{could} be used directly to determine the IFMR, by constructing
a posterior distribution where the initial masses \emph{per system}
are also free parameters to be sampled.
In practice however, for large sets of DWDs such as ours,
the posterior distribution becomes too highly dimensional to sample in finite time.

To ensure that the posterior distribution has as few free-parameters as possible
(i.e. the parameters defining the IFMR), we can instead marginalise over the initial masses
by integrating over them, i.e.
\begin{equation}
    P(\mathrm{DWD}|\mathrm{IFMR}) = \iint P(\mathrm{DWD}| \mathrm{IFMR}, M_{i1}, M_{i2}) \,\dd M_{i1}\,\dd M_{i2},
    \label{eq:like_int}
\end{equation}
recovering the desired likelihood in Equation~\eqref{eq:likelihood_explicit}. Conceptually,
the process of integration takes into account all possible combinations of $M_{i1}$ and $M_{i2}$
leading to a likelihood where the distribution of data depends only on the IFMR.
Even more rigorously, the integrand in Equation~\eqref{eq:like_int} can be multiplied
by a prior-distribution on $M_{i1}$ and $M_{i2}$ to weight the distribution by initial
masses that are more common, i.e. an initial mass function (IMF) as a prior.
In our implementation we use a Salpeter IMF \citep{salpeter55-1} with exponent
$\alpha=2.3$, i.e. the high mass part of the Kroupa IMF \citep{kroupa01-1}, and
apply this to both components.

In practice this integration step must be performed numerically, presenting its
own set of challenges. The integral must be calculated many times (specifically
the number of times the likelihood in Equation~\eqref{eq:likelihood_prod} is evaluated
multiplied by the number of DWDs in the sample), therefore computationally
expensive integration techniques such as Gaussian quadrature are not appropriate here.
Furthermore, for some specific DWD and IFMR sample, only a small region of the
$M_{i1}$-$M_{i2}$ plane will contribute any significant probability density, making
the choice of integration limits difficult (in order to avoid integrating over
regions of near-zero probability density).

The joint-distribution of final masses, however, \emph{is} known, and so for any
given IFMR, the corresponding region in the $M_{i1}$-$M_{i2}$ plane can be determined,
subject to the condition that the IFMR is monotonic (and thus invertible).
Therefore by performing a change of variables, we can instead perform the integration in the
$M_{f1}$-$M_{f2}$ plane, transforming the integral in Equation~\eqref{eq:like_int}
into 
\begin{equation}
    \iint P(\mathrm{DWD} | \mathrm{IFMR}, M_{i1}, M_{i2})
    \frac{\partial M_{i1}}{\partial M_{f1}} \frac{\partial M_{i2}}{\partial M_{f2}}
    \,\dd M_{f1}\,\dd M_{f2},
    \label{eq:like_int_Mf}
\end{equation}
where the Jacobian terms are simply the gradient of the inverse IFMR evaluated at $M_{f1}$ and $M_{f2}$,
and the $M_i$ are determined by plugging the $M_f$ into the inverse IFMR.
Since the joint distribution of $M_{f1}$ and $M_{f2}$ has already been measured from the data, we can
use Monte-Carlo samples from this distribution, restricting ourselves to a small
area in the $M_{f1}$-$M_{f2}$ plane (and the equivalent area in the $M_{i1}$-$M_{i2}$ plane) to perform the integral
in Equation~\eqref{eq:like_int_Mf}.

We achieve this numerical approach using the technique of importance sampling. In importance sampling,
one can evaluate the integral of a function $f(\mathbf{x})$ (where $f$ may be multivariate) by drawing
$N$ samples, $\mathbf{x}_i$, from a distribution $P(\mathbf{x}$). The integral may then be approximated as
\begin{equation}
    \int f(\mathbf{x})\,\dd \mathbf{x} \approx \frac{1}{N}\sum_{i}^N \frac{f(\mathbf{x}_i)}{P(\mathbf{x}_i)}.
    \label{eq:importance}
\end{equation}
Therefore importance sampling essentially turns the integral into a weighted mean, where --  
assuming that $P(\mathbf{x})$ has been chosen to be similar in shape to $f(\mathbf{x})$ --
the density of samples is highest close to where $f(\mathbf{x})$
is maximised, and few samples are placed at values of $\mathbf{x}$ that contribute
little to the integral of $f$.
Thus, the sample weights are simply $1/P(\mathbf{x}_i)$, and where $P$ is normalised over
the space $\mathbf{x}$.

\begin{figure}
    \includegraphics[width=\columnwidth]{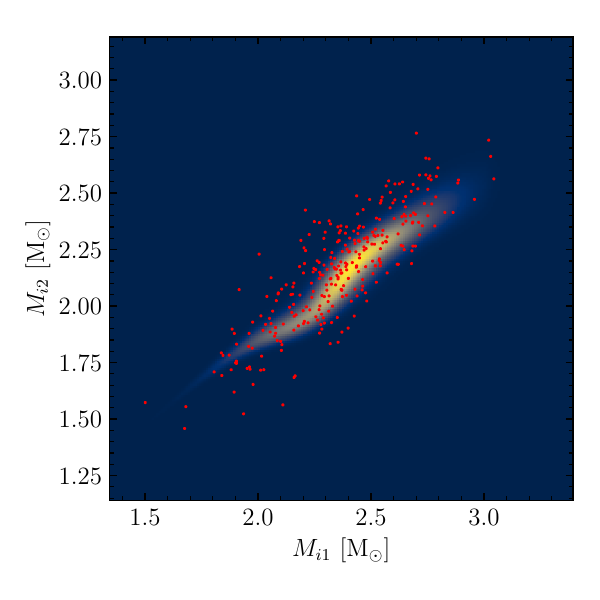}
    \caption{
    The likelihood given by Equation~\eqref{eq:likelihood_exp} in
    the $M_{i1}$-$M_{i2}$ plane, evaluated for
    \DWD{2007}{-}{3701} and a 3-segment mock IFMR. The red points show 300 samples
    drawn from the joint-distribution of $M_{f1}$ and $M_{f2}$ and then converted
    to initial masses via the inverse IFMR. These samples are subsequently used to
    numerically integrate the likelihood function.
    }
    \label{fig:Mi12_samples}
\end{figure}

In our case, we draw samples of $M_{f1}$ and $M_{f2}$ from a multivariate-normal
distribution using the mean and covariance matrix derived from our spectroscopic fits,
which can then be used to evaluate the double integral in Equation~\eqref{eq:like_int_Mf}.
We demonstrate this approach in Figure~\ref{fig:Mi12_samples},
where we use a 3-segment mock-IFMR to calculate the likelihood
for \DWD{2007}{-}{3701} in the $M_{f1}$-$M_{f2}$ plane.
The red points represent 300 random draws from the measured values of $M_{f1}$ and $M_{f2}$,
and their covariance matrix.
The likelihood is slightly narrower than the distribution of Monte-Carlo samples,
as it is further constrained by the measurement of \dtauWD. Although this means some
of the 300 samples have very low likelihood, evaluating
Equation~\eqref{eq:like_int} directly would require sampling the entire $M_{i1}$-$M_{i2}$ plane,
with almost all samples close to zero likelihood.
We reiterate that integration over the $M_{f1}$-$M_{f2}$ plane via importance sampling,
while drastically improving the computational efficiency of calculating the likelihood,
explicitly depends on the IFMR model being monotonic -- an assumption for which some recent works
suggest may not be justified (see Sections \ref{sec:nonmono} and \ref{sec:ifmr_comp}).

With a numerical recipe for calculating the integral in Equation~\eqref{eq:like_int_Mf},
we have everything we need to calculate the total likelihood in
Equation~\eqref{eq:likelihood_prod}. All that remains to determine the posterior distribution
on the IFMR, $P(\mathrm{IFMR}|\mathrm{DWDs})$ in Equation~\eqref{eq:posterior}, is
the prior distribution on the IFMR.
In the following sub-sections, we introduce extensions to this framework. The priors
on the associated hyper-parameters are defined therein. Specific choices of priors on
the IFMR and their effects on the results are given in Section~\ref{sec:results}. 

\subsection{Dealing with outliers}
\label{sec:outliers}

While some systems in our sample (e.g. \DWD{2223}{+}{2201}, Table~\ref{tab:dwds_fits})
clearly cannot obey an IFMR (i.e. two stars that formed together, evolving independently),
other outliers may be more subtle. It is therefore
insufficient to cherry-pick a-priori a sub-sample of our double white dwarfs
in order to fit the IFMR.
Instead we have implemented a mixture model \citep{hoggetal10-1} that naturally accounts for
the presence of outliers.

We first consider the likelihood given for a double white dwarf where both components
have evolved independently as single stars,
i.e. Equation~\ref{eq:likelihood_exp}, which we now refer to as $L_\mathrm{0}$.
For systems that have not followed this evolutionary path,
e.g. due to mergers or additional mass loss,
we make a simple modification to $L_\mathrm{0}$, instead assuming that the affected data
have been drawn from a distribution with a wider variance than implied by the
measured \dtauWD. Therefore in the covariance matrix, $\Sigma$,
we make the substitution $\sigma_{\dtauWD}^2 \mapsto
\sigma_{\dtauWD}^2 + \sigma_\mathrm{outlier}^2$,
where $\sigma_\mathrm{outlier}$ is a free parameter representing
the additional uncertainty in the cooling age difference.
We refer to this altered uncertainty as $L_\mathrm{outlier}$.
Finally these separate likelihoods are combined into a single likelihood, $L_\mathrm{total}$, where

\begin{equation}
    L_\mathrm{total} = (1-f_\mathrm{outlier})\times L_\mathrm{0}
    + f_\mathrm{outlier}\times L_\mathrm{outlier},
    \label{eq:like_outliers}
\end{equation}
and where $f_\mathrm{outlier}$ is a free-parameter representing the fraction of 
systems in our sample that are outliers.
This combined likelihood is used in all fits presented in later
sections. In terms of priors, we used a uniform distribution on $f_\mathrm{outlier}$
in the range 0 to 1, and a uniform distribution on
$\sigma_\mathrm{outlier}$ in the range 0 to 13.8\,Gyr.

\subsection{Dealing with underestimated uncertainties}
\label{sec:teff_logg_errs}

Uncertainties for spectroscopic fitting of DA white dwarfs are generally assumed
to be around a 1--2\,per\,cent in \Teff\ and a few 0.01\,dex in $\log g$
\citep{liebertetal05-1,vossetal07-1}.
However, when fitting high quality spectra such as ours (especially with the additional constraint
of \emph{Gaia} photometry), the formal uncertainties on the results are often found to be much
smaller than the numbers above.
Our analysis in Section~\ref{sec:results} is no exception with instances of \Teff\ uncertainties below $10$\,K
and $\log g$ errors of a few $0.001$\,dex (Table~\ref{tab:dwds_fits}).

The effect of ignoring such small uncertainties, is that the fitted IFMR may become overly constrained
by a few systems. Furthermore, in practice we found that this led to some objects that would consistently
yield a likelihood (Equation~\eqref{eq:likelihood_explicit}) of zero which would propagate through
to the total likelihood (Equation~\eqref{eq:likelihood_prod}). In the analysis of \citet{andrewsetal15-1},
the authors folded in \Teff\ relative uncertainties of 1.2\,percent, and $\log g$ uncertainties of
0.038\,dex according to \citet{liebertetal05-1}. While we could indeed adopt these same external uncertainties
and apply them to our analysis, we instead chose to estimate them as part of our fits to the IFMR,
including the relative \Teff\ uncertainty and $\log g$ uncertainty as free parameters to be
determined.

We implemented this in our model by adding these uncertainties in quadrature into the 
\Teff-$\log g$ covariance matrices for each system, at each step in the fit. These were then
converted to mass-\dtauWD\ covariance matrices via the mass-radius relations and
cooling age calculations of \citet{bedardetal20-1}.

To treat these extra parameters as part of our Bayesian framework also requires considering
their prior distributions. With no other
prior information, an uninformative Jeffreys prior can be used, i.e. $P(\sigma) \propto 1/\sigma$.
Fortunately we do have such extra information, as \DWD{1336}{-}{1620} was observed twice, with somewhat
different results on the \Teff\ and $\log g$, exceeding the internal fitting errors.

For $N$ observations of a quantity $x$ ($\ln\Teff$ or $\log g$) for both white dwarfs in a binary,
we consider the case where the values of $x$ are drawn from normal-distributions with
separate means ($\mu_A$ and $\mu_B$),
but a shared dispersion $\sigma$. Therefore the likelihood of our data
$\mathbf{x}$ is
\begin{equation}
    L(\mathbf{x}|\mu_A, \mu_B, \sigma) \propto \prod_i^N \sigma^{-2}
    \exp\left(-\frac{(x_{A,i}-\mu_A)^2}{2\sigma^2}
              -\frac{(x_{B,i}-\mu_B)^2}{2\sigma^2}
    \right).
\end{equation}
However, since we only care about the value of $\sigma$
we can marginalise over $\mu_{xA}$ and $\mu_{xB}$, yielding
\begin{equation}
    L(\mathbf{x}|\sigma) \propto \sigma^{-2(N-1)} \exp\left(-\frac{S}{2\sigma^2}\right),
\end{equation}
an inverse gamma-distribution in terms of $\sigma^2$ where
\begin{equation}
    S = \sum_{i=1}^{N} \left[(x_{A,i} - \langle x_A\rangle)^2
                           + (x_{B,i} - \langle x_B\rangle)^2 \right],
\end{equation}
and where $\langle x_A\rangle$ and $\langle x_B\rangle$ are the sample means.
Setting $N=2$ and multiplying by the same Jeffreys prior as before transforms
the likelihood into a posterior distribution on $\sigma$,
\begin{equation}
    P(\sigma|\mathbf{x}) \propto \sigma^{-3} \exp\left(-\frac{S}{2\sigma^2}\right),
    \label{eq:unc_prior}
\end{equation}
which in turn can be used as prior distribution on the uncertainty parameters when
fitting the IFMR.

For our systematic relative $\Teff$ uncertainty ($\sigma_T$),
and systematic $\log g$ uncertainty ($\sigma_g$),
our measured values of \DWD{1336}{-}{1620} led to $S_T=1.577\times10^{-4}$.
and $S_g=2.978\times10^{-3}$.
Consequently we found $\sigma_T=1.1_{-0.4}^{+1.1}$\,per\,cent and
$\sigma_g = 0.046_{-0.018}^{+0.046}$\,dex (median$\pm$16/84th percentiles),
with 95\,per\,cent highest density intervals of $0.3$--$4.0$\,per\,cent
and $0.015$--$0.172$\,dex, respectively.

The median values for repeat observations of \DWD{1336}{-}{1620}, are close to
those reported by \citet{liebertetal05-1} (see above).
Of course, systematic uncertainties in \Teff\ and $\log g$
can arise not only from the observations, but can also be stem from uncertainty in
the models. These priors only estimate uncertainty from the former, but by allowing
these additional variances to be free parameters, we allow additional sources
of uncertainty to be accounted for in the posterior distribution, should the data support it.

\subsection{IFMR functional form and parametrization}
Thus far we have referred to the IFMR as some function, but without specifying its form, nor
how it is parametrized.
A natural choice of function is piece-wise linear, since it is easy to enforce monotonicity,
as well as having a well defined inverse, and inverse gradient.

For an IFMR comprised of $N$ linear segments, this requires $N+1$ free parameters, though
the specific parametrization is implementation specific. For example the first segment
could be specified simply by a gradient and intercept, with subsequent segments specified only
by a gradient. Most published IFMRs are given in this format or similar
\citep[e.g.][]{williamsetal09-1,salarisetal09-1,cummingsetal18-1},
with each linear segment expressed in $y=mx+c$ form.
\citet{andrewsetal15-1} used an alternative formulation in their analysis,
specifying the angle of each segment makes in the $M_i$-$M_f$ plane, with a final free-parameter
specifying the perpendicular distance the extrapolated first segment makes with the origin.
In our analysis we simply chose to pre-specify a number of initial mass values, making
the corresponding final-mass values the free parameters, with a similar
approach also adopted by \citet{el-badryetal18-1}. This choice makes it easy
to check that the IFMR is monotonic, easy to invert the IFMR, and simple to calculate arbitrary
values between break points (via linear interpolation).

Rather than restrict ourselves to a few segments, with the initial-mass values of the break
points as free-parameters, we instead chose to use a much finer grid of fixed initial masses.
This allowed us to sample the IFMR across the entire initial-mass
range, thus permitting us to detect subtle features in the IFMR without imposing any
expectation on where they should occur. Furthermore, this would allow us to detect which regions
of the IFMR are well constrained by our data, and which regions remain uncertain.

After experimentation, we settled on an initial mass grid at 12 fixed values of initial mass
(11 segments), with grid points placed at 0.75--1.5\,\Msun\ 
in 0.25\,Msun\ steps, from 1.5--4.0\,\Msun\ in 0.50\,\Msun\ steps, from 4.0--6.0\,\Msun\ in 1.0\,\Msun\
steps, and with a final step at 8.0\,\Msun. This grid is used throughout for
the different fits described in the following subsections.

\subsection{Pre-white dwarf lifetimes}

The final ingredient required to constrain the IFMR is a relation between 
pre-white dwarf lifetimes and its dependence on initial-mass.
For this purpose, we used model grids from MIST
\citep[MESA Isochrones \& Stellar Tracks,][]{dotterMIST16,choiMIST16}
which themselves make use of MESA models
\citep[Modules for Experiments in Stellar Astrophysics,][]{paxtonetalMESA11,paxtonetalMESA13,paxtonetalMESA15,paxtonetalMESA18}.
Specifically pre-white dwarf lifetimes were determined from the Zero Age Main Sequence
(ZAMS) to the beginning of the Thermally-Pulsing AGB (TP-AGB),
with Solar metallicity (i.e. [Fe/H] = 0), and $v/v_\mathrm{crit}=0.4$.

Metallicity is known to affect both the pre-white dwarf lifetime and amount of mass of mass loss
occurring for a star of a given initial-mass \citep{mengetal08-1}, with higher metallicities
resulting in increased mass loss \citep{romeroetal15-1}. In cluster studies the progenitor metallicity
can be determined from the metallicity of other cluster members, and indeed can be found to
be non-Solar. While we cannot directly probe the metallicity for the progenitors of our DWD sample,
these objects are disk stars and therefore can be expected to have had values close to Solar
\citep{andrewsetal15-1}. While some of the systems in our sample must have total ages of many
Gyr, \citet{rebassa-mansergasetal21-1} recently investigated the age-metallicity relation
for a large sample of white dwarf-main sequence binaries,
finding no dependence of metallicity on their ages.
Therefore assuming Solar metallicity can be justified for the DWDs in our sample.
Of course, stars in the disc show a spread in metallicity either side of Solar,
however this represents a source of systematic uncertainty which we expect
to be absorbed by the parameters in our extended Bayesian model.

\subsection{Summary}

Here we briefly summarise the differences between our framework and the original framework
proposed by \citet{andrewsetal15-1}. Firstly Equation~\eqref{eq:likelihood_exp} accounts
for parameter covariances between $M_{f1}$, $M_{f2}$, and \dtauWD. This is particularly
important for our fits, since the two spectra of each DWD were fit simultaneously assuming
shared systematics. Even so, for systems with independent spectroscopic observations
(i.e. $M_{f1}$ and $M_{f2}$ are determined independently), \dtauWD\ will necessarily
covary with both masses, since cooling ages are sensitive to white dwarf masses.

Secondly, we make an important change to the importance sampling step.
\citet{andrewsetal15-1} chose to perform this integration over the space of
pre-WD lifetimes (see their Equation\,13).
In our opinion this choice adds unnecessary computational effort, as the
Jacobian will become the product of twice as many derivatives.
By performing the integration over initial masses,
as we have done in Equation~\eqref{eq:like_int},
we eliminate the need to calculate the gradient of the pre-WD lifetime function.
Subjectively, we also find the choice to integrate over initial-masses
to be conceptually simpler, since one generally imagines pre-WD lifetime to be a
function of the initial mass, rather than the other way around.

The other main differences to our model are the inclusions of outliers and unmodelled
uncertainty in the \Teff\ and $\log g$ of our fits. These are already 
discussed in detail in the preceding subsections.

\section{Results and discussion}
\label{sec:results}

\subsection{A Monotonic IFMR}
\label{sec:results_main}

For our main investigation of the IFMR, we chose a simple set of constraints. Chiefly,
we required the IFMR to be monotonic.
This choice allowed us to apply the importance-sampling integral
substitution in Equation~\ref{eq:like_int_Mf}, which
relies on the IFMR having both a well defined inverse and inverse
derivative.

In terms of other constraints, we only required the physically
realistic restriction of fractional mass-loss between 0 and 1, and that total
system ages were $< 13.8$\,Gyr (within the uncertainties). Specifically, we
did not enforce the final mass free-parameters (on the IFMR) to be below the Chandrasekhar-limit.
This is because, firstly, we would like to infer from our results at what
initial-mass is the Chandrasekhar-limit reached (if at all). Secondly, our highest
initial-mass point, at 8\,\Msun, is somewhat arbitrary, but since we require the IFMR
to be monotonic, limiting the final-mass points below 1.4\,\Msun, will therefore push all
preceding final-mass points down to lower values in an artificial way.
Nevertheless, we will investigate the effect of such a choice in Section~\ref{sec:results_Mch}.

We sampled the posterior distribution of our model described in Section~\ref{sec:model},
again using \texttt{emcee} \citep{foremanmackeyetal13-1} to perform an MCMC.
Due to the large number of free parameters (12 for the IFMR, plus 4 hyper-parameters),
we used an ensemble of 1000 walkers. For the importance sampling, we used 10\,000
samples per evaluation of the likelihood. We found that the fit converged in $\approx 10\,000$ steps
though we continued to 15\,000 steps to ensure a large number of samples could
be obtained post burn-in.

\begin{table}
    \centering
    \caption{\label{tab:resultsIFMR}
    Table of results for our three IFMR fits. The parameters in rows 5 onward correspond to the
    fixed initial masses with the tabulated values corresponding to the final masses at that point
    in the IFMR. Fit~1 refers to our primary fit with a monotonic IFMR. Fit~2 includes an additional
    prior where final mass samples are restricted below $M_\mathrm{Ch}=1.4$\,\Msun.
    Fit~3 instead includes the condition that mass-loss is also monotonic.
    }
    \begin{tabular}{lccc}
\hline
Parameter & Fit 1 & Fit 2 & Fit 3 \\
\hline
$f_\mathrm{outlier}$             & $0.59\pm0.21$             & $0.59_{-0.22}^{+0.23}$    & $0.59_{-0.22}^{+0.21}$    \smallskip\\
$\sigma_\mathrm{outlier}$ [Gyr]  & $0.70_{-0.22}^{+0.40}$    & $0.69_{-0.23}^{+0.41}$    & $0.72_{-0.23}^{+0.36}$    \smallskip\\
$\sigma_{T_\mathrm{eff}}$ [$\%$] & $1.36_{-0.63}^{+2.09}$    & $1.34_{-0.62}^{+2.23}$    & $1.38_{-0.65}^{+2.05}$    \smallskip\\
$\sigma_{\log g}$ [dex]          & $0.049_{-0.006}^{+0.007}$ & $0.049_{-0.006}^{+0.007}$ & $0.049\pm0.006$           \smallskip\\
\hline                                                                                      
$0.75$\,\Msun                    & $0.297_{-0.204}^{+0.182}$ & $0.295_{-0.200}^{+0.176}$ & $0.485_{-0.050}^{+0.045}$ \smallskip\\
$1.00$\,\Msun                    & $0.552_{-0.018}^{+0.015}$ & $0.551_{-0.019}^{+0.015}$ & $0.552_{-0.017}^{+0.015}$ \smallskip\\
$1.25$\,\Msun                    & $0.595_{-0.013}^{+0.011}$ & $0.594_{-0.012}^{+0.011}$ & $0.594_{-0.012}^{+0.011}$ \smallskip\\
$1.50$\,\Msun                    & $0.614_{-0.009}^{+0.008}$ & $0.613_{-0.008}^{+0.009}$ & $0.613_{-0.009}^{+0.009}$ \smallskip\\
$2.00$\,\Msun                    & $0.632_{-0.011}^{+0.013}$ & $0.631_{-0.011}^{+0.011}$ & $0.630_{-0.010}^{+0.013}$ \smallskip\\
$2.50$\,\Msun                    & $0.666_{-0.021}^{+0.027}$ & $0.657_{-0.017}^{+0.024}$ & $0.658_{-0.018}^{+0.024}$ \smallskip\\
$3.00$\,\Msun                    & $0.727_{-0.031}^{+0.036}$ & $0.711_{-0.030}^{+0.027}$ & $0.711_{-0.030}^{+0.027}$ \smallskip\\
$3.50$\,\Msun                    & $0.803_{-0.052}^{+0.056}$ & $0.760_{-0.029}^{+0.040}$ & $0.763_{-0.029}^{+0.034}$ \smallskip\\
$4.00$\,\Msun                    & $0.861_{-0.032}^{+0.033}$ & $0.835_{-0.041}^{+0.027}$ & $0.828_{-0.045}^{+0.030}$ \smallskip\\
$5.00$\,\Msun                    & $0.909_{-0.037}^{+0.134}$ & $0.875_{-0.024}^{+0.026}$ & $0.872_{-0.025}^{+0.026}$ \smallskip\\
$6.00$\,\Msun                    & $1.236_{-0.290}^{+2.305}$ & $0.912_{-0.032}^{+0.076}$ & $0.909_{-0.030}^{+0.059}$ \smallskip\\
$8.00$\,\Msun                    & $4.679_{-3.012}^{+2.311}$ & $1.101_{-0.102}^{+0.172}$ & $1.053_{-0.067}^{+0.107}$ \\
\hline
\end{tabular}

\end{table}

\begin{figure}
    \includegraphics[width=\columnwidth]{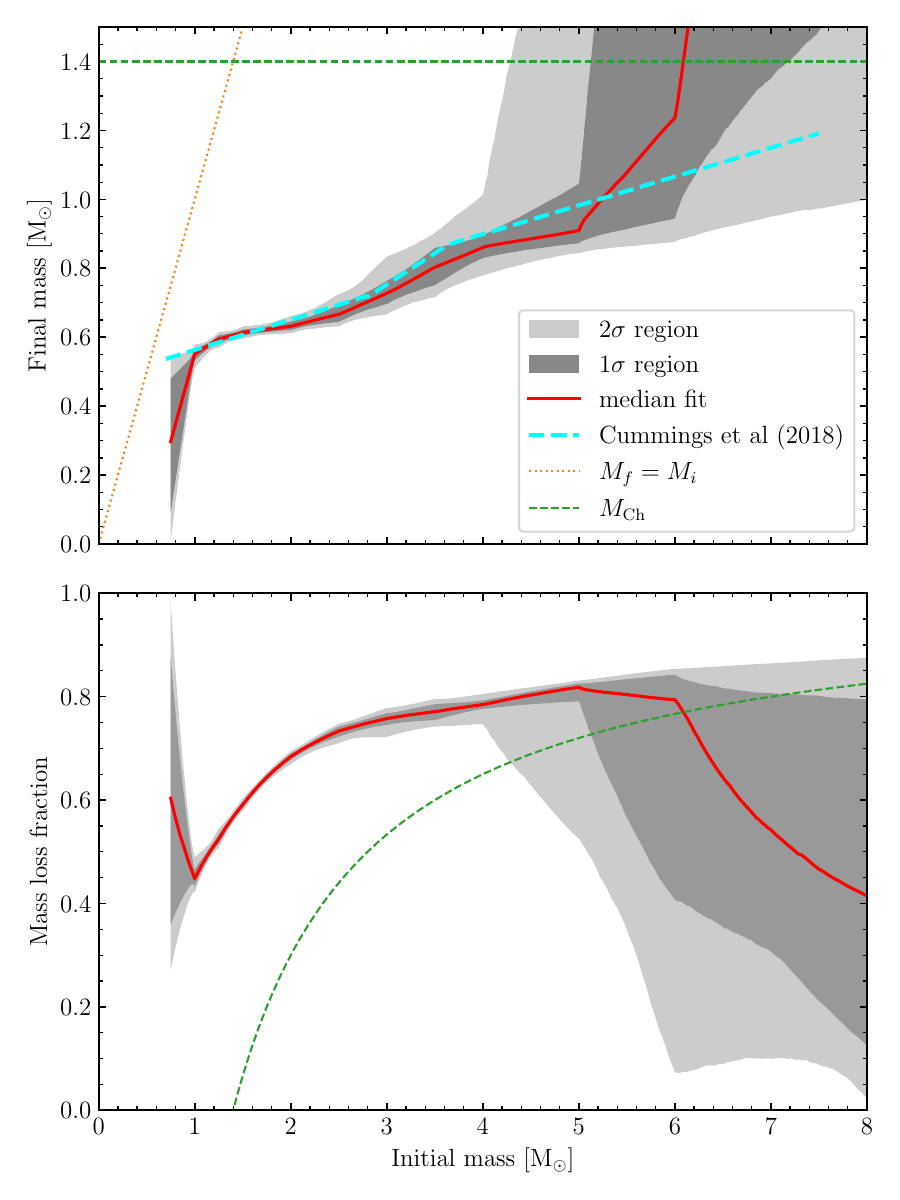}
    \caption{
    Our best fitting IFMR (red) from 0.75--8.0\,\Msun, with the constraints that the IFMR
    is monotonic, and mass loss is bounded between 0 and 1 (Fit~1). Furthermore the presence of
    outliers and underestimated \Teff\ and $\log g$ uncertainties are accounted for.
    The $1\sigma$ and $2\sigma$ uncertainty regions are indicated by the medium and light
    grey areas, respectively. Good agreement with the IFMR of \citet{cummingsetal18-1}
    to within $1\sigma$ is observed between 1--5\,\Msun.
    }
    \label{fig:IFMR_main}
\end{figure}

The resulting IFMR is shown in Figure~\ref{fig:IFMR_main}, along with the mass-loss fraction
occurring as a function of initial mass. This is compared with the IFMR from \citet{cummingsetal18-1}
(a wider comparison against other published IFMRs is performed in Section~\ref{sec:ifmr_comp}).
All results are given in the Fit~1 column of
Table~\ref{tab:resultsIFMR}. A corner plot of converged parameters is shown in
Figure~\ref{fig:cornerFit1}. In general we find good agreement between 1--5\,\Msun,
with extremely tight constraints for initial masses
of 1--2\,\Msun\ with $M_f$ uncertainties of 0.01--0.02\,\Msun. The steepening gradient occurs
in broadly the same region as that found by \citet{cummingsetal18-1}, though we found this
section of the IFMR to be somewhat wider being located at 2.5--4.0\,\Msun\ (compared with
2.85--3.60\,\Msun\ for \citealt{cummingsetal18-1}). From 4--5\,\Msun, we find our median result
is flatter than \citet{cummingsetal18-1}, though still within our widening $1\sigma$ contour.

For initial-masses above 6.0\,\Msun\ we have virtually no constraint on the IFMR due to the absence
of white dwarfs with final masses significantly above 1.0\,\Msun\ in our fitted sample.
Specifically in the mass-loss panel, the MCMC samples are seen to have filled the entire
available parameter space, subject to the condition of bounded mass-loss and a monotonic IFMR.
Similarly, for initial masses below 1.0\,\Msun\ we have effectively no data covering this
region, and so again the fit is very poor. This is most obvious in the mass-loss panel,
demonstrating that a wide range of mass-loss fractions are covered by our $2\sigma$ contour at
an initial mass of 0.75\,\Msun.

For the hyper-parameters, firstly, we found a particularly large (though poorly constrained)
$f_\mathrm{outlier}$ of $0.59\pm0.21$, and with a corresponding $\sigma_\mathrm{outlier}$ of
$0.70_{-0.22}^{+0.40}$\,Gyr. The interpretation of $f_\mathrm{outlier}$ is discussed in more
detail in Section~\ref{sec:mi_Poutlier}, but essentially is constrained only by a few
systems with high white dwarf masses. For the systematic uncertainty parameters, we found
$\sigma_{\Teff}=1.36_{-0.63}^{+2.09}$\,per\,cent, and $\sigma_{\log g}=0.049_{-0.006}^{+0.007}$\,dex.
For $\sigma_{\Teff}$, this remains similar to the prior distribution evaluated in
Section~\ref{sec:teff_logg_errs}, though with a slightly higher median value, and larger dispersion.
For $\sigma_{\log g}$ on the other hand, the distribution was found to be almost Gaussian,
with small uncertainty, indicating this parameter is particularly sensitive to our
complete set of data. 
These values are in close agreement to the $\sigma_{\Teff}=1.4$\,per\,cent
and $\sigma_{\log g}=0.042$\,dex found by \citet{genest+bergeron19-1}.
Converting these uncertainties into masses and cooling ages, we found
that typical white dwarf mass errors of 0.03\,\Msun, and cooling age relative uncertainties
of 5--20\,per\,cent. Importantly, for these four hyper-parameters, the corner-plot (Figure~\ref{fig:cornerFit1})
shows no obvious covariance with the IFMR parameters, meaning small changes to the hyper-parameters
do not strongly affect the IFMR itself.

\subsection{Chandrasekhar mass limit}
\label{sec:results_Mch}

In the previous section, it is clear that our IFMR is poorly constrained for $M_i > 5$\,\Msun,
due to an absence of data with $M_f > 1.0$\,\Msun. As a result, for $M_i$ in the range 5--8\,\Msun,
our posterior distribution easily fills the entire parameter space between our priors of a
monotonic IFMR prior and mass loss fractions between 0 and 1.
Naturally this means that many of the samples exceed the Chandrasekhar limit to an extreme degree.

Our code allows the Chandrasekhar-limit to easily be imposed as a prior on the $M_f$ values at
each fixed $M_i$. We therefore repeated our fit to the IFMR with this extra constraint enforced.
Because of the reduced phase-space our MCMC could explore, the model converged faster, only
taking about 6000 steps, though we ran to 10\,000 for the same reasons as before.

\begin{figure}
    \includegraphics[width=\columnwidth]{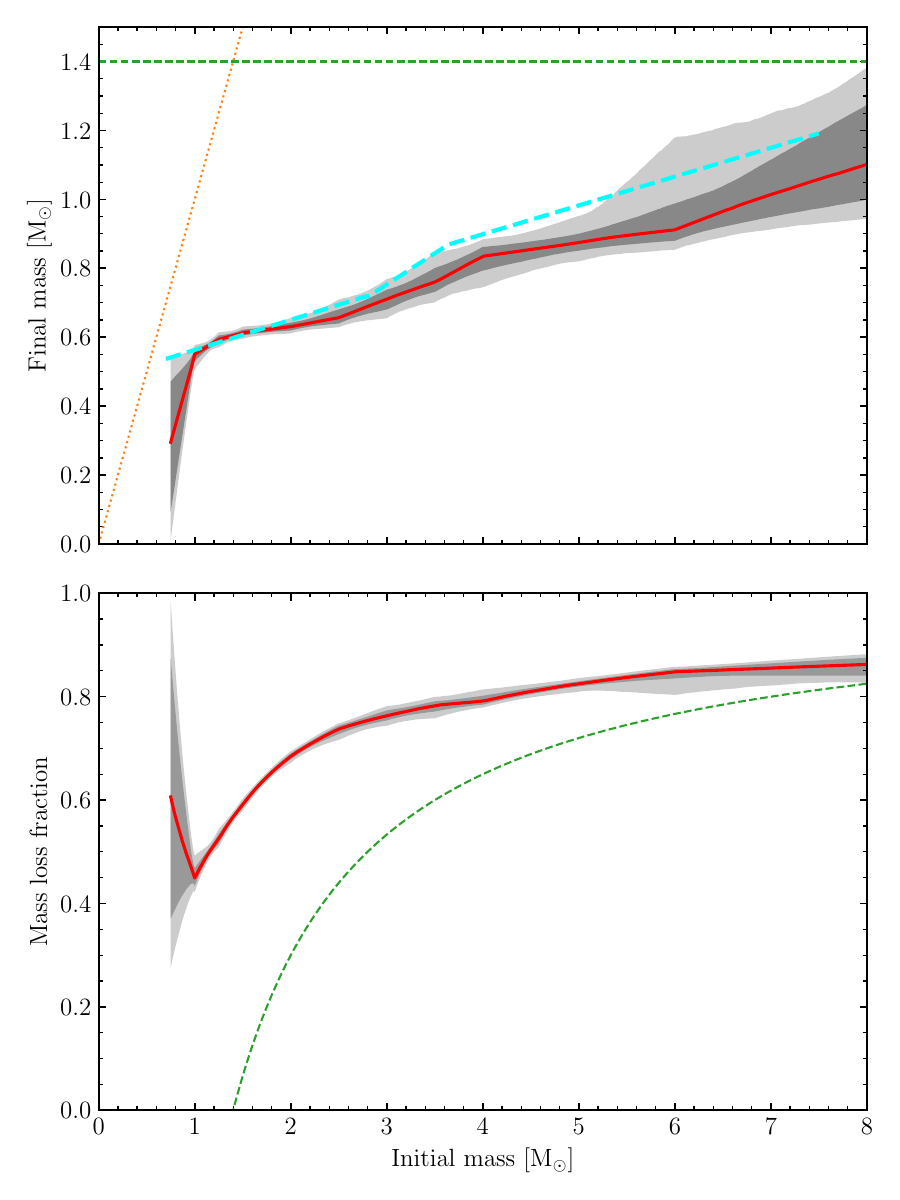}
    \caption{
    Our best fitting IFMR, including the additional constraint that all final
    mass points should be below 1.4\,\Msun\ (Fit~2). Figure elements have the same meaning
    as in Figure~\ref{fig:IFMR_main}.
    }
    \label{fig:IFMR_Ch}
\end{figure}

The resulting best fit is shown in Figure~\ref{fig:IFMR_Ch}, and tabulated in
column Fit~2 in Table~\ref{tab:resultsIFMR}. The four hyperparameters are virtually
unchanged, with differences only seen for the IFMR final mass parameters.
At the low mass end, where $M_i<2$\,\Msun,
essentially no difference is seen between this new IFMR and the one presented in
Figure~\ref{fig:IFMR_main} (Fit~1). Of course, at the high mass end, where the IFMR was
previously unconstrained, the $1\sigma$ and $2\sigma$ contours appear relatively tight
with $M_f=1.101_{-0.102}^{+0.172}$ at $M_i=8$\,\Msun. However, as predicted, this has
also had an affect on the intermediate range. Because we have imposed a monotonic IFMR,
forcing the high mass end to be below 1.4\,\Msun, has by extension also pushed down
the final mass values at intermediate initial masses by about $1\sigma$. The result is
that this IFMR now disagrees with the \citet{cummingsetal18-1} IFMR by more than $2\sigma$
between initial masses of 3.5--5.5\,\Msun. We therefore recommend using the IFMR
(Fit~1) over this one (Fit~2), and only for initial masses below 5\,\Msun.

\subsection{Monotonic mass loss}
\label{sec:results_monoML}

Inspecting Figure~\ref{fig:IFMR_main}, it is apparent that over the well constrained range (1--5\,\Msun), 
mass-loss also appears to be monotonic with respect to initial mass.
This trend only breaks down at higher masses where we are
unconstrained by data, and so our MCMC samples the entire parameter space. We therefore introduced the additional assumption that not only the IFMR is monotonic, but so also is mass-loss as a function of initial mass. To investigate the effect of imposing this restriction, we repeated the fit from Section~\ref{sec:results_main}, the same
number of walkers (1000) and marginalisation samples (10\,000) were used running for 10\,000 steps.
The only change compared to Fit~1 was that we imposed a prior constraining the mass loss fraction
to also be monotonic.

\begin{figure}
    \includegraphics[width=\columnwidth]{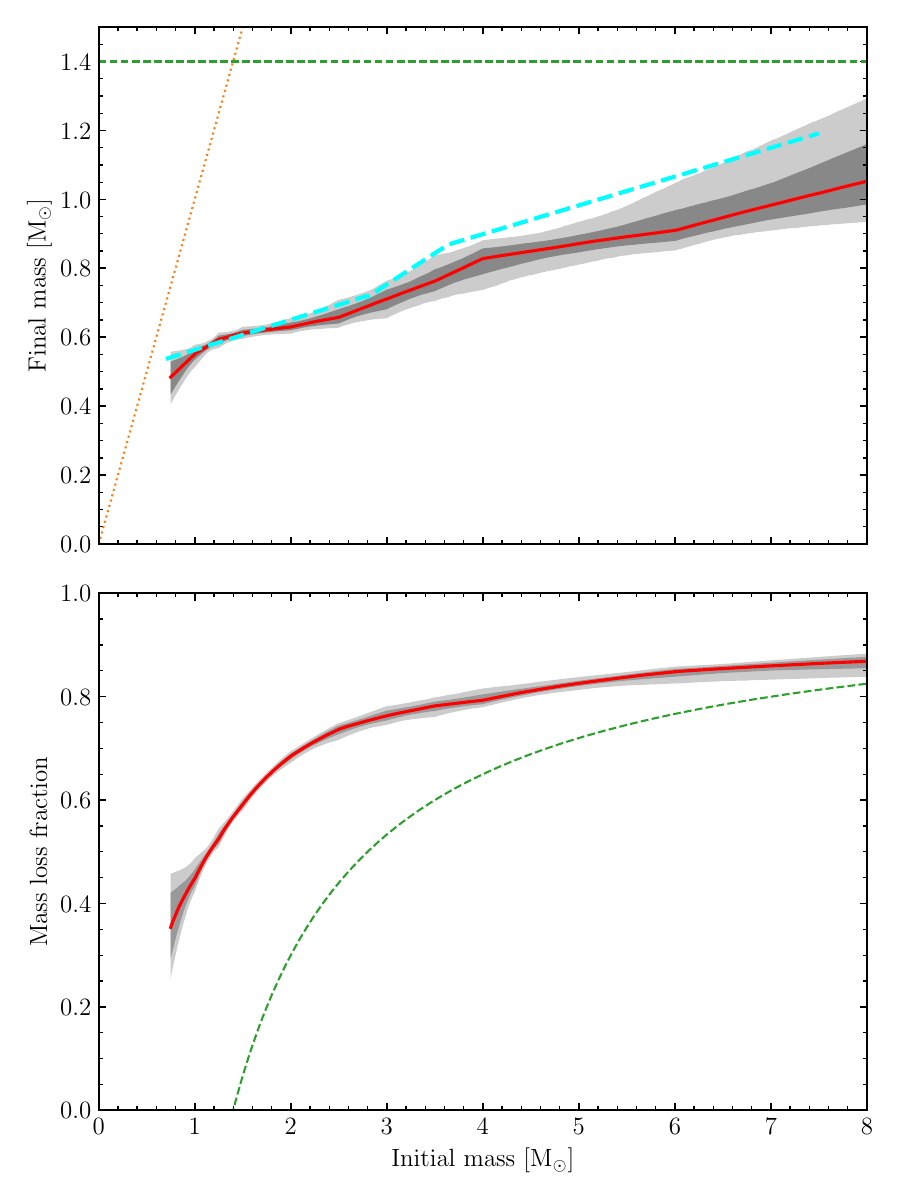}
    \caption{
    Our best fitting IFMR including the constraint that mass-loss is also monotonic with initial mass (Fit~3).
    Figure elements have the same meaning as in Figure~\ref{fig:IFMR_main}.
    }
    \label{fig:IFMR_MonoML}
\end{figure}

The resulting IFMR is shown in Figure~\ref{fig:IFMR_MonoML},
and with the results tabulated as Fit~3.
Compared to our results with non-monotonic mass-loss (Figure~\ref{fig:IFMR_main}),
the IFMR is pushed down substantially at the high-mass end (even more so than in Fit~2),
with a final mass value of $1.06_{-0.07}^{+0.11}$\,\Msun\ at 8.0\,\Msun. Therefore, an IFMR with
this assumption is unable to produce near Chandrasekhar-mass white dwarfs through single-star evolution.
Furthermore, between 3.0--6.5\,\Msun, we find more than $2\sigma$ disagreement with the IFMR of
\citet{cummingsetal18-1}, implying that the constraint of monotonic mass-loss is unlikely.
Consequently it is possible that the mass-loss fraction does
in fact peak somewhere between 5--8\,\Msun. Evidently this result, and that of Fit~2,
demonstrate that care must be taken when choosing priors, as these can have an unexpectedly large
effect on the results.

At the low mass end, between 0.75--1.0\,\Msun, this IFMR is more tightly constrained than
in either Fit~1 or Fit~2, as IFMRs with high mass loss at low initial-mass are ruled out, 
with the median line appearing to tend towards 0.
Although, very poorly constrained by our data (only our priors), this result is somewhat realistic
as the lowest mass stars -- all of which are still on the main sequence given their long lifetimes --
will evolve directly into white dwarfs without significant mass-loss once their hydrogen fuel
is depleted, due to the reduced ability or inability to ignite helium burning.
Therefore, users of our results may wish to use the parameters from Fit~3 when estimating
the initial masses of low mass white dwarfs, though these results are clearly unrealistic
at higher masses, where parameters from Fit~1 should be used instead.

\subsection{Non-monotonic IFMR}
\label{sec:nonmono}

Although we restrict ourselves here to monotonic IFMRs only, we note that our code does permit
non-monotonic IFMRs to be fitted. However, this requires changing the domain of integration
from the final-final mass plane to the initial-initial mass plane, where it is not possible
to restrict the area of integration to a region smaller than the whole plane up to 8\,\Msun\ for
both components. Therefore in order to maintain precision
of the integrals the number of integration samples should be increased by a factor 100 or more.

Given these computational requirements, we consider such a task beyond
the scope of this work at present. However we note that recent work by
\citet{marigoetal20-1} and \citet{marigoetal22-1}
suggest that the IFMR may indeed be non-monotonic
over a small mass range (1.8--2.2\,\Msun), and may be responsible for the formation of carbon stars.
A more detailed comparison with this result is given in the following section.

\subsection{Comparison with other IFMRs}
\label{sec:ifmr_comp}

\begin{figure*}
    \includegraphics[width=\textwidth]{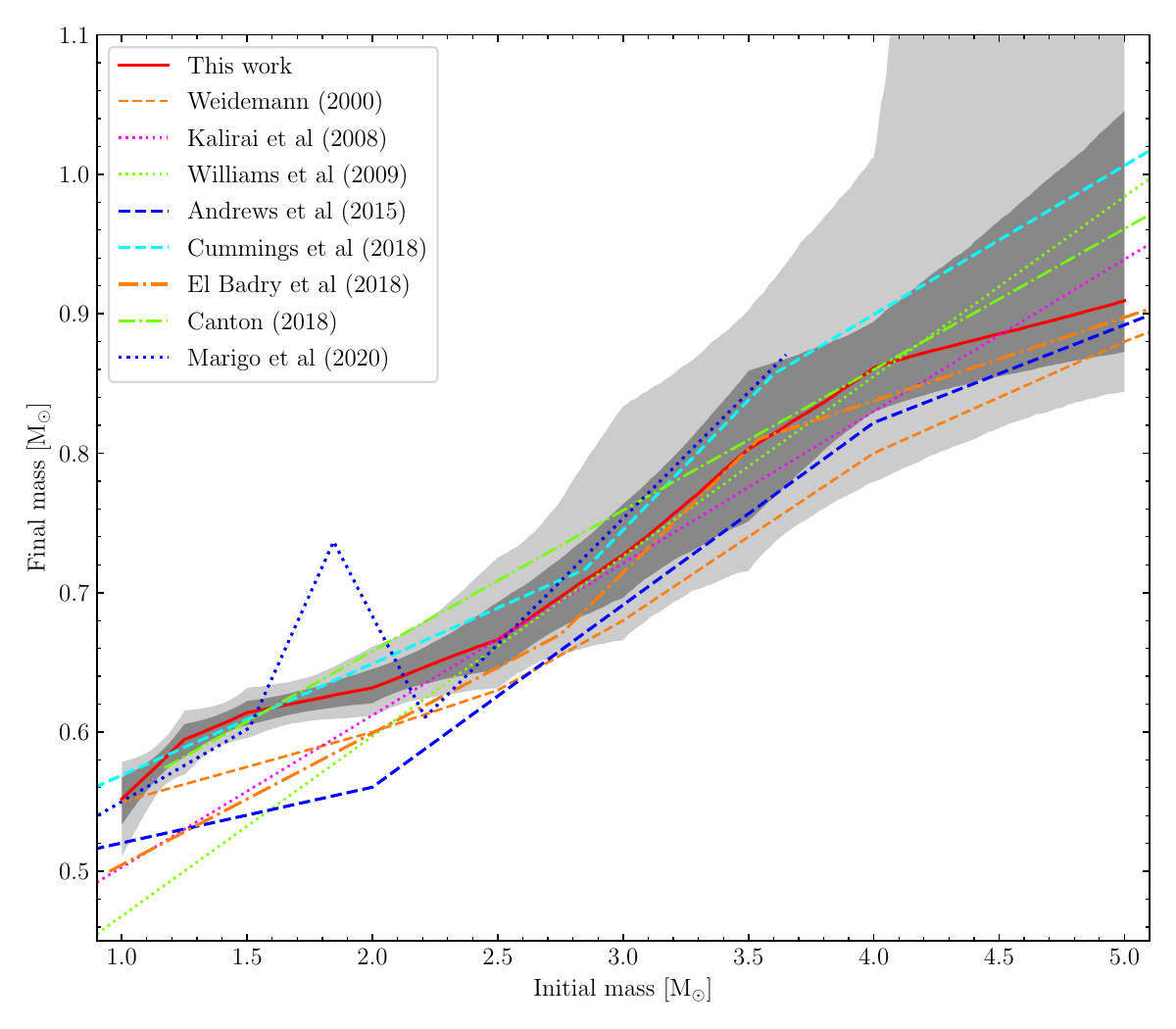}
    \caption{
    Comparison of our best fitting IFMR (Fit~1, red line) to those from other works, focusing on the range
    1--5\,\Msun. The grey regions have the same meaning as in Figure~\ref{fig:IFMR_main}. Note that the
    uncertainties on the comparison IFMRs are not shown.
    }
    \label{fig:IFMR_multi}
\end{figure*}

In Sections~\ref{sec:results_main}--\ref{sec:results_monoML}, we presented our different IFMR fits
while providing a comparison only to the IFMR of \citet{cummingsetal18-1}. Of course many other works
in recent years have presented other IFMRs, which we also wish to compare to.
In Figure~\ref{fig:IFMR_multi},
we compare our best fitting results to a variety of published IFMRs over the range 1--5\,\Msun, where
our results are well constrained.
This includes several cluster based IFMRs
\citep{weidemann00-1,kaliraietal08-1,williamsetal09-1,cummingsetal18-1,cantonPhD18},
the \emph{Gaia} population based IFMR of \citet{el-badryetal18-1},
and the fiducial model of \citet{andrewsetal15-1} using DWDs.
As with our own IFMR, many of these works also make the assumption of a
monotonic IFMR \citep{el-badryetal18-1, andrewsetal15-1}, or are comprised of only a few
linear segments and so do not result in negative gradients over the wide
initial-mass range covered by each segment.
We also show the non-monotonic IFMR of \citet{marigoetal20-1}.

Over the range 2.5--5.0\,\Msun, we observe agreement with all comparison IFMRs, within our 2$\sigma$ shaded region.
Above initial masses of 3.0\,\Msun, the comparison IFMRs show even better agreement,
with the majority within the 1$\sigma$ shaded region.
At these higher masses, we generally have less data to constrain the IFMR, so it is perhaps no surprise that we find
the greatest agreement where our constraints are weakest. Even so, at $M_i=4.0$\,\Msun, our final masses are still
constrained to within $\pm0.033$\,\Msun.

The largest disagreement is observed between $1.0$--$2.5$\,\Msun, depending on the IFMR. For the cluster results
of \citet{cummingsetal18-1} and \citet{cantonPhD18}, we continue to find reasonable agreement over this range, 
noting that the latter consists of a single linear relation. Compared to the other cluster IFMRs
\citep{weidemann00-1,kaliraietal08-1,williamsetal09-1}, we find final masses that are about 0.05\,\Msun\ higher. However,
 these are the oldest studies we show; the two more recent cluster analyses are consistent with our results,
which we take that as a sign that our results are accurate in this range.

We also find a similar level of disagreement with the result of \citet{el-badryetal18-1}, which was based
on modelling the distribution of DA white dwarfs in the \emph{Gaia} DR2 Hertzsprung-Russel diagram.
At the low mass end, their IFMR has similar level of uncertainty reported, and closely follows the results
of \citet{kaliraietal08-1} over the whole initial mass range shown.

Somewhat surprisingly, at $M_i=2.0$\,\Msun, our results show the largest disagreement
with \citet{andrewsetal15-1}, despite using the same general methodology.
Even at higher masses, our results remain about $1\sigma$ above those of \citet{andrewsetal15-1}.
Of course our results have benefited from a larger sample size,
with parallaxes and precise photometry from \emph{Gaia} DR3 also providing more accurate spectroscopic parameters.
Since we have taken great care to account for sources of systematic uncertainties in both the spectroscopic fits
and in the IFMR fitting, we expect that our IFMR ought to be more accurate, and with realistic uncertainties.

Finally we observe clear disagreement with \citet{marigoetal20-1}, where their IFMR
peaks near 1.8\,\Msun. This non-monotonic IFMR was invoked following the discovery of cluster white dwarfs
with low initial masses (1.6--2.1\,\Msun). In particular, a white dwarf in NGC\,752 was found with
a lower final mass, but higher initial mass than members of R-147 and NGC\,7789, implying a downwards
trend in the IFMR, before turning upwards again.
Since our fits assume a monotonic IFMR as a prior, our model cannot reproduce this feature.
However, if the IFMR kink was present in our sample, we would expect our resulting IFMR to
have been forced upwards through the middle of the triangular feature, rather than across its base.
This is especially true, given that the 0.6--0.75\,\Msun\ range in final masses is covered by a large
fraction of our DWD sample (Figure~\ref{fig:mdist}). Given that the work by \citet{marigoetal20-1}
required observations of particularly old clusters, we can only speculate that those cluster members
may not be representative of the disk stars in our analysis.
As discussed in Section~\ref{sec:nonmono}, in principle our code has the ability to fit
non-monotonic IFMRs, but at greatly increased computational cost, opening
up the possibility to investigate the reported IFMR kink with a larger number of DWDs in the future.

\subsection{Determining progenitor masses of white dwarfs}
\label{sec:prog_mass}

We consider the common case of deriving progenitor masses from
the masses of isolated white dwarfs and an IFMR. The simplest approach is to draw normally distributed
samples of the white dwarf mass from a measurement and its uncertainty, and then
use the inverse-IFMR to determine the corresponding initial mass. This can be repeated
over all IFMR samples to account for uncertainty in the IFMR.

A better alternative is to consider the likelihood of a final mass, given its uncertainty,
an initial mass, and an IFMR,
\begin{equation}
    P(M_f | \mathrm{IFMR}, M_i, \sigma_{M_f}) \propto \frac{1}{\sigma_{M_f}} \exp\left(-\frac{1}{2}\left[\frac{M_f-\mathrm{IFMR}(M_i)}{\sigma_{M_f}}\right]^2\right).
    \label{eq:like_mf}
\end{equation}
This likelihood can be turned into a posterior distribution on $M_i$ by multiplying by a prior distribution on $M_i$.
In the absence of data to constrain $M_i$, the only assumption that can be made is that the star
was drawn from the mass distribution of all main sequence stars, i.e. the IMF.
We therefore adopt the high-mass part of the IMF (valid for $M_i > 0.5$\,\Msun) as a prior,
so that $P(M_i) \propto M_i ^{-2.3}$ and
\begin{equation}
    P(M_i | \mathrm{IFMR}, M_f, \sigma_{M_f}) \propto M_i^{-2.3}
    P(M_f | \mathrm{IFMR}, M_i, \sigma_{M_f}).
    \label{eq:post_mi}
\end{equation}
However, this distribution depends on a specific IFMR (as opposed to a distribution of IFMRs),
and so we must marginalise over the entire space of IFMRs,
i.e. samples from the posterior in Section~\ref{sec:results_main}.
Numerically, this equates to taking the arithmetic mean over IFMR samples
\begin{equation}
    P(M_i | M_f, \sigma_{M_f}) =
    \frac{1}{N} \sum_k^N P(M_i | \mathrm{IFMR}_k, M_f, \sigma_{M_f}),
    \label{eq:post_mi2}
\end{equation}
where the summand must be normalised over the range 0.75 to 8\,\Msun,
to provide correct weighting over each $P(M_i | \mathrm{IFMR}_k, M_f, \sigma_{M_f})$.
Since the resulting distribution is one dimensional, it is simple to draw $M_i$ samples from the
posterior without resorting to more complex methods such as MCMC.

\begin{figure}
    \includegraphics[width=\columnwidth]{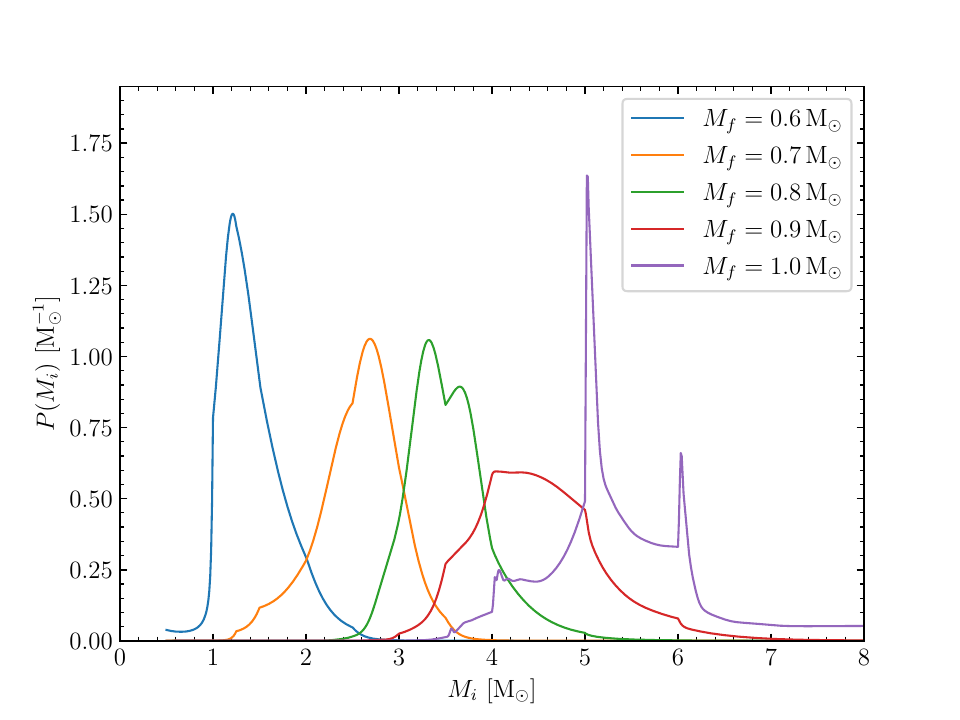}
    \caption{
    Posterior distributions of initial mass for a range of final masses,
    and with assumed final mass uncertainties of $0.03$\,\Msun. Marginalisation
    over IFMRs uses the final MCMC step of our fit in Section~\ref{sec:results_main},
    with 1000 samples.
    }
    \label{fig:mi_distr}
\end{figure}

Posterior distributions from our Bayesian approach are shown in Figure~\ref{fig:mi_distr}, using
a final mass uncertainty of 0.03\,\Msun\
(typical for our DWD sample, when systematic uncertainties are accounted for). 
The distributions for $M_f = 0.6$ and $0.7$\,\Msun, appear particularly well-behaved given
that they correspond to the most tightly constrained part of our IFMR. While the
distributions for $M_f=0.8$ and $0.9$\,\Msun\ appear further from normal, they are still
adequate to determine 16th, 50th, and 84th percentiles and draw conclusions about the initial
mass. For $M_f=1.0$\,\Msun, sharp features appear in the posterior, resulting from a lack of
constraining data, and poor sampling by our IFMR break points.
We therefore recommend using our IFMR results over the range $0.53$--$0.95$\,\Msun\ 
(for lower masses, the implied median pre-WD lifetimes exceed the age of the universe).

In comparison to the simple approach described earlier, we found the Bayesian approach
yields more robust results. For example, using $M_f=0.60\pm0.03$\,\Msun, the simple approach
gives $M_i=1.34_{-0.23}^{+0.67}$\,\Msun, which is similar to the $1.36_{-0.24}^{+0.42}$\,\Msun\
found for the Bayesian method (Figure~\ref{fig:mi_distr}), but with slightly
worse uncertainty. However, inspecting the distribution of $M_i$ samples, the simple
approach also shows spikes or discontinuities at each break point in the IFMR
(similar to what is seen for $M_f=1.0$\,\Msun\ in Figure~\ref{fig:mi_distr}),
which are not present for the Bayesian approach in this mass
range. We provide a tool within our \textsc{python} package for calculating these distributions
from white dwarf masses.

We conclude by noting that Equation~\ref{eq:like_mf} relies on the IFMR only in the foward direction,
making no reference to the inverse-IFMR,
and so this approach is equally valid for non-monotonic IFMRs.
While the resulting $M_i$ distribution will inevitably be multi-modal,
these modes will be weighted by the IMF.

\subsection{Initial masses and outlier probabilities for double white dwarfs}
\label{sec:mi_Poutlier}

Within our Bayesian framework for fitting the IFMR (Section~\ref{sec:model}), we marginalised
over initial masses in order to keep the number of free parameters to an acceptable level.
However, it is still possible to calculate these after the fit. However, we cannot simply
apply the methodology from Section~\ref{sec:prog_mass} to each component separately.
Firstly because $M_{i1}$ and $M_{i2}$ may have correlated uncertainties, secondly
because we also have the constraint on cooling ages and their difference, and thirdly
because we must consider the effect of the derived hyper-parameters on our sample.

To account for these caveats, we modify Equation~\ref{eq:post_mi2} to
\begin{equation}
    P(M_{i1},M_{i2}|\mathrm{DWD},\Sigma) = \frac{1}{N} \sum_k^N
    P(M_{i1},M_{i2}|\theta_k, \mathrm{IFMR}_k, \mathrm{DWD},\Sigma),
\end{equation}
where $\theta_k$ is a vector of hyperparameters, and $\Sigma$ is the covariance matrix of the DWD
parameters. We use the same $M_f$ priors and likelihood as in Sections~\ref{sec:model} and \ref{sec:outliers}.
While the resulting distribution is now 2D, it is still possible to sample the $M_{f1}$-$M_{f2}$ plane
on a fine grid, and calculate the posterior probability at each point. $M_{i1}$-$M_{i2}$ samples
can then be drawn from this grid weighted according to the posterior. We give results for the estimated
$M_{i1}$, $M_{i2}$ values and their covariance in Table~\ref{tab:m_tau_p}. 

Similarly, we can determine the probability that a specific DWD from our sample is an outlier,
$P_\mathrm{outlier}$, after having fitted our IFMR.
This requires converged samples of the IFMR and the hyper-parameters, $\theta$,
from our fit. For each sample we calculate the outlier likelihood, $L_\mathrm{outlier}$
and the total likelihood $L_\mathrm{total}$ (defined according to Equation~\eqref{eq:like_outliers}),
but where both likelihoods have been marginalised over the $M_{i1}$-$M_{i2}$ plane
(using importance sampling from Section~\ref{sec:model}).
Finally, to determine the probability that a specific system is an outlier, we simply require
marginalising over our samples of fitted parameters according to
\begin{equation}
    P_\mathrm{outlier} = \frac{1}{N} \sum_k^N
    \frac{f_{\mathrm{outlier},k}\times L_{\mathrm{outlier}}(\mathrm{DWD}|\theta_k,\mathrm{IFMR}_k,\Sigma)}
    {L_{\mathrm{total}}(\mathrm{DWD}|\theta_k,\mathrm{IFMR}_k,\Sigma)},
\end{equation}
where $f_\mathrm{outlier}$ is the hyper-parameter corresponding to the fraction of systems that are outliers.
These $P_\mathrm{outlier}$ will have values close to 1, if the two components in a system are not coeval,
i.e. $\dtauWD \neq -\dtauMS$ considering the uncertainties on both \dtauWD\ and \dtauMS.
This necessarily means that we require high precision on both \dtauWD\ and \dtauMS\ in order
to be confident that a system is an outlier, or that both its components are coeval.
We provide our estimate for $P_\mathrm{outlier}$ in Table~\ref{tab:m_tau_p}.

The first thing to notice is that the vast majority of systems have $P_\mathrm{outlier}$ within
one per\,cent of the mean value of $f_\mathrm{outlier}$. This is because for \emph{most} systems,
the uncertainty on the observed \dtauWD\ dominates the 0.7\,Gyr of additional variance found for outlier
systems (particularly once the systematics on \Teff\ and $\log g$ are considered),
and so our methodology cannot confidently conclude whether additional variance is required.
This begs the question of how $f_\mathrm{outlier}$ should be interpreted, and why its value is so high.

\begin{figure}
    \includegraphics[width=\columnwidth]{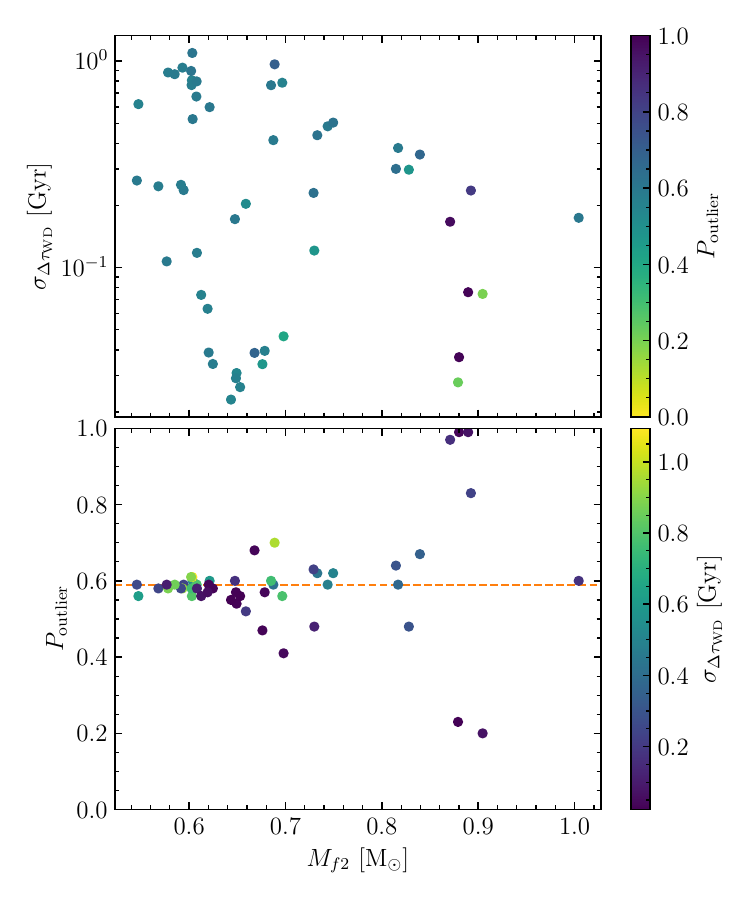}
    \caption{Outlier probabilities as for our 52 DWDs used to fit the IFMR. This is shown as a function
    of the mass of the heaviest component in the binary, $M_{f2}$, and the uncertainty in the cooling
    age difference (including \Teff\ and $\log g$ systematics), $\sigma_{\Delta\tau_\mathrm{WD}}$.
    In the bottom panel, $P_\mathrm{outlier}$ and $\sigma_{\Delta\tau_\mathrm{WD}}$ are inverted,
    and the median value of $f_\mathrm{outlier}$ is shown by the dashed orange line.
    }
    \label{fig:Mf2_dtau_p}
\end{figure}

We demonstrate where this outlier probability arises from in Figure~\ref{fig:Mf2_dtau_p},
where we show the outlier probabilities for each system as a function of the more massive
white dwarf in the pair, $M_{f2}$, and the uncertainty on \dtauWD\, with assumed \Teff\
and $\log g$ systematic uncertainties of 1.4\,per\,cent and 0.052\,dex included, respectively.
Across most of the parameter space, the $P_\mathrm{outlier}$ are almost identical to
the central value found for $f_\mathrm{outlier}$ (0.59). Only six systems
stand out from the crowd. These are \DWD{0101}{-}{1629}, \DWD{1215}{+}{0948}
\DWD{2115}{-}{0741}, and \DWD{2223}{+}{2201} with high $P_\mathrm{outlier}$; and
\DWD{1313}{+}{2030} and \DWD{1336}{-}{1620} with low $P_\mathrm{outlier}$.
Inspecting the parameters of the high outlier probability systems it is easy to
see how this classification was made, for example \DWD{0101}{-}{1629}, the more
massive component has been cooling for about 490\,Myr, whereas the lower mass
companion has a cooling age of 720\,Myr -- the inverse of what is expected
for two white dwarfs born at the same time, but evolving independently as single stars.

All six of these systems (confident outliers, and confident non-outliers) have one thing in common:
the masses of the heavier component, which are all between $0.87$--$0.91$\,\Msun.
While objects resulting from mergers are expected in this mass range, this does
not explain why the most confident non-outliers are also found here. It is certainly
true that for the more massive systems, the \dtauWD\ uncertainties appear substantially
smaller, with 240\,Myr for the worst of the six (\DWD{1215}{+}{0948}). Instead we speculate
that only for higher masses\footnotemark\ do we have enough precision on both \dtauWD\ \emph{and}
\dtauMS\ to make any distinction between outliers and non-outliers. Therefore
the outlier fraction for all systems, $f_\mathrm{outlier}$
is largely down to this small subset of six leading to such a high  (and poorly
constrained) value of $0.59_{-0.21}^{+0.22}$.
Inspecting the bottom panel of Figure~\ref{fig:Mf2_dtau_p}, we note that a few systems
also show slightly more distinct $P_\mathrm{outlier}$ (both above and below 0.59)
for white dwarf masses near 0.7\,\Msun. At this mass range though,
the balance is slightly towards systems with lower outlier probabilities.

\footnotetext{We note that \DWD{0215}{+}{1821} which has the highest mass of all systems in our sample,
has an outlier probability that is not distinct from $f_\mathrm{outlier}$. The uncertainty on \dtauWD
is just under 200\,Myr, and so one might expect that this object has a well established outlier
probability. However, the 1\,\Msun\ component corresponds to a poorly constrained part of our IFMR,
restricting the precision on the pre-WD lifetime.}

For the four high $P_\mathrm{outlier}$ systems, even if there is a selection bias
towards high outlier/non-outlier confidence near 0.9\,\Msun,
we conclude that the massive components in those four systems are
candidate merger products from within former hierarchical triple systems.
To rule out an alternative hypothesis of binaries formed via captures in multi-star
interactions,
we looked at the distribution of $P_\mathrm{outlier}$ against projected sky separation.
All four high probability outliers are separated by only 400--1200\,au making
captured binaries extremely unlikely, and lending further weight to our explanation of mergers.
The widest system in our fitted sample, \DWD{2351}{-}{1601} at $27\,600\pm700$\,au,
does have a slightly high $P_\mathrm{outlier}$ of 0.68, raising the prospect that this system
could have formed via a capture event.

This approach to outlier probabilities provided one final opportunity to investigate
the possibility of a non-monotonic IFMR as identified by \citet{marigoetal20-1}. We generated
synthetic binaries with initial masses and system ages covering a wide range of values.
These were converted to final masses using the IFMR of \citet{marigoetal20-1}, and the
pre-WD lifetimes (and hence cooling ages) calculated from the same MIST models as before.
We then determined the outlier probabilities using the same method as above (i.e. using
our own IFMR results to make this determination). We found that the outlier probability could
be maximised by placing the component with lower initial mass at the peak of the
\citet{marigoetal20-1} IFMR, i.e. at $M_i=1.85\,\Msun$, while also requiring short cooling
ages to reduce the uncertainties on \dtauWD. With the initial mass of the other component
$\geq 2.10\,\Msun$, we were able to produce outlier probabilities of $\simeq 0.90$
(though we were unable to produce the $\geq 0.97$ probabilities found for our most extreme outliers),
demonstrating that our methodology can be sensitive to outliers produced by a non-monotonic IFMR.

As a final exercise we selected systems from our sample with both component masses between
0.6--0.75\,\Msun\ (Table\,\ref{tab:dwds_fits}),
i.e. covering initial masses of 1.5--3.0\,\Msun\ from the
\citet{marigoetal20-1} IFMR, and with at least one of the cooling ages below 300\,Myr. 
This selection yielded 12 systems, all with outlier probabilities (Table\,\ref{tab:m_tau_p})
close to the median of 0.59, with the highest two values at 0.62
(\DWD{0120}{-}{1622} and \DWD{0309}{+}{1505}). Since these systems do not have
outlier probabilities well above the 0.59 median or approaching the 0.90 found from our
synthetic systems, we conclude that our IFMR and DWD sample do not provide evidence
for a non-monotonic IFMR.

\section{Conclusions and future work}
\label{sec:conc}

We observed 90 DWDs using FORS2 spectroscopy. Of these, we were able to use 52 
DA+DA, DA+DC, and DC+DC pairs to constrain the IFMR assuming a monotonic
piecewise-linear functional form. Furthermore we were able to
do this including the contribution of un-modelled uncertainties in their \Teff\
and $\log g$, while also establishing that several systems in our sample are outliers,
potentially suggesting the mergers taking place within former hierarchical triple systems.

While the vast majority of our binary sample can be fitted well with hydrogen dominated model spectra,
some of the white dwarfs clearly have helium dominated atmospheres, with the most obvious of
these having DB and DZ spectral classifications. Additionally some DC white dwarfs, while at first ambiguous,
turned out to incompatible with a hydrogen dominated atmosphere at the derived temperature
(i.e. hydrogen Balmer lines would otherwise have been present). In the future, we intend to also include
these systems by fitting models with helium dominated atmospheres instead.
However, it should be noted that a first investigation of the IFMR for hydrogen-poor white dwarfs
suggests that these objects have higher mass progenitors, i.e. that they follow a separate
IFMR altogether \citep{barnettetal21-1}

Looking further ahead, our sample is still relatively small, containing only
52 pairs used to constrain the IFMR, and only for final masses up to
1.0\,\Msun. This means that our IFMR is only well
constrained for initial masses below $5$--$6$\,\Msun. Above that range,
our fits merely reflect our choice of priors. Of course,
identifying ``well behaved'' systems containing
ultra-massive white dwarfs where both components are non-magnetic DAs,
presents its own observational challenges. Fortunately, the entire
\emph{Gaia} white dwarf sample contains over 1200 wide pairs\footnotemark\ --
more than an order of magnitude increase over the 90 we have observed so far.
Over the next decade, many of these will be observed in multi-fibre
spectroscopic surveys such as WEAVE, DESI, SDSS V, and 4MOST.
Taking advantage of these upcoming systems will allow us to constrain the IFMR
to even greater precision in the future.

A caveat to our implementation of the Bayesian framework is the time-complexity.
Na\"ively, Equation~\eqref{eq:likelihood_prod} implies a linear time-complexity in the number of
white dwarfs, $N$, i.e. $\mathcal{O}(N)$. However, the importance sampling step to
calculate the integral in Equation~\eqref{eq:like_int_Mf} limits the relative precision
of this quantity for each DWD. Therefore, to maintain a constant relative precision
in the total likelihood (Equation~\eqref{eq:likelihood_prod}), the number of integration
samples (per system) must also increase linearly with $N$,
increasing the overall time complexity to $\mathcal{O}(N^2)$. For our fitted
sample of $N=52$, this does not currently pose an issue, with the main
IFMR fit taking a few days on a modern 10-core desktop machine, but will become
problematic for samples with $N>100$.
Therefore, alternative integration techniques that can reduce
this time complexity should be investigated in advance of the large number
of systems that will be observed in the coming years.

\footnotetext{Estimated from our own exploration of the
\citet{gentilefusilloetal21-2} \emph{Gaia} DR3 white dwarf catalogue.}

\section*{Acknowledgements}

We thank the anonymous referee for their feedback which improved the quality
of this manuscript.
MAH acknowledges useful conversations with Pier-Emmanuel Tremblay regarding 3D corrections to 1D atmospheric models.
MAH and SL were supported by grant ST/V000853/1 from the Science and Technology Facilities Council (STFC).
SCP acknowledges the support of a STFC Ernest Rutherford Fellowship.
For the purpose of open access, the authors has applied a
creative commons attribution (CC BY) licence to any author accepted
manuscript version arising.
%
%

\section*{Data Availability}

The spectroscopic data obtained in this work can be found from the ESO Science Archive Facility
with Program IDs 0103.D-0718 and 109.231B. The MCMC chains from our IFMR fits will be made available
alongside our python package, and also available upon reasonable request to the authors.



\bibliographystyle{mnras}
\bibliography{aamnem99,aabib} 

\begin{thebibliography}{}
\makeatletter
\relax
\def\mn@urlcharsother{\let\do\@makeother \do\$\do\&\do\#\do\^\do\_\do\%\do\~}
\def\mn@doi{\begingroup\mn@urlcharsother \@ifnextchar [ {\mn@doi@}
  {\mn@doi@[]}}
\def\mn@doi@[#1]#2{\def\@tempa{#1}\ifx\@tempa\@empty \href
  {http://dx.doi.org/#2} {doi:#2}\else \href {http://dx.doi.org/#2} {#1}\fi
  \endgroup}
\def\mn@eprint#1#2{\mn@eprint@#1:#2::\@nil}
\def\mn@eprint@arXiv#1{\href {http://arxiv.org/abs/#1} {{\tt arXiv:#1}}}
\def\mn@eprint@dblp#1{\href {http://dblp.uni-trier.de/rec/bibtex/#1.xml}
  {dblp:#1}}
\def\mn@eprint@#1:#2:#3:#4\@nil{\def\@tempa {#1}\def\@tempb {#2}\def\@tempc
  {#3}\ifx \@tempc \@empty \let \@tempc \@tempb \let \@tempb \@tempa \fi \ifx
  \@tempb \@empty \def\@tempb {arXiv}\fi \@ifundefined
  {mn@eprint@\@tempb}{\@tempb:\@tempc}{\expandafter \expandafter \csname
  mn@eprint@\@tempb\endcsname \expandafter{\@tempc}}}

\bibitem[\protect\citeauthoryear{{Alam} et~al.,}{{Alam}
  et~al.}{2015}]{alametal15-1}
{Alam} S.,  et~al., 2015, \mn@doi [ApJS] {10.1088/0067-0049/219/1/12}, \href
  {http://adsabs.harvard.edu/abs/2015ApJS..219...12A} {219, 12}

\bibitem[\protect\citeauthoryear{{Andrews}, {Ag{\"u}eros}, {Gianninas},
  {Kilic}, {Dhital}  \& {Anderson}}{{Andrews} et~al.}{2015}]{andrewsetal15-1}
{Andrews} J.~J.,  {Ag{\"u}eros} M.~A.,  {Gianninas} A.,  {Kilic} M.,  {Dhital}
  S.,   {Anderson} S.~F.,  2015, \mn@doi [ApJ] {10.1088/0004-637X/815/1/63},
  \href {https://ui.adsabs.harvard.edu/abs/2015ApJ...815...63A} {815, 63}

\bibitem[\protect\citeauthoryear{{Barnett}, {Williams}, {B{\'e}dard}  \&
  {Bolte}}{{Barnett} et~al.}{2021}]{barnettetal21-1}
{Barnett} J.~W.,  {Williams} K.~A.,  {B{\'e}dard} A.,   {Bolte} M.,  2021,
  \mn@doi [AJ] {10.3847/1538-3881/ac1423}, \href
  {https://ui.adsabs.harvard.edu/abs/2021AJ....162..162B} {162, 162}

\bibitem[\protect\citeauthoryear{{Barrientos} \& {Chanam{\'e}}}{{Barrientos} \&
  {Chanam{\'e}}}{2021}]{barrientosetal21-1}
{Barrientos} M.,  {Chanam{\'e}} J.,  2021, \mn@doi [ApJ]
  {10.3847/1538-4357/ac2f49}, \href
  {https://ui.adsabs.harvard.edu/abs/2021ApJ...923..181B} {923, 181}

\bibitem[\protect\citeauthoryear{{B{\'e}dard}, {Bergeron}, {Brassard}  \&
  {Fontaine}}{{B{\'e}dard} et~al.}{2020}]{bedardetal20-1}
{B{\'e}dard} A.,  {Bergeron} P.,  {Brassard} P.,   {Fontaine} G.,  2020, arXiv
  e-prints, \href {https://ui.adsabs.harvard.edu/abs/2020arXiv200807469B} {p.
  arXiv:2008.07469}

\bibitem[\protect\citeauthoryear{{Bergeron}, {Dufour}, {Fontaine}, {Coutu},
  {Blouin}, {Genest-Beaulieu}, {B{\'e}dard}  \& {Rolland}}{{Bergeron}
  et~al.}{2019}]{bergeronetal19-1}
{Bergeron} P.,  {Dufour} P.,  {Fontaine} G.,  {Coutu} S.,  {Blouin} S.,
  {Genest-Beaulieu} C.,  {B{\'e}dard} A.,   {Rolland} B.,  2019, \mn@doi [ApJ]
  {10.3847/1538-4357/ab153a}, \href
  {https://ui.adsabs.harvard.edu/abs/2019ApJ...876...67B} {876, 67}

\bibitem[\protect\citeauthoryear{{Bloecker}}{{Bloecker}}{1995}]{bloecker95-1}
{Bloecker} T.,  1995, A\&A, \href
  {https://ui.adsabs.harvard.edu/abs/1995A&A...297..727B} {297, 727}

\bibitem[\protect\citeauthoryear{{Camisassa} et~al.,}{{Camisassa}
  et~al.}{2019}]{camisassaetal19-1}
{Camisassa} M.~E.,  et~al., 2019, \mn@doi [A\&A] {10.1051/0004-6361/201833822},
  \href {https://ui.adsabs.harvard.edu/abs/2019A&A...625A..87C} {625, A87}

\bibitem[\protect\citeauthoryear{{Canton}}{{Canton}}{2018}]{cantonPhD18}
{Canton} P.,  2018, PhD thesis, University of Oklahoma, Norman

\bibitem[\protect\citeauthoryear{{Casewell}, {Dobbie}, {Napiwotzki},
  {Burleigh}, {Barstow}  \& {Jameson}}{{Casewell}
  et~al.}{2009}]{casewelletal09-1}
{Casewell} S.~L.,  {Dobbie} P.~D.,  {Napiwotzki} R.,  {Burleigh} M.~R.,
  {Barstow} M.~A.,   {Jameson} R.~F.,  2009, \mn@doi [MNRAS]
  {10.1111/j.1365-2966.2009.14593.x}, \href {2009MNRAS.395.1795C} {395, 1795}

\bibitem[\protect\citeauthoryear{{Catal{\'a}n}, {Isern}, {Garc{\'{\i}}a-Berro},
  {Ribas}, {Allende Prieto}  \& {Bonanos}}{{Catal{\'a}n}
  et~al.}{2008}]{catalanetal08-1}
{Catal{\'a}n} S.,  {Isern} J.,  {Garc{\'{\i}}a-Berro} E.,  {Ribas} I.,
  {Allende Prieto} C.,   {Bonanos} A.~Z.,  2008, \mn@doi [A\&A]
  {10.1051/0004-6361:20078111}, \href {2008A&A...477..213C} {477, 213}

\bibitem[\protect\citeauthoryear{{Choi}, {Dotter}, {Conroy}, {Cantiello},
  {Paxton}  \& {Johnson}}{{Choi} et~al.}{2016}]{choiMIST16}
{Choi} J.,  {Dotter} A.,  {Conroy} C.,  {Cantiello} M.,  {Paxton} B.,
  {Johnson} B.~D.,  2016, \mn@doi [ApJ] {10.3847/0004-637X/823/2/102}, \href
  {https://ui.adsabs.harvard.edu/abs/2016ApJ...823..102C} {823, 102}

\bibitem[\protect\citeauthoryear{{Cummings}, {Kalirai}, {Tremblay},
  {Ramirez-Ruiz}  \& {Choi}}{{Cummings} et~al.}{2018}]{cummingsetal18-1}
{Cummings} J.~D.,  {Kalirai} J.~S.,  {Tremblay} P.~E.,  {Ramirez-Ruiz} E.,
  {Choi} J.,  2018, \mn@doi [ApJ] {10.3847/1538-4357/aadfd6}, \href
  {https://ui.adsabs.harvard.edu/abs/2018ApJ...866...21C} {866, 21}

\bibitem[\protect\citeauthoryear{{Cummings}, {Kalirai}, {Choi}, {Georgy},
  {Tremblay}  \& {Ramirez-Ruiz}}{{Cummings} et~al.}{2019}]{cummingsetal19-1}
{Cummings} J.~D.,  {Kalirai} J.~S.,  {Choi} J.,  {Georgy} C.,  {Tremblay}
  P.~E.,   {Ramirez-Ruiz} E.,  2019, \mn@doi [ApJ Lett.]
  {10.3847/2041-8213/aafc2d}, \href
  {https://ui.adsabs.harvard.edu/abs/2019ApJ...871L..18C} {871, L18}

\bibitem[\protect\citeauthoryear{{Dobbie}, {Napiwotzki}, {Burleigh},
  {Williams}, {Sharp}, {Barstow}, {Casewell}  \& {Hubeny}}{{Dobbie}
  et~al.}{2009}]{dobbieetal09-1}
{Dobbie} P.~D.,  {Napiwotzki} R.,  {Burleigh} M.~R.,  {Williams} K.~A.,
  {Sharp} R.,  {Barstow} M.~A.,  {Casewell} S.~L.,   {Hubeny} I.,  2009,
  \mn@doi [MNRAS] {10.1111/j.1365-2966.2009.14688.x}, \href
  {2009MNRAS.395.2248D} {395, 2248}

\bibitem[\protect\citeauthoryear{{Dotter}}{{Dotter}}{2016}]{dotterMIST16}
{Dotter} A.,  2016, \mn@doi [ApJS] {10.3847/0067-0049/222/1/8}, \href
  {https://ui.adsabs.harvard.edu/abs/2016ApJS..222....8D} {222, 8}

\bibitem[\protect\citeauthoryear{{El-Badry} \& {Rix}}{{El-Badry} \&
  {Rix}}{2018}]{el-badryetal18-2}
{El-Badry} K.,  {Rix} H.-W.,  2018, \mn@doi [MNRAS] {10.1093/mnras/sty2186},
  \href {https://ui.adsabs.harvard.edu/abs/2018MNRAS.480.4884E} {480, 4884}

\bibitem[\protect\citeauthoryear{{El-Badry}, {Rix}  \& {Weisz}}{{El-Badry}
  et~al.}{2018}]{el-badryetal18-1}
{El-Badry} K.,  {Rix} H.-W.,   {Weisz} D.~R.,  2018, \mn@doi [ApJ]
  {10.3847/2041-8213/aaca9c}, \href
  {https://ui.adsabs.harvard.edu/#abs/2018ApJ...860L..17E} {860, L17}

\bibitem[\protect\citeauthoryear{{Ferrario}, {Wickramasinghe}, {Liebert}  \&
  {Williams}}{{Ferrario} et~al.}{2005}]{ferrarioetal05-1}
{Ferrario} L.,  {Wickramasinghe} D.,  {Liebert} J.,   {Williams} K.~A.,  2005,
  \mn@doi [MNRAS] {10.1111/j.1365-2966.2005.09244.x}, \href
  {2005MNRAS.361.1131F} {361, 1131}

\bibitem[\protect\citeauthoryear{{Fields}, {Farmer}, {Petermann}, {Iliadis}  \&
  {Timmes}}{{Fields} et~al.}{2016}]{fieldsetal16-1}
{Fields} C.~E.,  {Farmer} R.,  {Petermann} I.,  {Iliadis} C.,   {Timmes} F.~X.,
   2016, \mn@doi [ApJ] {10.3847/0004-637X/823/1/46}, \href
  {https://ui.adsabs.harvard.edu/abs/2016ApJ...823...46F} {823, 46}

\bibitem[\protect\citeauthoryear{{Fontaine}, {Brassard}  \&
  {Bergeron}}{{Fontaine} et~al.}{2001}]{fontaineetal01-1}
{Fontaine} G.,  {Brassard} P.,   {Bergeron} P.,  2001, PASP, \href
  {2001PASP..113..409F} {113, 409}

\bibitem[\protect\citeauthoryear{{Foreman-Mackey}, {Hogg}, {Lang}  \&
  {Goodman}}{{Foreman-Mackey} et~al.}{2013}]{foremanmackeyetal13-1}
{Foreman-Mackey} D.,  {Hogg} D.~W.,  {Lang} D.,   {Goodman} J.,  2013, \mn@doi
  [PASP] {10.1086/670067}, \href
  {http://adsabs.harvard.edu/abs/2013PASP..125..306F} {125, 306}

\bibitem[\protect\citeauthoryear{{Gaia Collaboration} et~al.,}{{Gaia
  Collaboration} et~al.}{2021}]{gaiaEDR3-collab-1}
{Gaia Collaboration} et~al., 2021, \mn@doi [A\&A]
  {10.1051/0004-6361/202039657}, \href
  {https://ui.adsabs.harvard.edu/abs/2021A&A...649A...1G} {649, A1}

\bibitem[\protect\citeauthoryear{{Genest-Beaulieu} \&
  {Bergeron}}{{Genest-Beaulieu} \& {Bergeron}}{2014}]{genestetal14-1}
{Genest-Beaulieu} C.,  {Bergeron} P.,  2014, \mn@doi [ApJ]
  {10.1088/0004-637X/796/2/128}, \href
  {http://adsabs.harvard.edu/abs/2014ApJ...796..128G} {796, 128}

\bibitem[\protect\citeauthoryear{{Genest-Beaulieu} \&
  {Bergeron}}{{Genest-Beaulieu} \& {Bergeron}}{2019}]{genest+bergeron19-1}
{Genest-Beaulieu} C.,  {Bergeron} P.,  2019, \mn@doi [ApJ]
  {10.3847/1538-4357/aafac6}, \href
  {https://ui.adsabs.harvard.edu/abs/2019ApJ...871..169G} {871, 169}

\bibitem[\protect\citeauthoryear{{Gentile Fusillo} et~al.,}{{Gentile Fusillo}
  et~al.}{2019}]{gentilefusilloetal19-1}
{Gentile Fusillo} N.~P.,  et~al., 2019, \mn@doi [MNRAS]
  {10.1093/mnras/sty3016}, \href
  {https://ui.adsabs.harvard.edu/abs/2019MNRAS.482.4570G} {482, 4570}

\bibitem[\protect\citeauthoryear{{Gentile Fusillo} et~al.,}{{Gentile Fusillo}
  et~al.}{2021}]{gentilefusilloetal21-2}
{Gentile Fusillo} N.~P.,  et~al., 2021, \mn@doi [MNRAS]
  {10.1093/mnras/stab2672}, \href
  {https://ui.adsabs.harvard.edu/abs/2021MNRAS.508.3877G} {508, 3877}

\bibitem[\protect\citeauthoryear{{Green}, {Schlafly}, {Zucker}, {Speagle}  \&
  {Finkbeiner}}{{Green} et~al.}{2019}]{greenetal19-1}
{Green} G.~M.,  {Schlafly} E.,  {Zucker} C.,  {Speagle} J.~S.,   {Finkbeiner}
  D.,  2019, \mn@doi [ApJ] {10.3847/1538-4357/ab5362}, \href
  {https://ui.adsabs.harvard.edu/abs/2019ApJ...887...93G} {887, 93}

\bibitem[\protect\citeauthoryear{{Heintz}, {Hermes}, {El-Badry}, {Walsh}, {van
  Saders}, {Fields}  \& {Koester}}{{Heintz} et~al.}{2022}]{heintzetal22-1}
{Heintz} T.~M.,  {Hermes} J.~J.,  {El-Badry} K.,  {Walsh} C.,  {van Saders}
  J.~L.,  {Fields} C.~E.,   {Koester} D.,  2022, \mn@doi [ApJ]
  {10.3847/1538-4357/ac78d9}, \href
  {https://ui.adsabs.harvard.edu/abs/2022ApJ...934..148H} {934, 148}

\bibitem[\protect\citeauthoryear{{Hogg}, {Bovy}  \& {Lang}}{{Hogg}
  et~al.}{2010}]{hoggetal10-1}
{Hogg} D.~W.,  {Bovy} J.,   {Lang} D.,  2010, \mn@doi [arXiv e-prints]
  {10.48550/arXiv.1008.4686}, \href
  {https://ui.adsabs.harvard.edu/abs/2010arXiv1008.4686H} {p. arXiv:1008.4686}

\bibitem[\protect\citeauthoryear{{Hollands}, {Tremblay}, {G{\"a}nsicke},
  {Gentile-Fusillo}  \& {Toonen}}{{Hollands} et~al.}{2018}]{hollandsetal18-2}
{Hollands} M.~A.,  {Tremblay} P.-E.,  {G{\"a}nsicke} B.~T.,  {Gentile-Fusillo}
  N.~P.,   {Toonen} S.,  2018, \mn@doi [MNRAS] {10.1093/mnras/sty2057}, \href
  {https://ui.adsabs.harvard.edu/abs/2018MNRAS.480.3942H} {480, 3942}

\bibitem[\protect\citeauthoryear{{Horne}}{{Horne}}{1986}]{horne86-1}
{Horne} K.,  1986, PASP, \href {1986PASP...98..609H} {98, 609}

\bibitem[\protect\citeauthoryear{{Kalirai}, {Hansen}, {Kelson}, {Reitzel},
  {Rich}  \& {Richer}}{{Kalirai} et~al.}{2008}]{kaliraietal08-1}
{Kalirai} J.~S.,  {Hansen} B.~M.~S.,  {Kelson} D.~D.,  {Reitzel} D.~B.,  {Rich}
  R.~M.,   {Richer} H.~B.,  2008, \mn@doi [ApJ] {10.1086/527028}, \href
  {2008ApJ...676..594K} {676, 594}

\bibitem[\protect\citeauthoryear{{Koester}}{{Koester}}{2010}]{koester10-1}
{Koester} D.,  2010, Memorie della Societa Astronomica Italiana,, \href
  {2010MmSAI..81..921K} {81, 921}

\bibitem[\protect\citeauthoryear{{Koester}}{{Koester}}{2013}]{koester13-1}
{Koester} D.,  2013, {White Dwarf Stars}.
Springer, p.~559, \mn@doi{10.1007/978-94-007-5615-1_11}

\bibitem[\protect\citeauthoryear{{Kroupa}}{{Kroupa}}{2001}]{kroupa01-1}
{Kroupa} P.,  2001, \mn@doi [MNRAS] {10.1046/j.1365-8711.2001.04022.x}, \href
  {https://ui.adsabs.harvard.edu/abs/2001MNRAS.322..231K} {322, 231}

\bibitem[\protect\citeauthoryear{{Liebert}, {Bergeron}  \& {Holberg}}{{Liebert}
  et~al.}{2005}]{liebertetal05-1}
{Liebert} J.,  {Bergeron} P.,   {Holberg} J.~B.,  2005, ApJS, \href
  {2005ApJS..156...47L} {156, 47}

\bibitem[\protect\citeauthoryear{{Lindegren} et~al.,}{{Lindegren}
  et~al.}{2018}]{gaiaDR2-collab-2}
{Lindegren} L.,  et~al., 2018, \mn@doi [A\&A] {10.1051/0004-6361/201832727},
  \href {https://ui.adsabs.harvard.edu/abs/2018A&A...616A...2L} {616, A2}

\bibitem[\protect\citeauthoryear{{Marigo} \& {Girardi}}{{Marigo} \&
  {Girardi}}{2007}]{marigo+girardi07-1}
{Marigo} P.,  {Girardi} L.,  2007, \mn@doi [A\&A] {10.1051/0004-6361:20066772},
  \href {https://ui.adsabs.harvard.edu/abs/2007A&A...469..239M} {469, 239}

\bibitem[\protect\citeauthoryear{{Marigo} et~al.,}{{Marigo}
  et~al.}{2020}]{marigoetal20-1}
{Marigo} P.,  et~al., 2020, \mn@doi [Nat Astron.] {10.1038/s41550-020-1132-1},
  \href {https://ui.adsabs.harvard.edu/abs/2020NatAs...4.1102M} {4, 1102}

\bibitem[\protect\citeauthoryear{{Marigo} et~al.,}{{Marigo}
  et~al.}{2022}]{marigoetal22-1}
{Marigo} P.,  et~al., 2022, \mn@doi [ApJS] {10.3847/1538-4365/ac374a}, \href
  {https://ui.adsabs.harvard.edu/abs/2022ApJS..258...43M} {258, 43}

\bibitem[\protect\citeauthoryear{{Meng}, {Chen}  \& {Han}}{{Meng}
  et~al.}{2008}]{mengetal08-1}
{Meng} X.,  {Chen} X.,   {Han} Z.,  2008, \mn@doi [A\&A]
  {10.1051/0004-6361:20078841}, \href
  {https://ui.adsabs.harvard.edu/abs/2008A&A...487..625M} {487, 625}

\bibitem[\protect\citeauthoryear{{Mestel}}{{Mestel}}{1952}]{mestel52-1}
{Mestel} L.,  1952, \mn@doi [MNRAS] {10.1093/mnras/112.6.583}, \href
  {http://adsabs.harvard.edu/abs/1952MNRAS.112..583M} {112, 583}

\bibitem[\protect\citeauthoryear{{Paxton}, {Bildsten}, {Dotter}, {Herwig},
  {Lesaffre}  \& {Timmes}}{{Paxton} et~al.}{2011}]{paxtonetalMESA11}
{Paxton} B.,  {Bildsten} L.,  {Dotter} A.,  {Herwig} F.,  {Lesaffre} P.,
  {Timmes} F.,  2011, \mn@doi [ApJS] {10.1088/0067-0049/192/1/3}, \href
  {https://ui.adsabs.harvard.edu/abs/2011ApJS..192....3P} {192, 3}

\bibitem[\protect\citeauthoryear{{Paxton} et~al.,}{{Paxton}
  et~al.}{2013}]{paxtonetalMESA13}
{Paxton} B.,  et~al., 2013, \mn@doi [ApJS] {10.1088/0067-0049/208/1/4}, \href
  {https://ui.adsabs.harvard.edu/abs/2013ApJS..208....4P} {208, 4}

\bibitem[\protect\citeauthoryear{{Paxton} et~al.,}{{Paxton}
  et~al.}{2015}]{paxtonetalMESA15}
{Paxton} B.,  et~al., 2015, \mn@doi [ApJS] {10.1088/0067-0049/220/1/15}, \href
  {https://ui.adsabs.harvard.edu/abs/2015ApJS..220...15P} {220, 15}

\bibitem[\protect\citeauthoryear{{Paxton} et~al.,}{{Paxton}
  et~al.}{2018}]{paxtonetalMESA18}
{Paxton} B.,  et~al., 2018, \mn@doi [ApJS] {10.3847/1538-4365/aaa5a8}, \href
  {https://ui.adsabs.harvard.edu/abs/2018ApJS..234...34P} {234, 34}

\bibitem[\protect\citeauthoryear{{Perpiny{\`a}-Vall{\`e}s},
  {Rebassa-Mansergas}, {G{\"a}nsicke}, {Toonen}, {Hermes}, {Gentile Fusillo}
  \& {Tremblay}}{{Perpiny{\`a}-Vall{\`e}s} et~al.}{2019}]{perpinyaetal19-1}
{Perpiny{\`a}-Vall{\`e}s} M.,  {Rebassa-Mansergas} A.,  {G{\"a}nsicke} B.~T.,
  {Toonen} S.,  {Hermes} J.~J.,  {Gentile Fusillo} N.~P.,   {Tremblay} P.~E.,
  2019, \mn@doi [MNRAS] {10.1093/mnras/sty3149}, \href
  {https://ui.adsabs.harvard.edu/abs/2019MNRAS.483..901P} {483, 901}

\bibitem[\protect\citeauthoryear{{Rebassa-Mansergas}, {Rybicka}, {Liu}, {Han}
  \& {Garc{\'{\i}}a-Berro}}{{Rebassa-Mansergas}
  et~al.}{2015}]{rebassa-mansergasetal15-1}
{Rebassa-Mansergas} A.,  {Rybicka} M.,  {Liu} X.-W.,  {Han} Z.,
  {Garc{\'{\i}}a-Berro} E.,  2015, \mn@doi [MNRAS] {10.1093/mnras/stv1399},
  \href {http://adsabs.harvard.edu/abs/2015MNRAS.452.1637R} {452, 1637}

\bibitem[\protect\citeauthoryear{{Rebassa-Mansergas}
  et~al.,}{{Rebassa-Mansergas} et~al.}{2021}]{rebassa-mansergasetal21-1}
{Rebassa-Mansergas} A.,  et~al., 2021, \mn@doi [MNRAS]
  {10.1093/mnras/stab1559}, \href
  {https://ui.adsabs.harvard.edu/abs/2021MNRAS.505.3165R} {505, 3165}

\bibitem[\protect\citeauthoryear{{Romero}, {Campos}  \& {Kepler}}{{Romero}
  et~al.}{2015}]{romeroetal15-1}
{Romero} A.~D.,  {Campos} F.,   {Kepler} S.~O.,  2015, \mn@doi [MNRAS]
  {10.1093/mnras/stv848}, \href
  {https://ui.adsabs.harvard.edu/abs/2015MNRAS.450.3708R} {450, 3708}

\bibitem[\protect\citeauthoryear{{Salaris}, {Serenelli}, {Weiss}  \& {Miller
  Bertolami}}{{Salaris} et~al.}{2009}]{salarisetal09-1}
{Salaris} M.,  {Serenelli} A.,  {Weiss} A.,   {Miller Bertolami} M.,  2009,
  \mn@doi [ApJ] {10.1088/0004-637X/692/2/1013}, \href {2009ApJ...692.1013S}
  {692, 1013}

\bibitem[\protect\citeauthoryear{{Salpeter}}{{Salpeter}}{1955}]{salpeter55-1}
{Salpeter} E.~E.,  1955, ApJ, \href {1955ApJ...121..161S} {121, 161}

\bibitem[\protect\citeauthoryear{{Sion}, {Greenstein}, {Landstreet}, {Liebert},
  {Shipman}  \& {Wegner}}{{Sion} et~al.}{1983}]{sionetal83-1}
{Sion} E.~M.,  {Greenstein} J.~L.,  {Landstreet} J.~D.,  {Liebert} J.,
  {Shipman} H.~L.,   {Wegner} G.~A.,  1983, \mn@doi [ApJ] {10.1086/161036},
  \href {http://adsabs.harvard.edu/abs/1983ApJ...269..253S} {269, 253}

\bibitem[\protect\citeauthoryear{{Tassoul}, {Fontaine}  \& {Winget}}{{Tassoul}
  et~al.}{1990}]{tassouletal90-1}
{Tassoul} M.,  {Fontaine} G.,   {Winget} D.~E.,  1990, \mn@doi [ApJS]
  {10.1086/191420}, \href
  {https://ui.adsabs.harvard.edu/abs/1990ApJS...72..335T} {72, 335}

\bibitem[\protect\citeauthoryear{{Tremblay}, {Ludwig}, {Steffen}  \&
  {Freytag}}{{Tremblay} et~al.}{2013}]{tremblayetal13-1}
{Tremblay} P.-E.,  {Ludwig} H.-G.,  {Steffen} M.,   {Freytag} B.,  2013,
  \mn@doi [A\&A] {10.1051/0004-6361/201322318}, \href
  {http://adsabs.harvard.edu/abs/2013A%26A...559A.104T} {559, A104}

\bibitem[\protect\citeauthoryear{{Voss}, {Koester}, {Napiwotzki}, {Christlieb}
  \& {Reimers}}{{Voss} et~al.}{2007}]{vossetal07-1}
{Voss} B.,  {Koester} D.,  {Napiwotzki} R.,  {Christlieb} N.,   {Reimers} D.,
  2007, \mn@doi [A\&A] {10.1051/0004-6361:20077285}, \href
  {2007A&A...470.1079V} {470, 1079}

\bibitem[\protect\citeauthoryear{{Weidemann}}{{Weidemann}}{1987}]{weidemann87-1}
{Weidemann} V.,  1987, A\&A, \href {1987A&A...188...74W} {188, 74}

\bibitem[\protect\citeauthoryear{{Weidemann}}{{Weidemann}}{2000}]{weidemann00-1}
{Weidemann} V.,  2000, A\&A, \href {2000A&A...363..647W} {363, 647}

\bibitem[\protect\citeauthoryear{{Weidemann} \& {Koester}}{{Weidemann} \&
  {Koester}}{1983}]{weidemann+koester83-1}
{Weidemann} V.,  {Koester} D.,  1983, A\&A, \href
  {https://ui.adsabs.harvard.edu/abs/1983A&A...121...77W} {121, 77}

\bibitem[\protect\citeauthoryear{{Williams}, {Bolte}  \& {Koester}}{{Williams}
  et~al.}{2009}]{williamsetal09-1}
{Williams} K.~A.,  {Bolte} M.,   {Koester} D.,  2009, \mn@doi [ApJ]
  {10.1088/0004-637X/693/1/355}, \href {2009ApJ...693..355W} {693, 355}

\makeatother
\end{thebibliography}




\appendix

\section{Additional tables and figures}

\onecolumn
\newpage

\begin{longtable}{lccccccc}
    \caption{\label{tab:dwds_obs} Coordinates, spectral types (SpT), and \emph{Gaia} DR3 data for
    our entire DWD sample that were followed up with FORS2. The ``run(s)'' column specifies which observing
    run (or runs) each DWD was observed in.}\\
    \hline
    WD\,J & J2016 coordinates & SpT & parallax [mas] & $G$ [mag] & $B_p$ [mag] & $R_p$ [mag] & run(s) \\
    \hline
    \endfirsthead
    \caption{continued}\\
    \hline
    WD\,J & J2016 coordinates & SpT & parallax [mas] & $G$ [mag] & $B_p$ [mag] & $R_p$ [mag] & run(s) \\
    \hline
    \endhead
    \hline
    \endfoot
    0002$+$0733A & 00:02:16.07$+$07:33:49.3 & DAH & 12.349±0.138 & 17.946±0.003 & 18.098±0.013 & 17.707±0.015 & \multirow{2}{*}{C} \\*
0002$+$0733B & 00:02:15.27$+$07:33:58.5 & DA & 12.115±0.121 & 17.724±0.003 & 17.897±0.013 & 17.445±0.011 & \smallskip \\
0007$-$1605A & 00:07:34.39$-$16:05:14.1 & DC & 12.526±0.547 & 19.942±0.005 & 20.593±0.086 & 19.241±0.054 & \multirow{2}{*}{C} \\*
0007$-$1605B & 00:07:35.02$-$16:05:32.6 & DA & 12.337±0.045 & 16.152±0.003 & 16.162±0.004 & 16.157±0.008 & \smallskip \\
0022$-$7232A & 00:22:29.86$-$72:32:08.5 & DAZ & 11.251±0.088 & 17.686±0.003 & 17.817±0.009 & 17.461±0.014 & \multirow{2}{*}{E} \\*
0022$-$7232B & 00:22:33.21$-$72:32:32.6 & DC & 11.178±0.254 & 19.419±0.004 & 19.867±0.060 & 18.900±0.033 & \smallskip \\
0023$+$0643A & 00:23:36.32$+$06:43:20.2 & DA & 8.458±0.094 & 17.416±0.003 & 17.474±0.007 & 17.346±0.011 & \multirow{2}{*}{E} \\*
0023$+$0643B & 00:23:35.64$+$06:43:25.2 & DA & 8.839±0.172 & 18.448±0.003 & 18.624±0.016 & 18.158±0.024 & \smallskip \\
\end{longtable}

\small
\begin{longtable}{lccccccc}
    \caption{\label{tab:dwds_fits} Best fit parameters for our DWD sample. Note that the
    \dtauWD\ account for correlated uncertainties,
    which cannot be determined from the individual age uncertainties alone.} \\
    \hline
    WD\,J & \Teff\ [K] & $\log g$ & $M_f$ [\Msun] & $\tau_\mathrm{WD}$ [Gyr] & \dtauWD\ [Gyr] & $r_v$ [\kms] & $E(B-V)$\\
    \hline
    \endfirsthead
    \caption{continued} \\
    \hline
    WD\,J & \Teff\ [K] & $\log g$ & $M_f$ [\Msun] & $\tau_\mathrm{WD}$ [Gyr] & \dtauWD\ [Gyr] & $r_v$ [\kms] & $E(B-V)$\\
    \hline
    \endhead
    \hline
    \endfoot
    \hline \\[-4mm]
    \caption*{Notes:
    ($a$) Magnetism affected quality of fit.
    ($b$) DAZ spectrum is fitted with a DA model, though the narrow \Ion{Ca}{ii} lines do not 
    affect the atmospheric parameters.
    ($c$) Best fit indicates extremely low mass.
    ($d$) DC spectrum cannot be fit correctly with DA model.
    ($e$) Object appears twice with separate parameters
    ($f$) Zeeman splitting detected, but does not significantly affect the quality
    of the fit.
    }
    \endlastfoot
    0002$+$0733A$^a$ & $6826\pm37$ & $7.941\pm0.026$ & $0.557\pm0.015$ & $1.538\pm0.044$ & \multirow{2}{*}{$0.229\pm0.023$} & \multirow{2}{*}{$9.0\pm4.4$} &  \\*
0002$+$0733B & $7364\pm30$ & $7.965\pm0.021$ & $0.572\pm0.012$ & $1.311\pm0.032$ &  &  &  \smallskip \\
0007$-$1605A & $4953\pm14$ & $8.391\pm0.012$ & $0.836\pm0.009$ & $9.513\pm0.037$ & \multirow{2}{*}{$9.183\pm0.036$} & \multirow{2}{*}{$11.6\pm2.4$} &  \\*
0007$-$1605B & $12785\pm39$ & $8.001\pm0.006$ & $0.608\pm0.003$ & $0.330\pm0.003$ &  &  &  \smallskip \\
0022$-$7232A$^b$ & $8004\pm21$ & $7.984\pm0.014$ & $0.585\pm0.008$ & $1.089\pm0.018$ & \multirow{2}{*}{$-3.550\pm0.202$} & \multirow{2}{*}{$13.2\pm3.0$} &  \\*
0022$-$7232B & $5094\pm12$ & $7.952\pm0.019$ & $0.552\pm0.011$ & $4.641\pm0.215$ &  &  &  \smallskip \\
0023$+$0643A & $10238\pm30$ & $7.959\pm0.015$ & $0.577\pm0.009$ & $0.565\pm0.009$ & \multirow{2}{*}{$-0.645\pm0.019$} & \multirow{2}{*}{$12.8\pm3.5$} &  \\*
0023$+$0643B & $7603\pm32$ & $7.966\pm0.020$ & $0.573\pm0.012$ & $1.211\pm0.026$ &  &  &  \smallskip \\
\end{longtable}

\normalsize
\begin{longtable}{lcccccc}
    \caption{\label{tab:m_tau_p} Derived parameters for each of the 52 DWDs from our IFMR fit. $\rho_{M_i}$ is the
    correlation between the two initial mass values.
    } \\
    \hline
    WD\,J & $M_{i,\mathrm{A}}$ [\Msun] & $M_{i,\mathrm{B}}$ [\Msun] & $\rho_{M_i}$ & $\tau_\mathrm{preWD,A}$ [Gyr] & $\tau_\mathrm{preWD,B}$ [Gyr] & $p_\mathrm{outlier}$ \\
    \hline
    \endfirsthead
    \caption{continued} \\
    \hline
    WD\,J & $M_{i,\mathrm{A}}$ [\Msun] & $M_{i,\mathrm{B}}$ [\Msun] & $\rho_{M_i}$ & $\tau_\mathrm{preWD,A}$ [Gyr] & $\tau_\mathrm{preWD,B}$ [Gyr] & $p_\mathrm{outlier}$ \\
    \hline
    \endhead
    \hline
    \endfoot
    \hline \\[-4mm]
    \endlastfoot
    0022$-$7232AB & $1.07_{-0.06}^{+0.09}$ & $1.28_{-0.14}^{+0.32}$ & $+0.61$ & $9.0_{-2.2}^{+2.1}$ & $4.9_{-2.5}^{+2.3}$ & $0.59$ \smallskip\\
0023$+$0643AB & $1.21_{-0.13}^{+0.25}$ & $1.25_{-0.15}^{+0.33}$ & $+0.93$ & $6.0_{-2.8}^{+3.0}$ & $5.3_{-2.8}^{+3.0}$ & $0.59$ \smallskip\\
0052$+$1353AB & $1.57_{-0.25}^{+0.29}$ & $1.65_{-0.30}^{+0.40}$ & $+0.84$ & $2.5_{-1.0}^{+1.9}$ & $2.2_{-0.8}^{+2.0}$ & $0.59$ \smallskip\\
0101$-$1629AB & $4.29_{-0.59}^{+0.71}$ & $2.54_{-0.37}^{+0.35}$ & $+0.29$ & $0.174_{-0.056}^{+0.084}$ & $0.76_{-0.24}^{+0.41}$ & $0.99$ \smallskip\\
0104$+$2120AB & $1.57_{-0.22}^{+0.35}$ & $1.45_{-0.16}^{+0.30}$ & $+0.50$ & $2.6_{-1.1}^{+1.6}$ & $3.2_{-1.4}^{+1.5}$ & $0.61$ \smallskip\\
0109$-$1042AB & $1.52_{-0.25}^{+0.40}$ & $1.09_{-0.06}^{+0.05}$ & $+0.62$ & $2.8_{-1.3}^{+2.2}$ & $8.4_{-1.3}^{+2.0}$ & $0.58$ \smallskip\\
0120$-$1622AB & $1.35_{-0.06}^{+0.11}$ & $3.02_{-0.36}^{+0.36}$ & $-0.09$ & $4.1_{-0.9}^{+0.7}$ & $0.46_{-0.13}^{+0.20}$ & $0.62$ \smallskip\\
0215$+$1821AB & $5.21_{-0.60}^{+0.85}$ & $1.59_{-0.07}^{+0.17}$ & $+0.03$ & $0.108_{-0.032}^{+0.038}$ & $2.5_{-0.6}^{+0.4}$ & $0.60$ \smallskip\\
\end{longtable}

\begin{figure}
    \includegraphics[width=\textwidth]{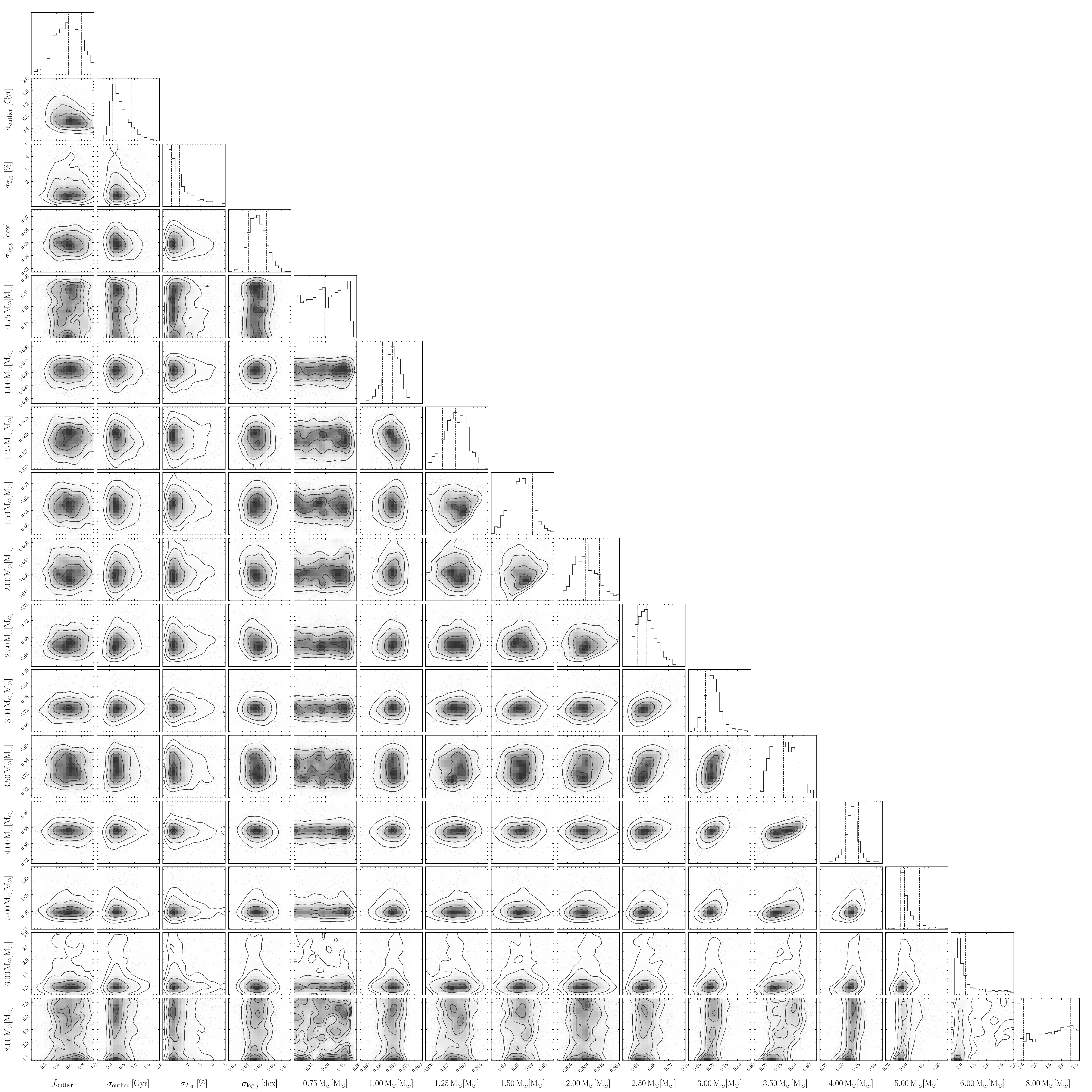}
    \caption{
    Corner plot for our main fit with a monotonic IFMR (Fit~1). The first 4 parameters
    correspond to the model hyper-parameters, with those thereafter corresponding
    to the final-masses at the given fixed initial-masses. The vertical dashed lines in the
    1D histograms correspond to the 16th, 50th and 84th percentiles, i.e. the same values
    given in Table~\ref{tab:resultsIFMR}.
    }
    \label{fig:cornerFit1}
\end{figure}

\begin{figure}
    \includegraphics[width=\textwidth]{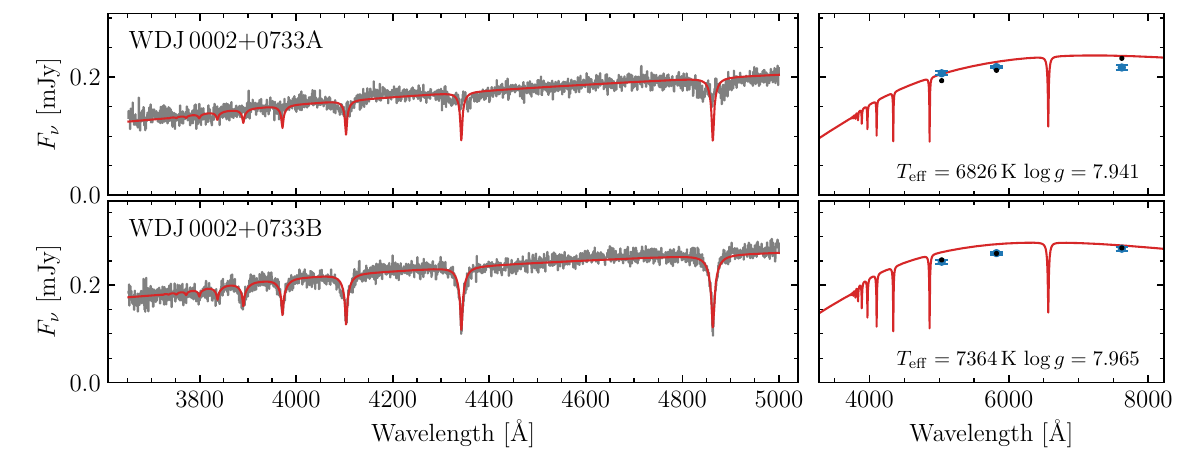}
    \includegraphics[width=\textwidth]{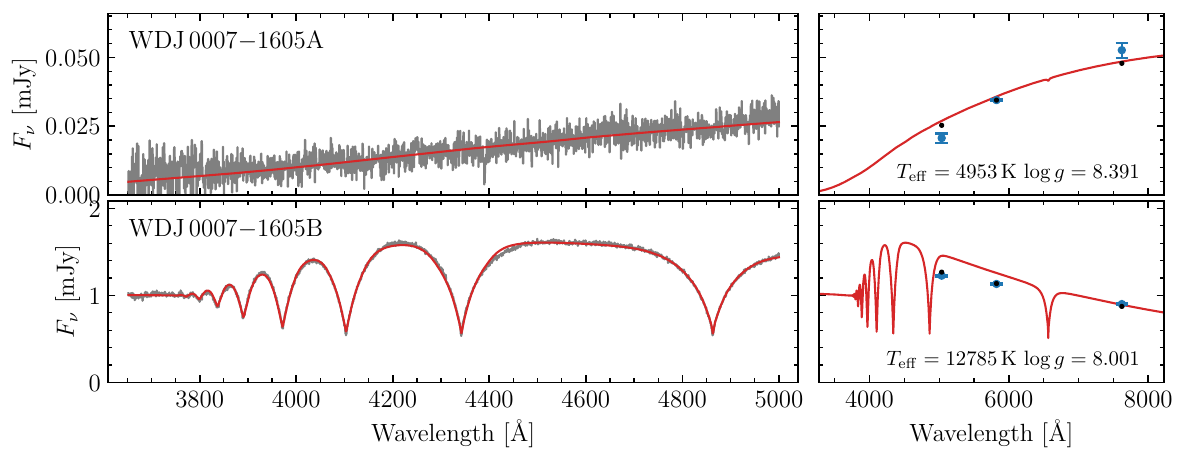}
    \includegraphics[width=\textwidth]{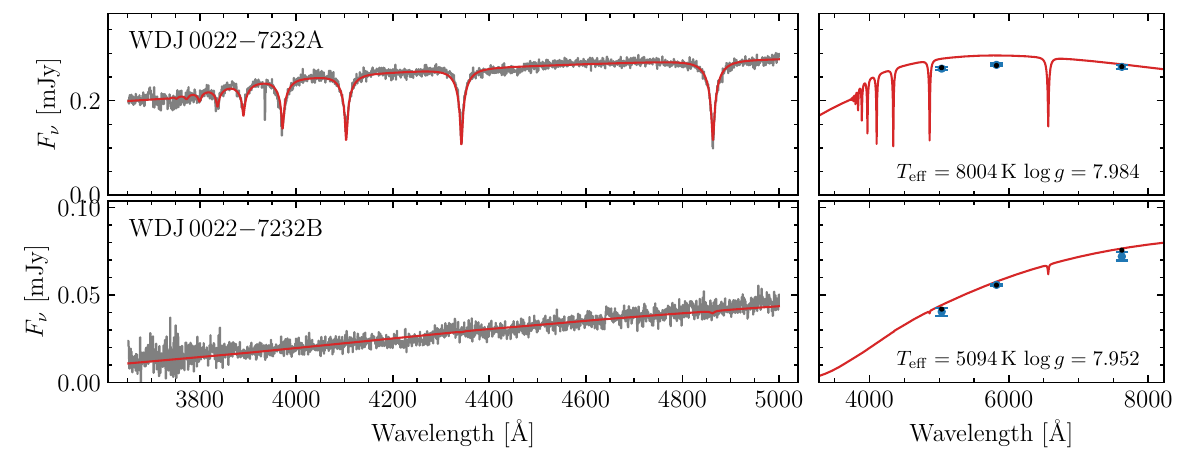}
    \caption{
        Combined spectroscopic and photometric fits to our double white dwarf sample.
        The left panels show the best fitting models to the FORS2 spectra, where the
        data have had their fluxes corrected against the models. The \emph{Gaia} fluxes
        are shown in the right hand panels (blue points) with the corresponding
        synthetic values (black points). The models include the effects of interstellar
        reddening where appropriate. Parameter uncertainties can be found in
        Table~\ref{tab:dwds_fits}.
    }
    \label{fig:fits01}
\end{figure}

\begin{figure}
    \includegraphics[width=\textwidth]{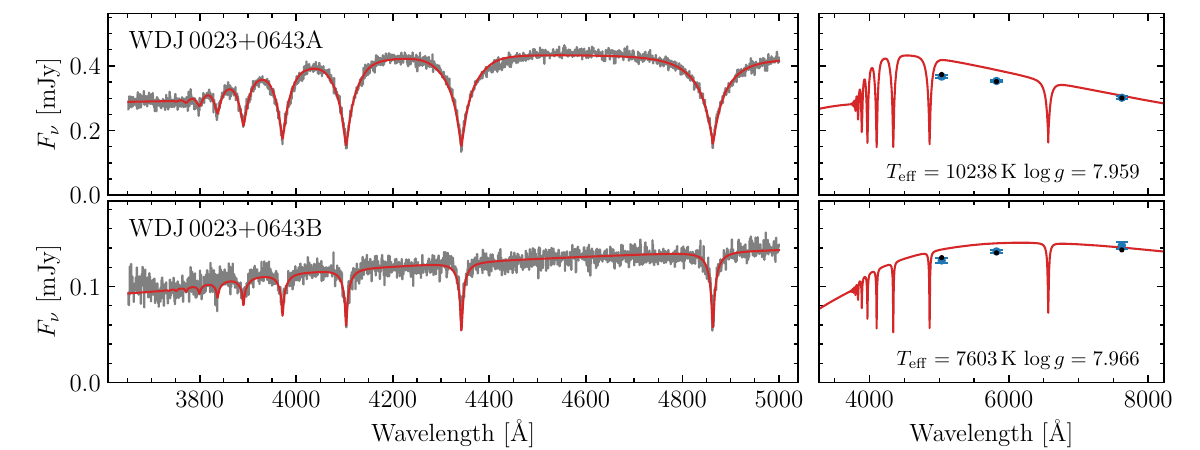}
    \includegraphics[width=\textwidth]{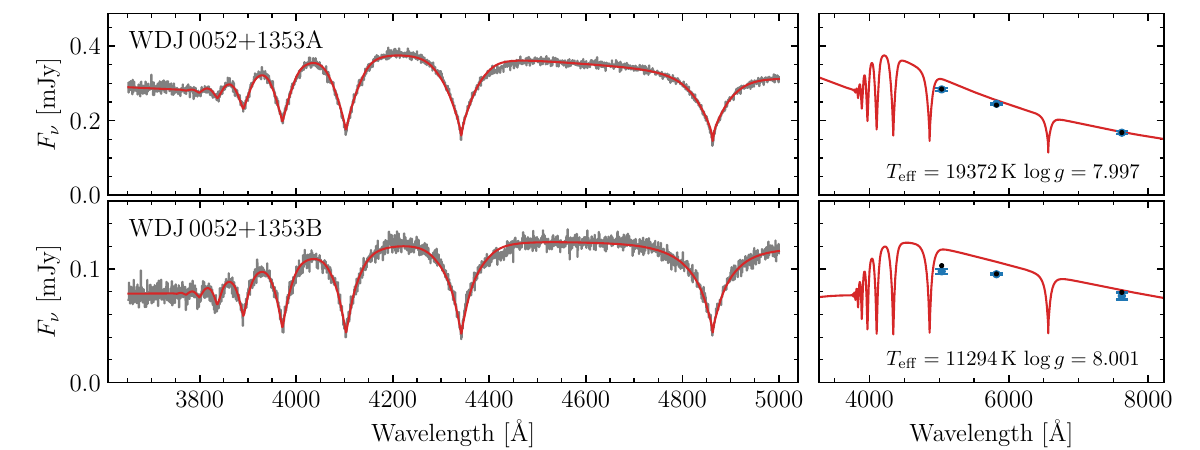}
    \includegraphics[width=\textwidth]{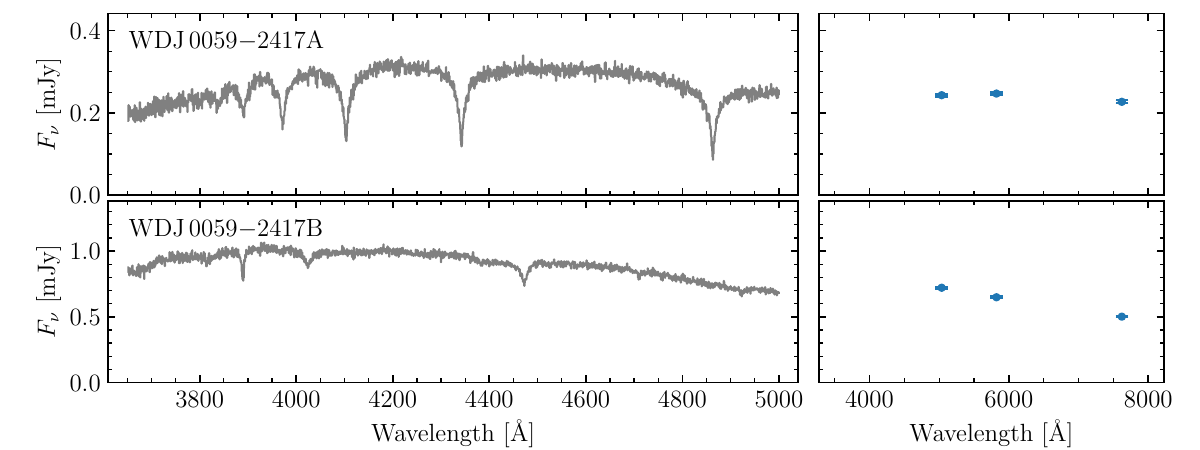}
    \caption{
        Same as Figure~\ref{fig:fits01}, continued.
    }
    \label{fig:fits02}
\end{figure}

\begin{figure}
    \includegraphics[width=\textwidth]{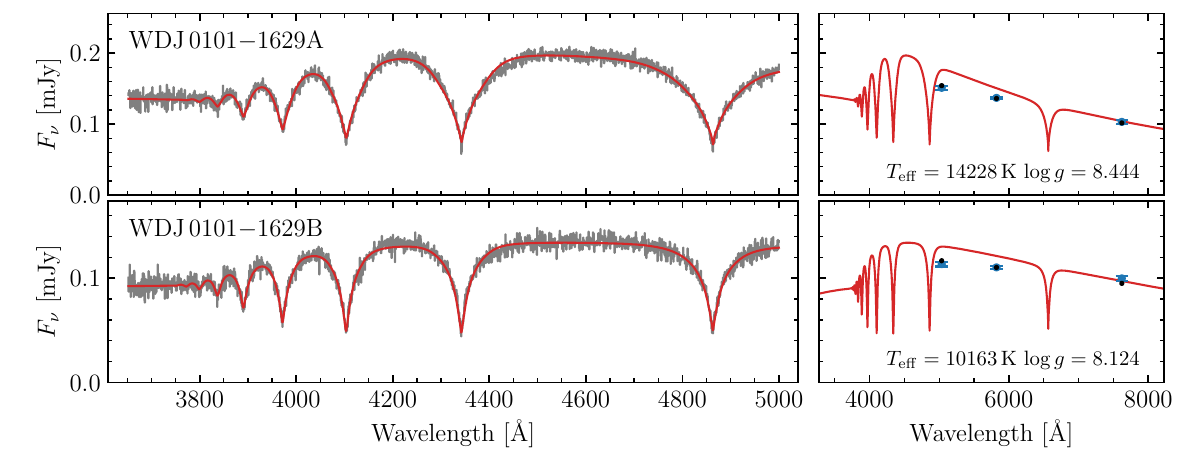}
    \includegraphics[width=\textwidth]{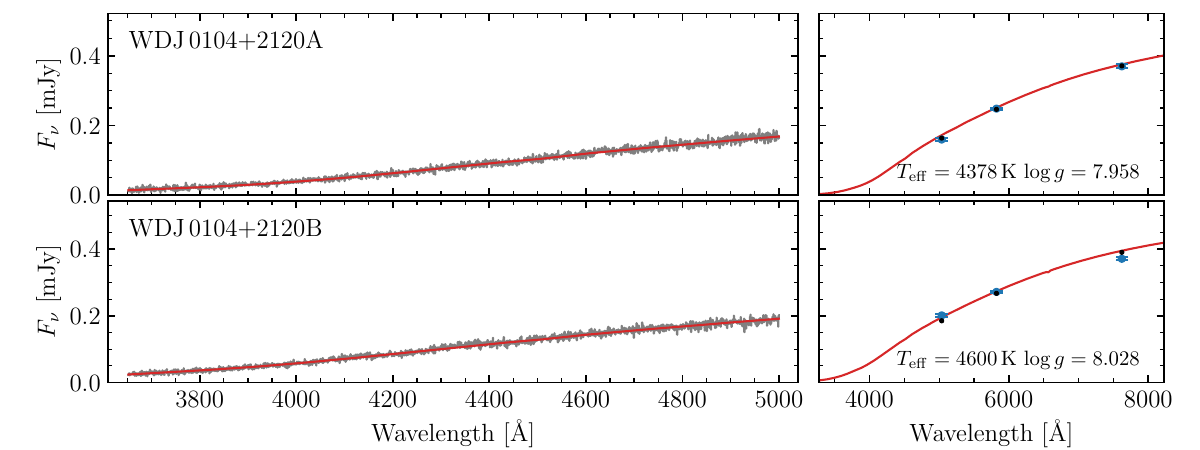}
    \includegraphics[width=\textwidth]{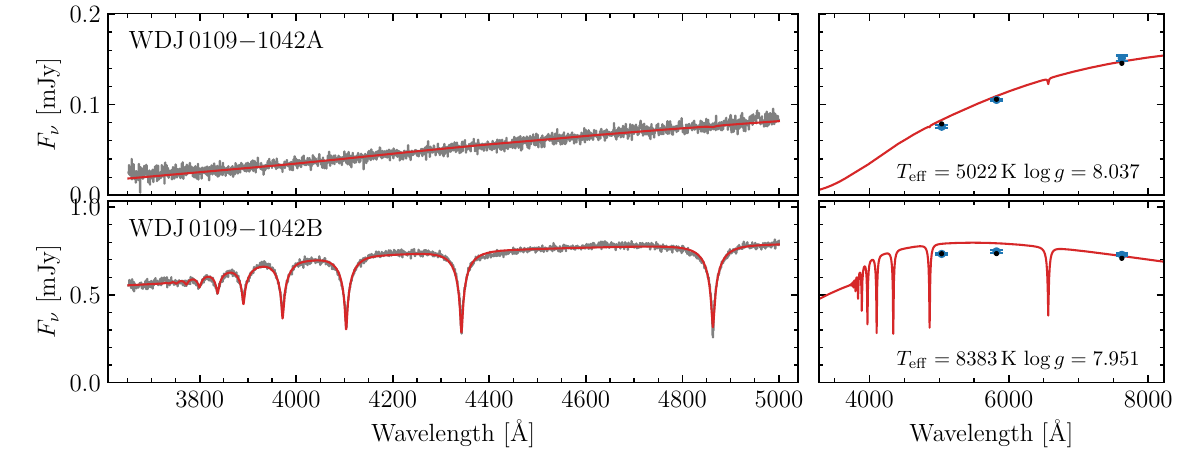}
    \caption{
        Same as Figure~\ref{fig:fits01}, continued.
    }
    \label{fig:fits03}
\end{figure}

\begin{figure}
    \includegraphics[width=\textwidth]{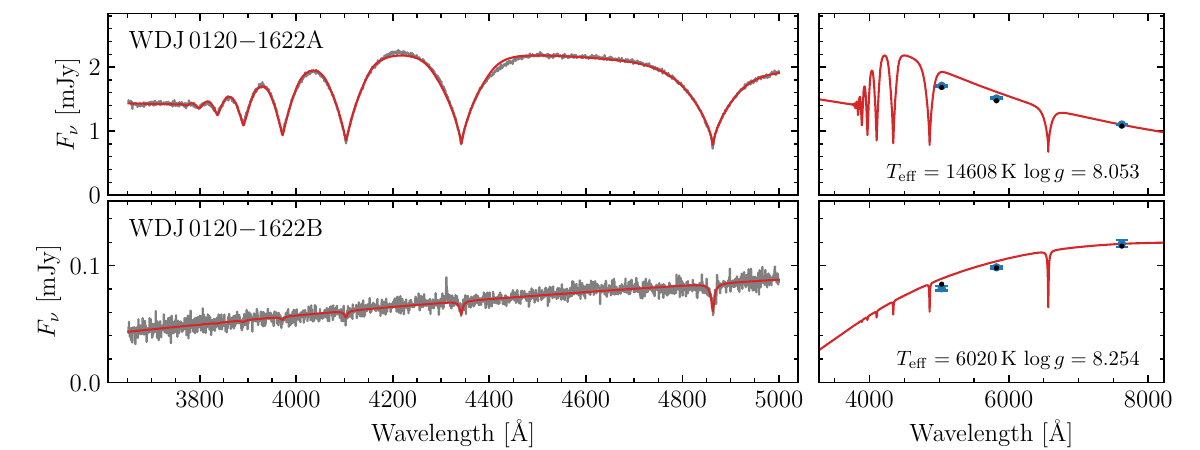}
    \includegraphics[width=\textwidth]{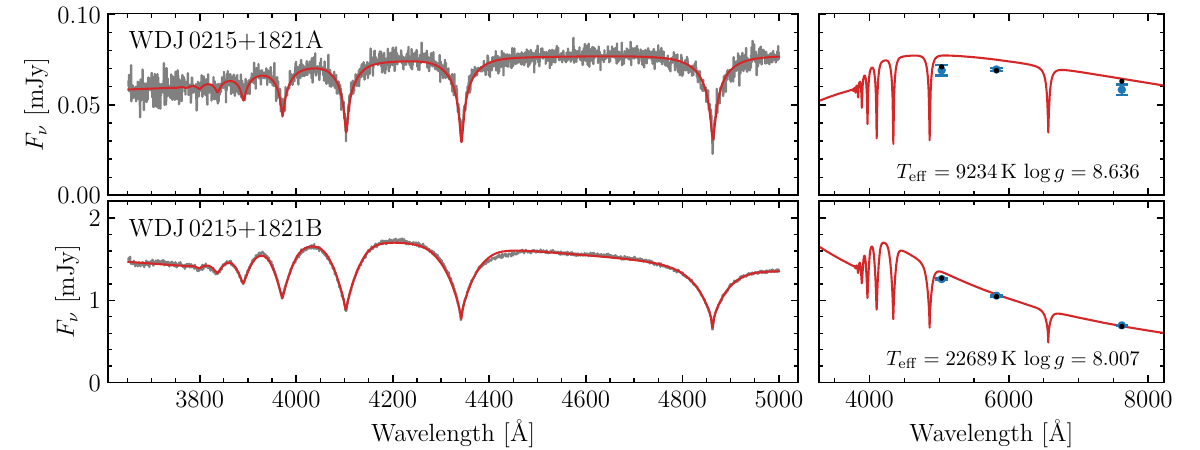}
    \includegraphics[width=\textwidth]{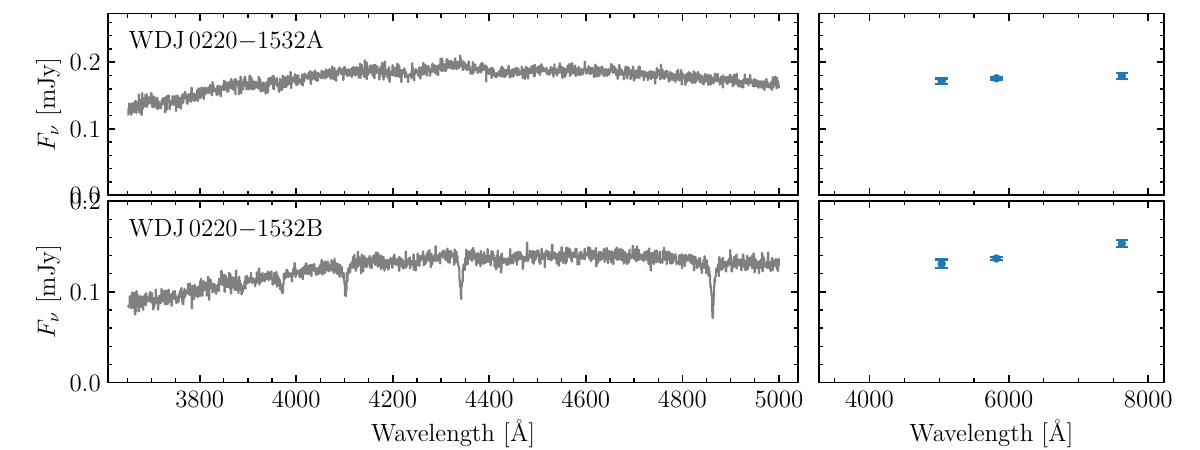}
    \caption{
        Same as Figure~\ref{fig:fits01}, continued.
    }
    \label{fig:fits04}
\end{figure}

\begin{figure}
    \includegraphics[width=\textwidth]{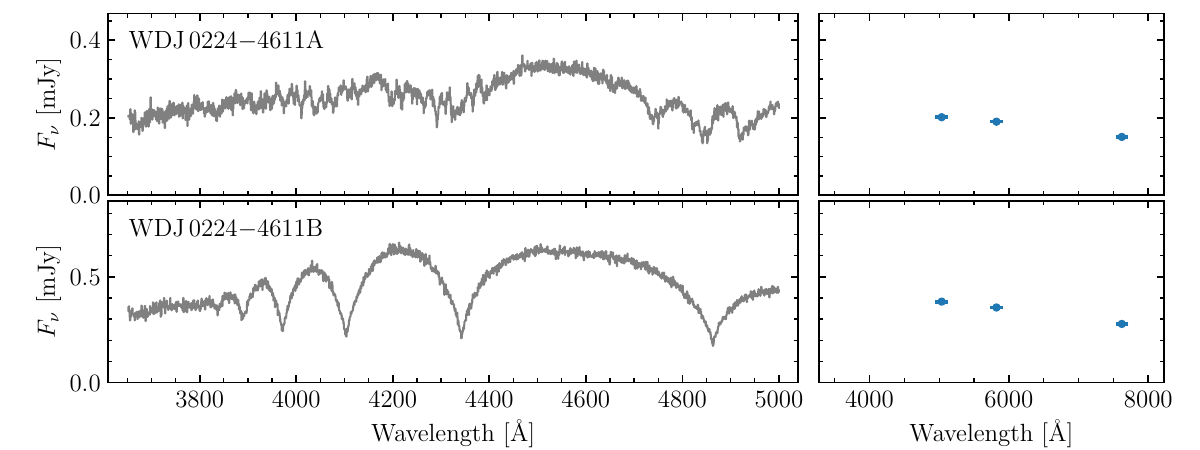}
    \includegraphics[width=\textwidth]{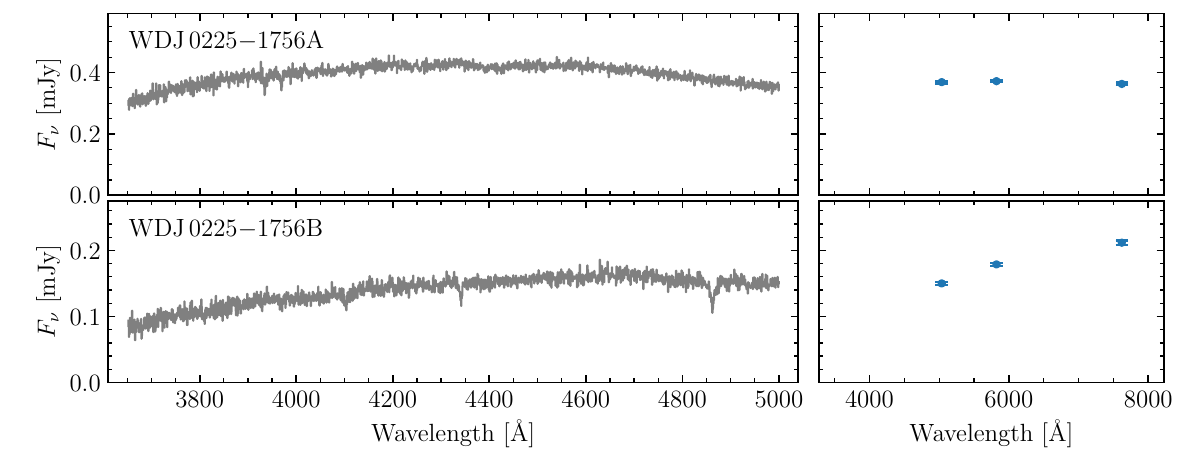}
    \includegraphics[width=\textwidth]{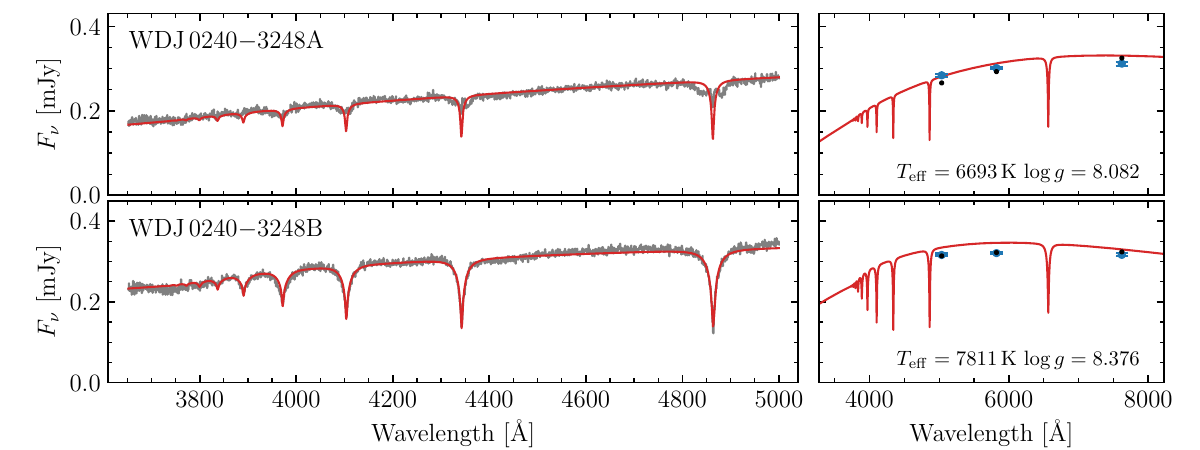}
    \caption{
        Same as Figure~\ref{fig:fits01}, continued.
    }
    \label{fig:fits05}
\end{figure}

\begin{figure}
    \includegraphics[width=\textwidth]{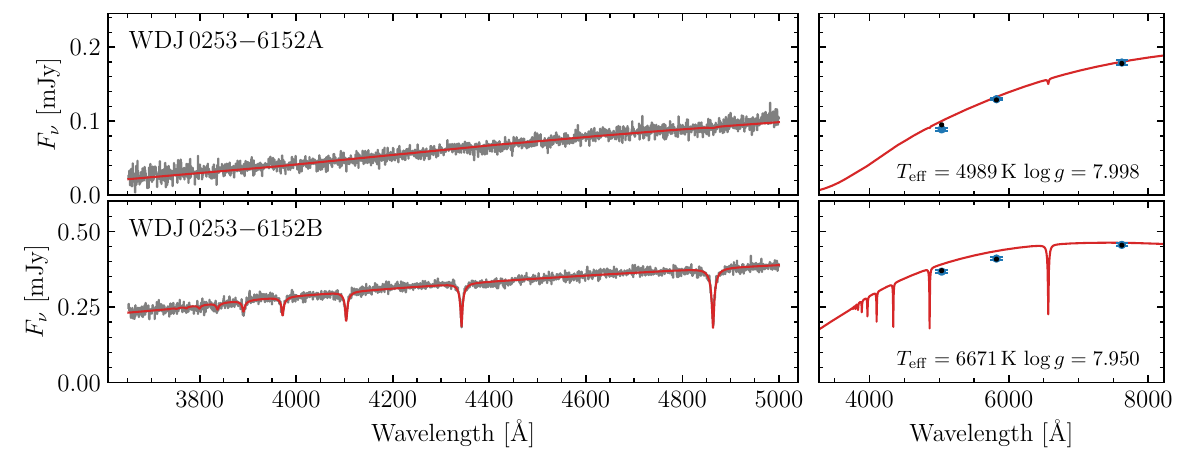}
    \includegraphics[width=\textwidth]{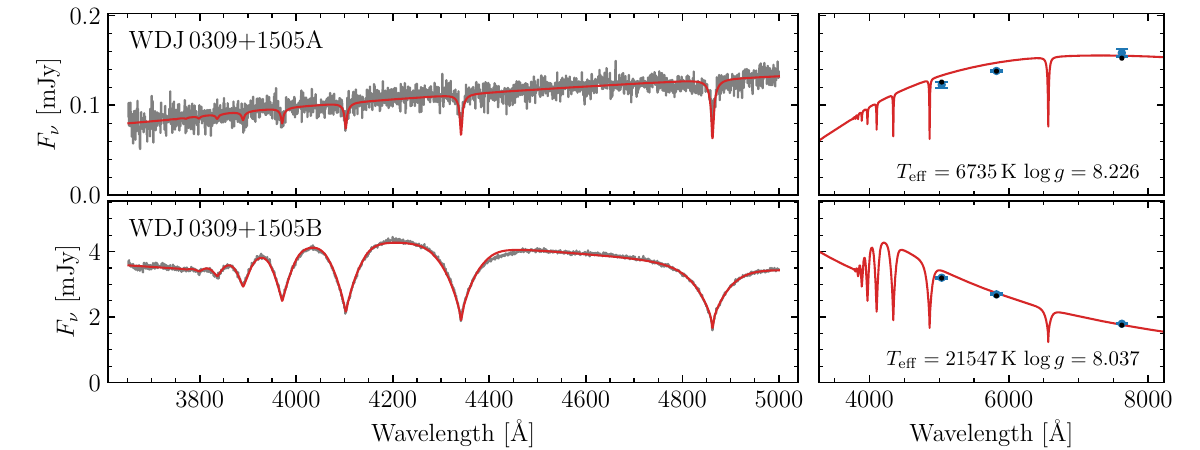}
    \includegraphics[width=\textwidth]{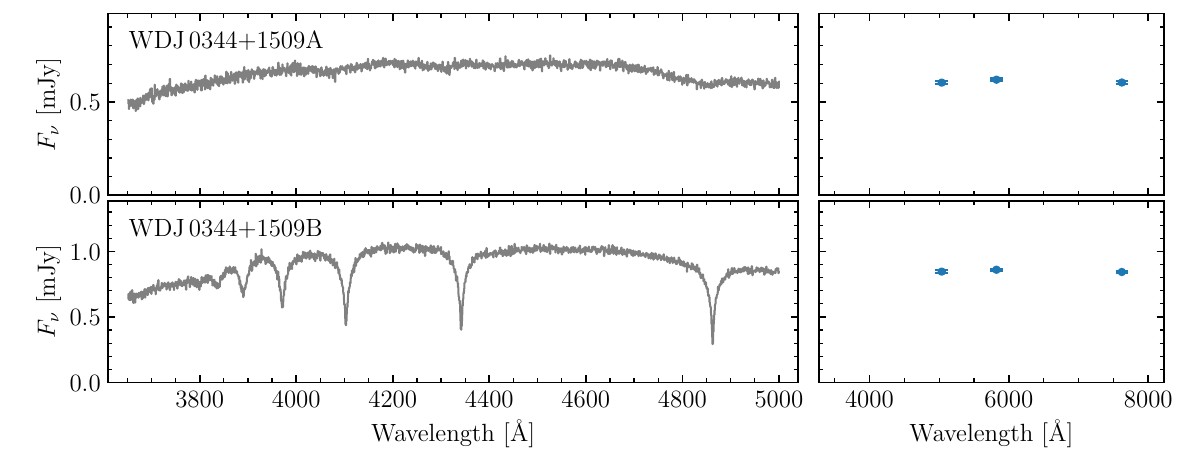}
    \caption{
        Same as Figure~\ref{fig:fits01}, continued.
    }
    \label{fig:fits06}
\end{figure}

\begin{figure}
    \includegraphics[width=\textwidth]{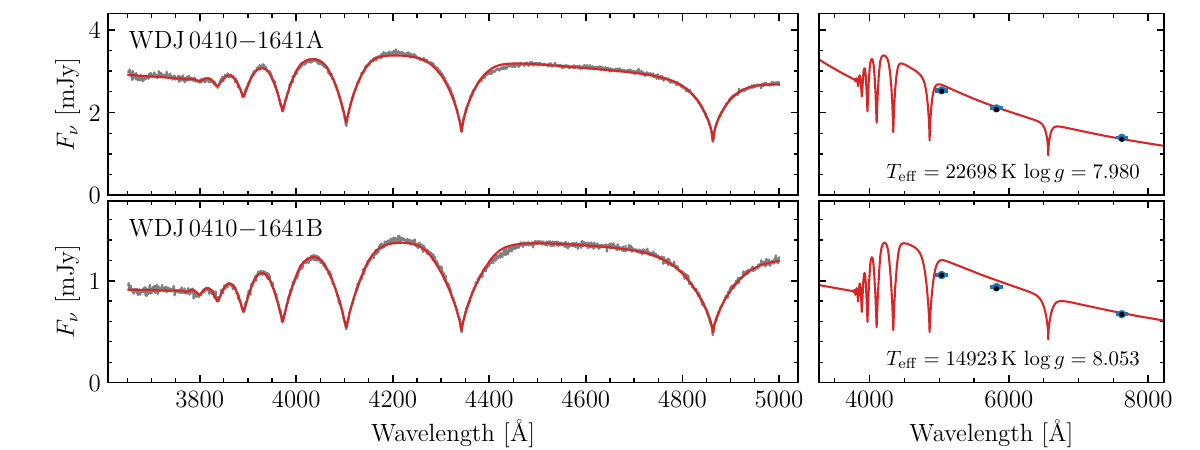}
    \includegraphics[width=\textwidth]{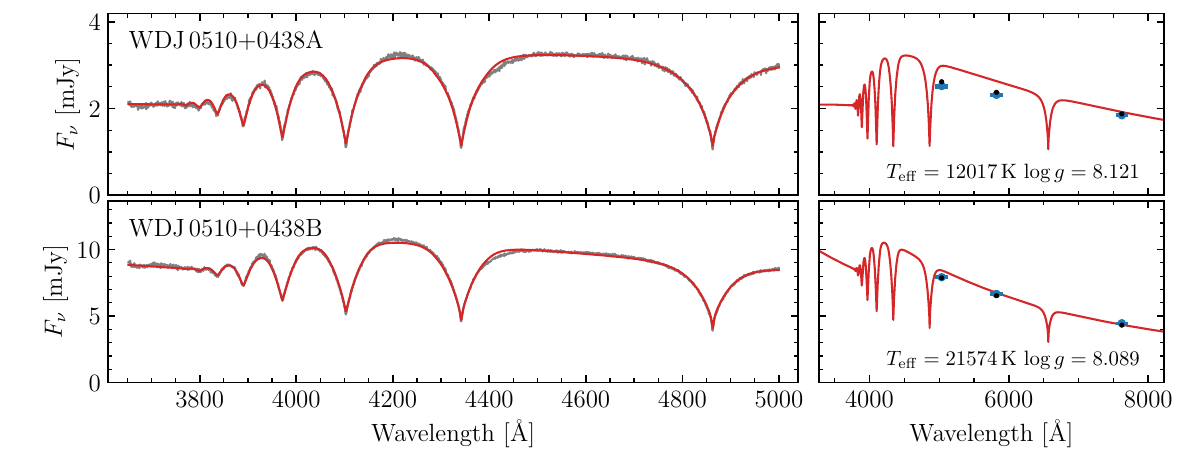}
    \includegraphics[width=\textwidth]{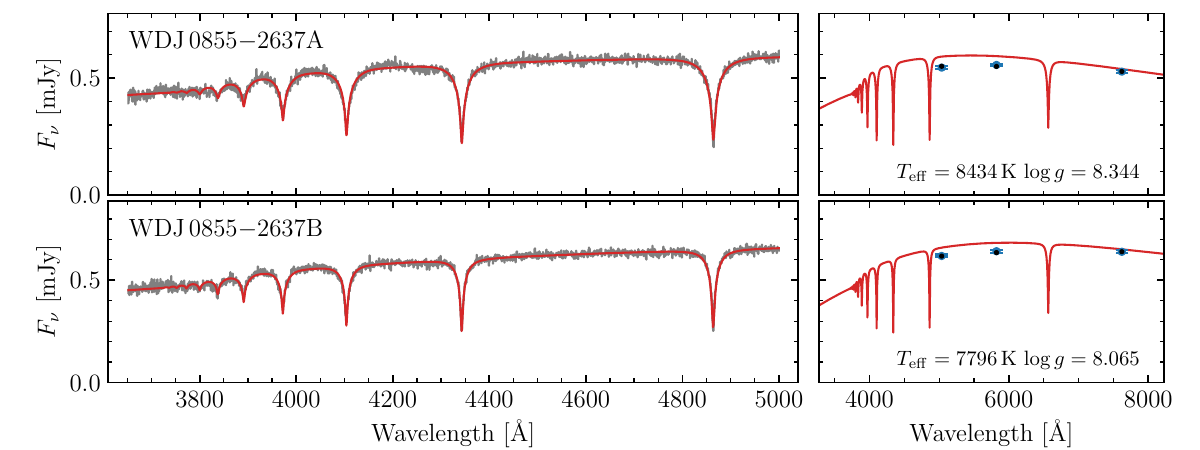}
    \caption{
        Same as Figure~\ref{fig:fits01}, continued.
    }
    \label{fig:fits07}
\end{figure}

\begin{figure}
    \includegraphics[width=\textwidth]{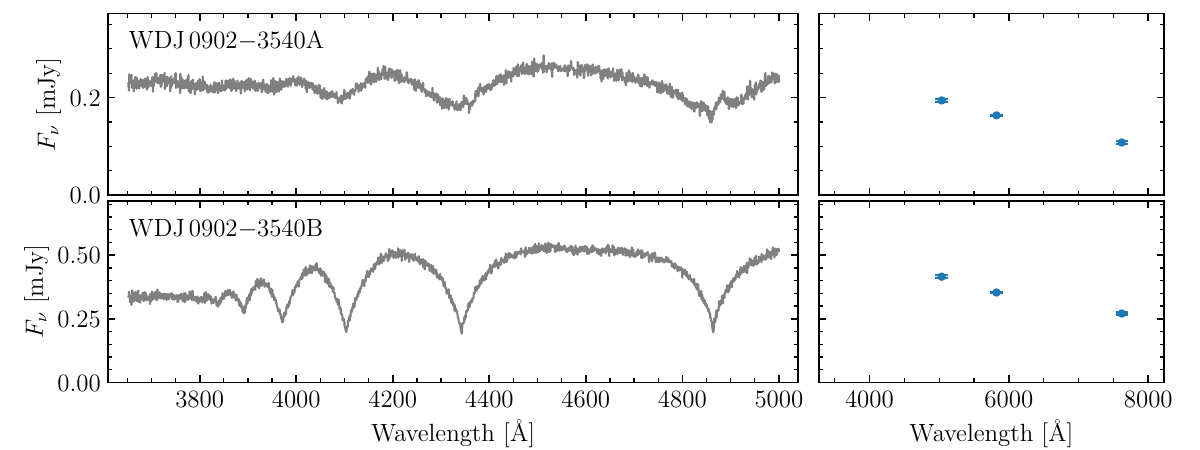}
    \includegraphics[width=\textwidth]{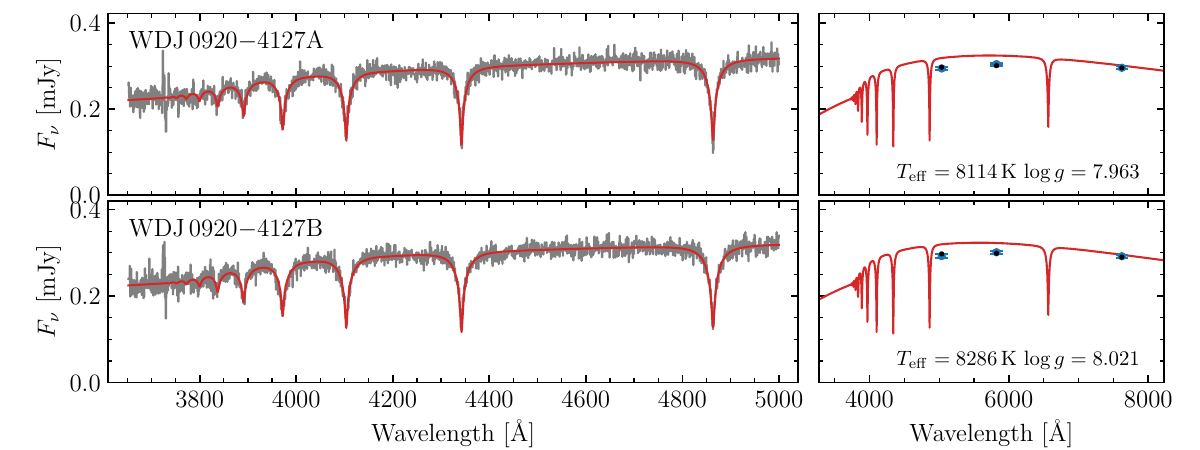}
    \includegraphics[width=\textwidth]{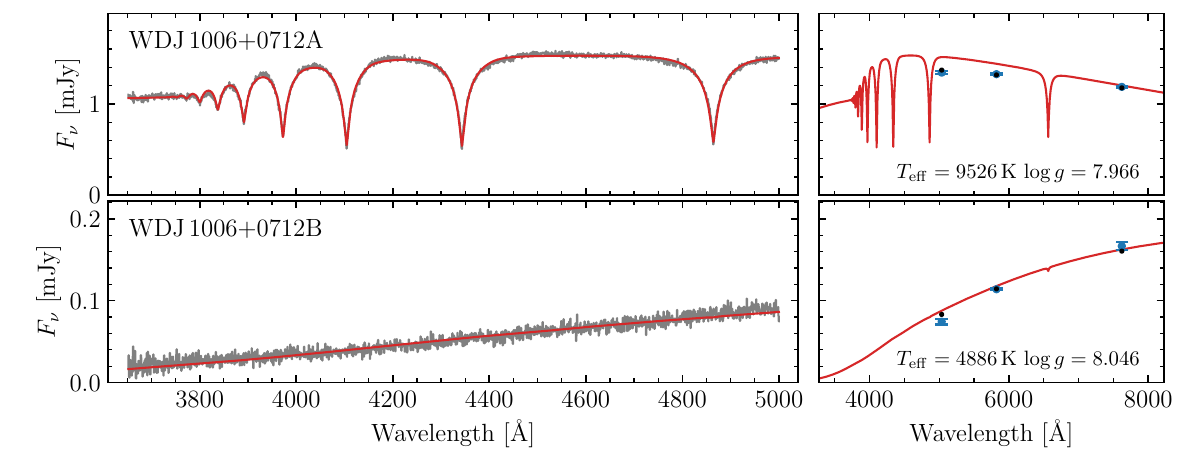}
    \caption{
        Same as Figure~\ref{fig:fits01}, continued.
    }
    \label{fig:fits08}
\end{figure}

\begin{figure}
    \includegraphics[width=\textwidth]{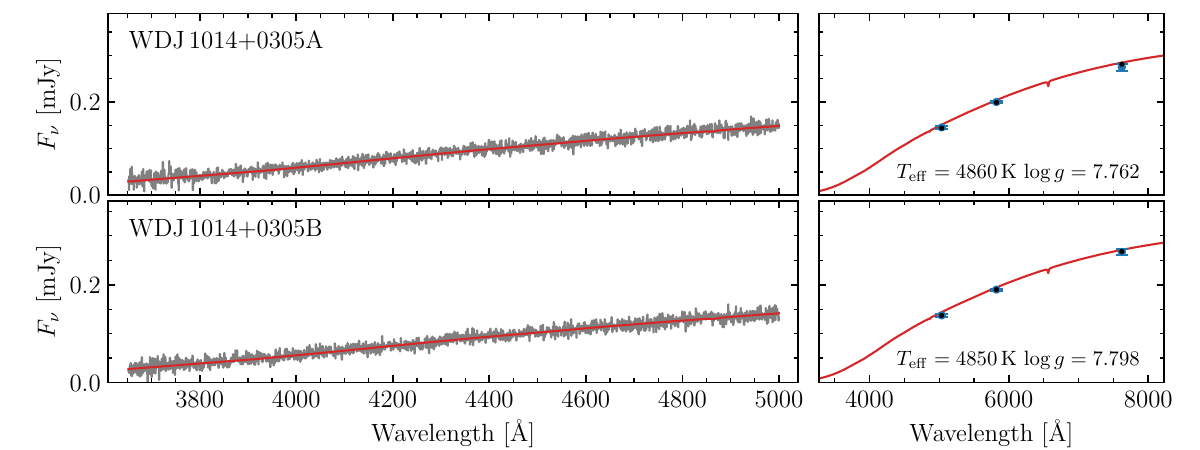}
    \includegraphics[width=\textwidth]{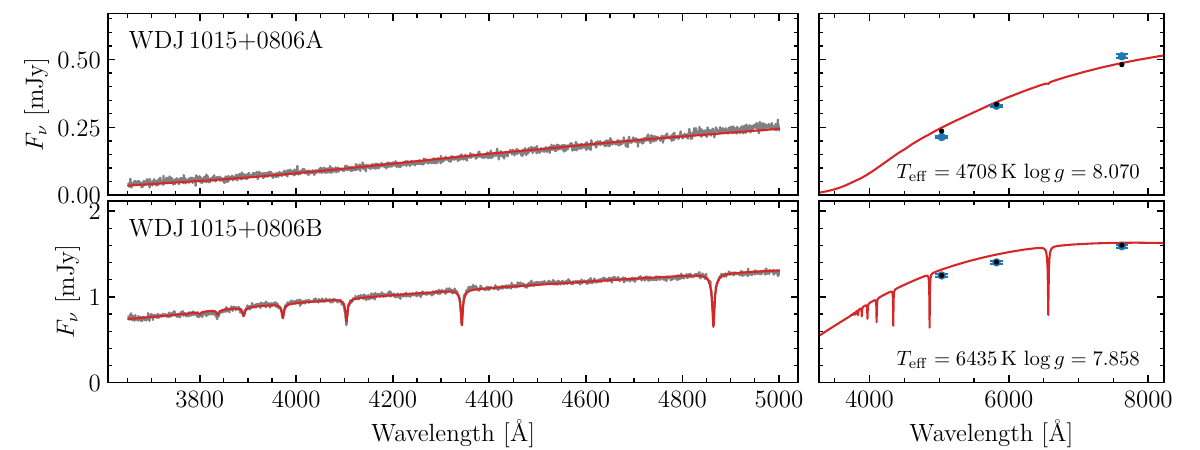}
    \includegraphics[width=\textwidth]{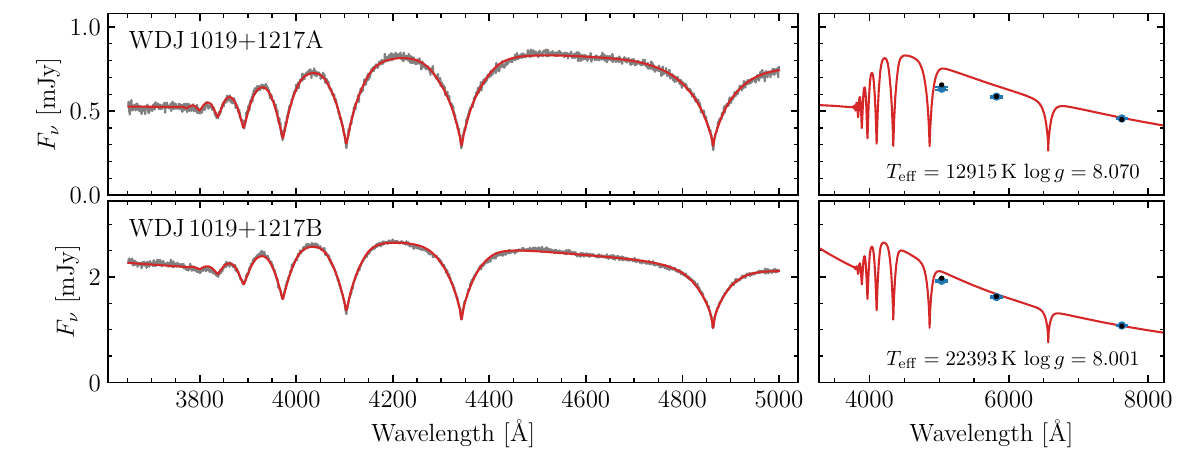}
    \caption{
        Same as Figure~\ref{fig:fits01}, continued.
    }
    \label{fig:fits09}
\end{figure}

\begin{figure}
    \includegraphics[width=\textwidth]{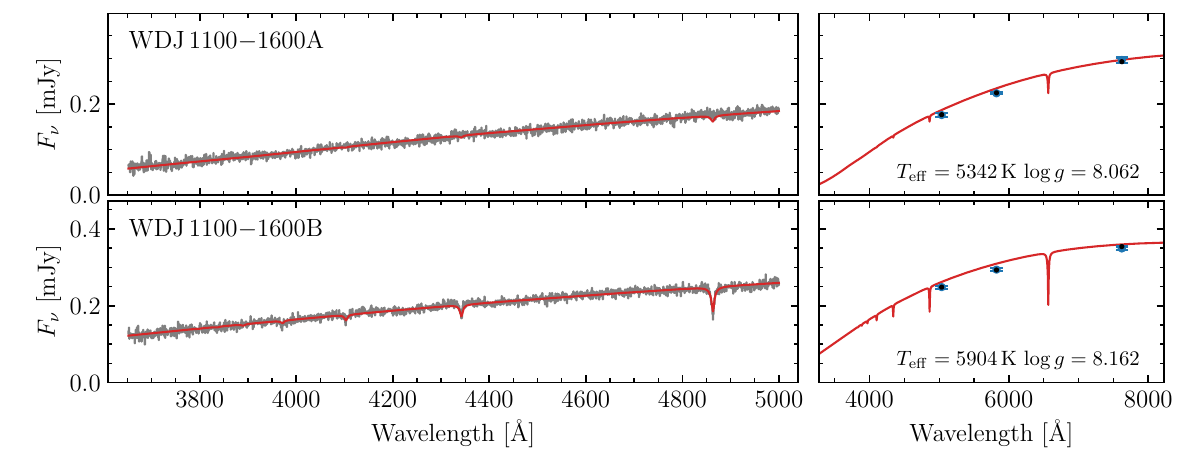}
    \includegraphics[width=\textwidth]{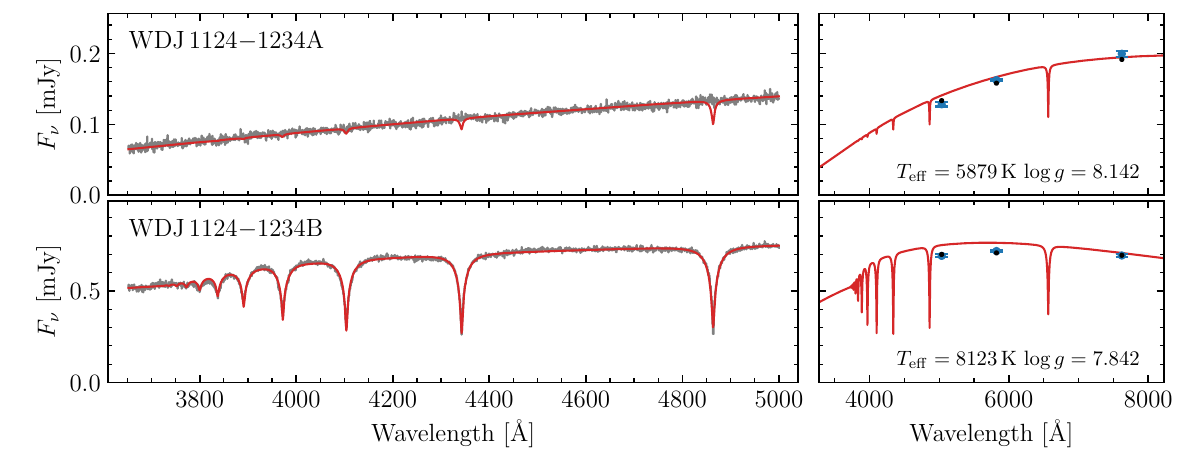}
    \includegraphics[width=\textwidth]{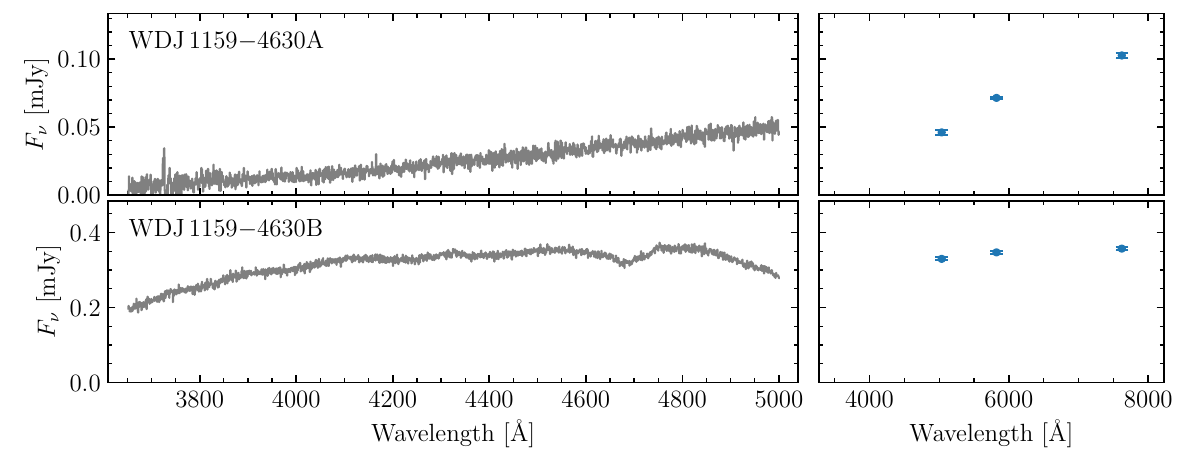}
    \caption{
        Same as Figure~\ref{fig:fits01}, continued.
    }
    \label{fig:fits10}
\end{figure}

\begin{figure}
    \includegraphics[width=\textwidth]{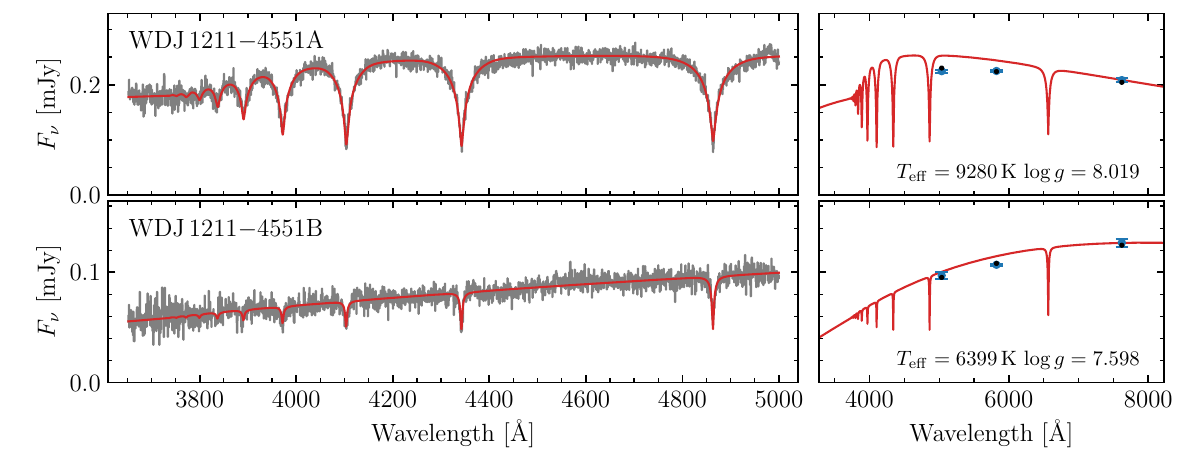}
    \includegraphics[width=\textwidth]{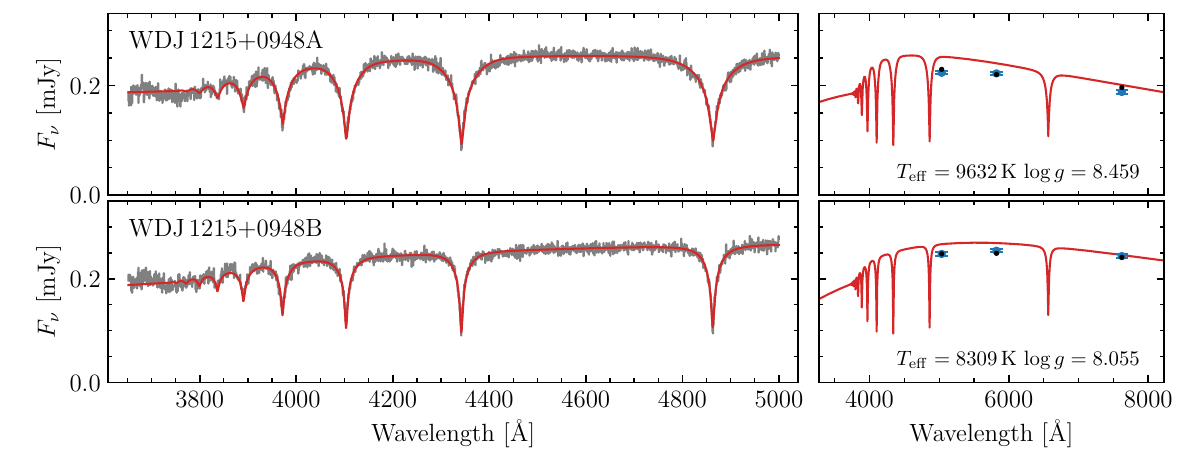}
    \includegraphics[width=\textwidth]{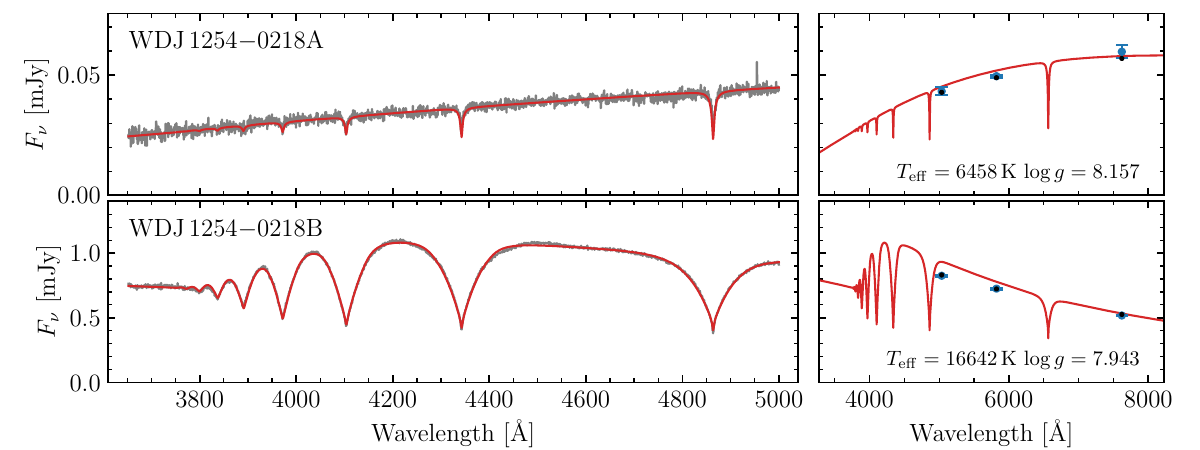}
    \caption{
        Same as Figure~\ref{fig:fits01}, continued.
    }
    \label{fig:fits11}
\end{figure}

\begin{figure}
    \includegraphics[width=\textwidth]{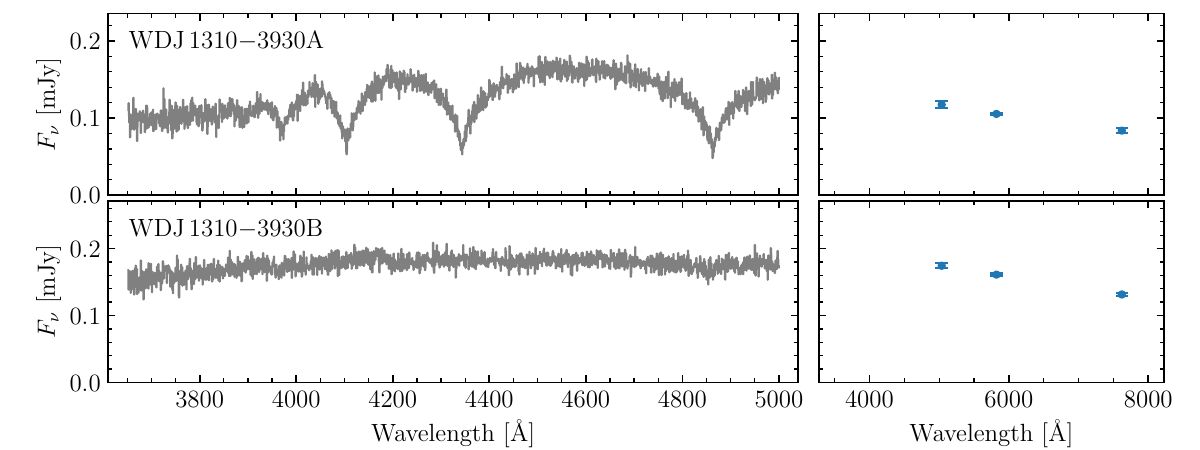}
    \includegraphics[width=\textwidth]{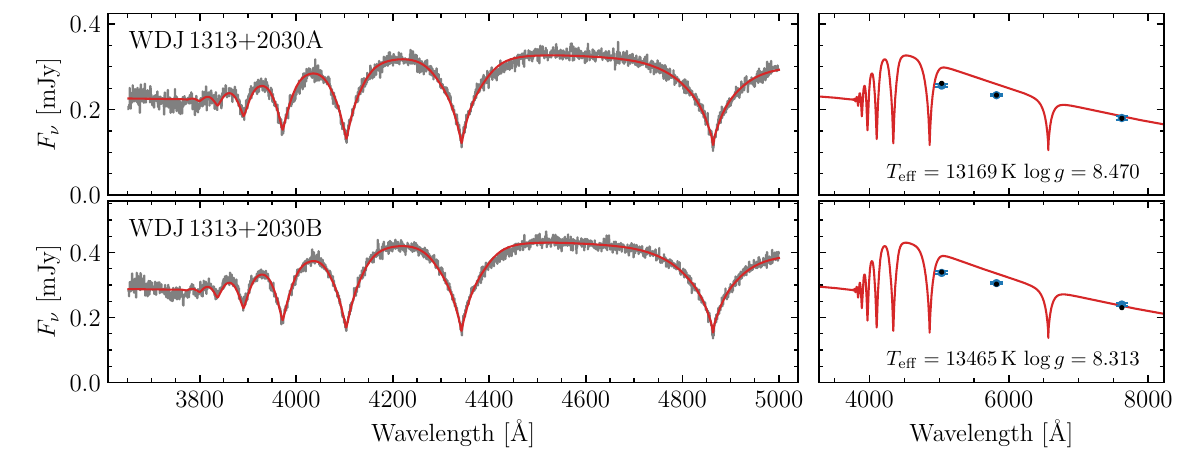}
    \includegraphics[width=\textwidth]{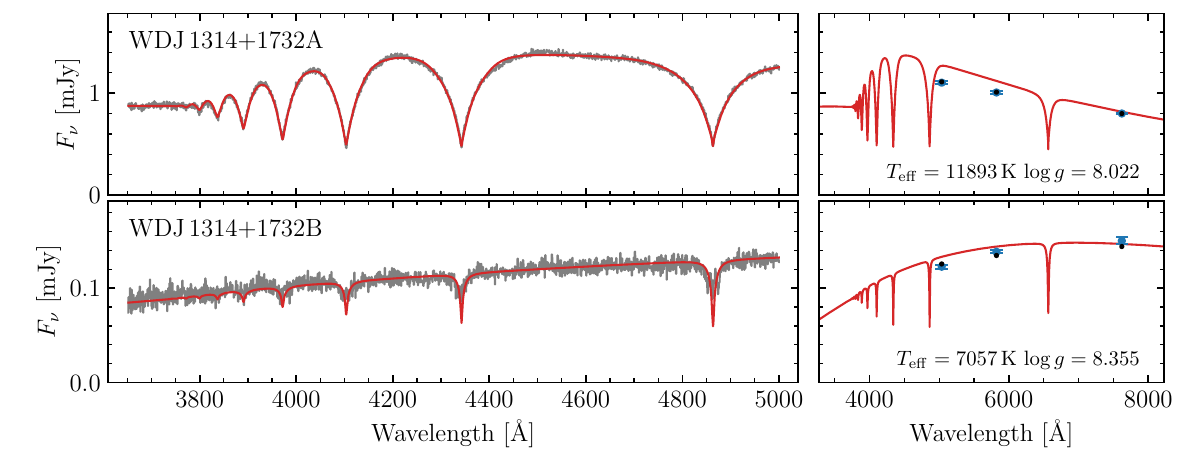}
    \caption{
        Same as Figure~\ref{fig:fits01}, continued.
    }
    \label{fig:fits12}
\end{figure}

\begin{figure}
    \includegraphics[width=\textwidth]{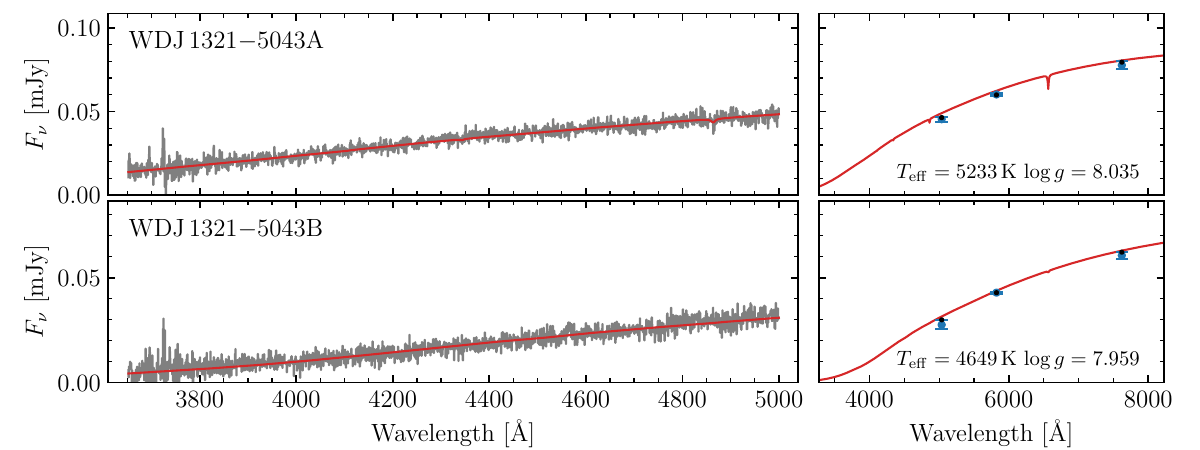}
    \includegraphics[width=\textwidth]{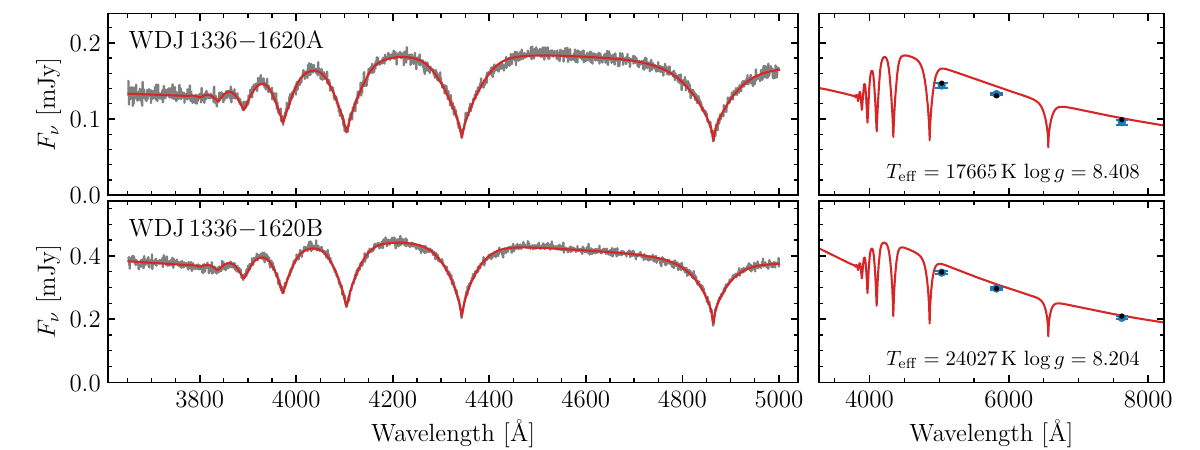}
    \includegraphics[width=\textwidth]{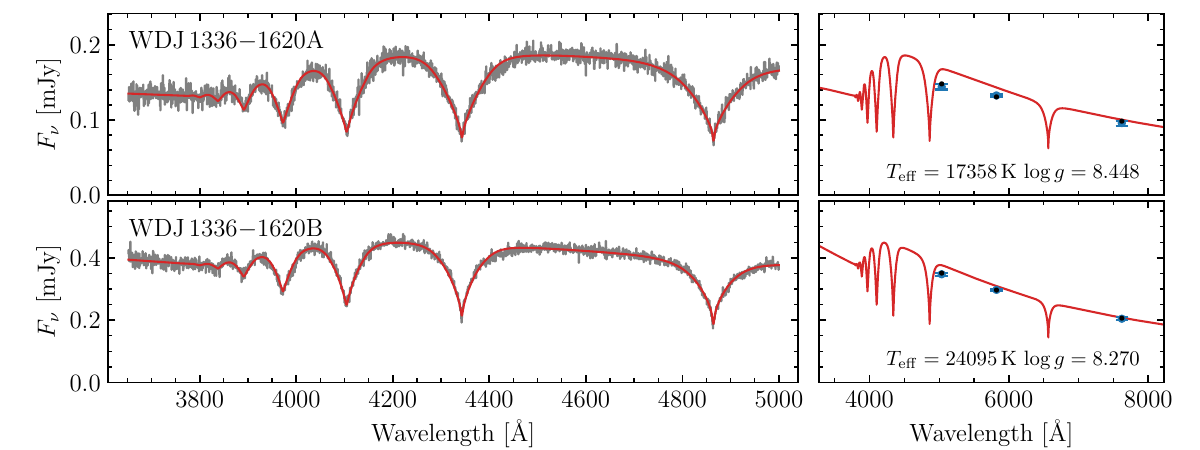}
    \caption{
        Same as Figure~\ref{fig:fits01}, continued.
    }
    \label{fig:fits13}
\end{figure}

\begin{figure}
    \includegraphics[width=\textwidth]{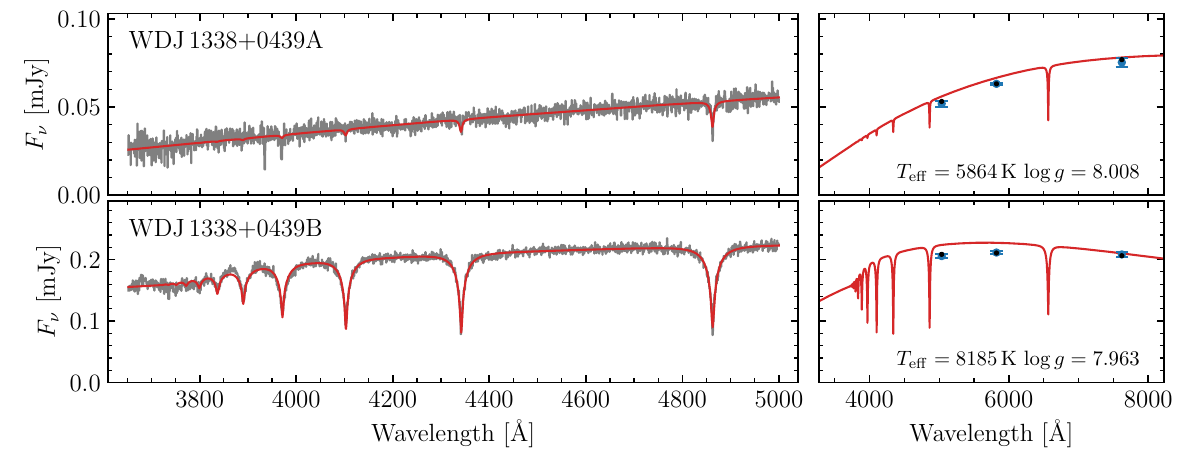}
    \includegraphics[width=\textwidth]{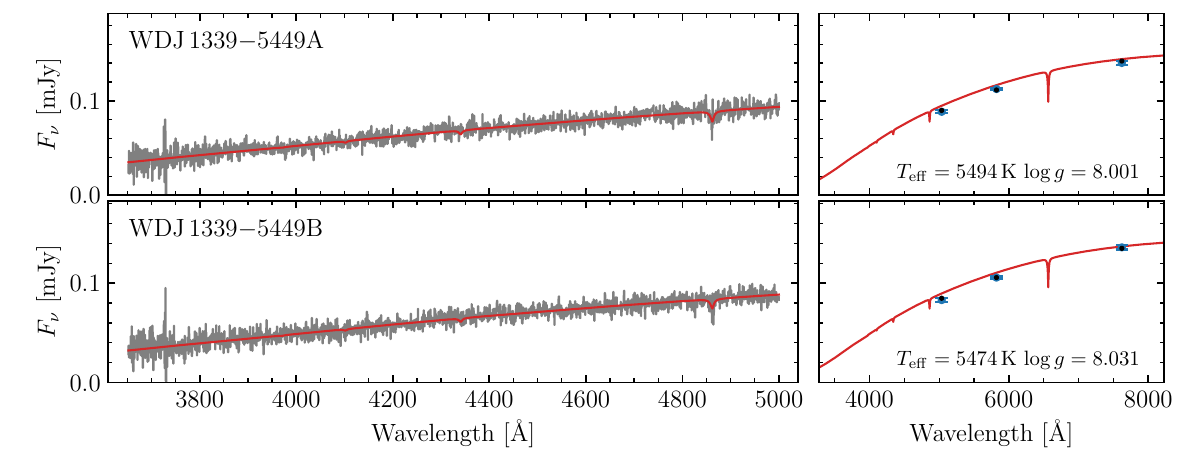}
    \includegraphics[width=\textwidth]{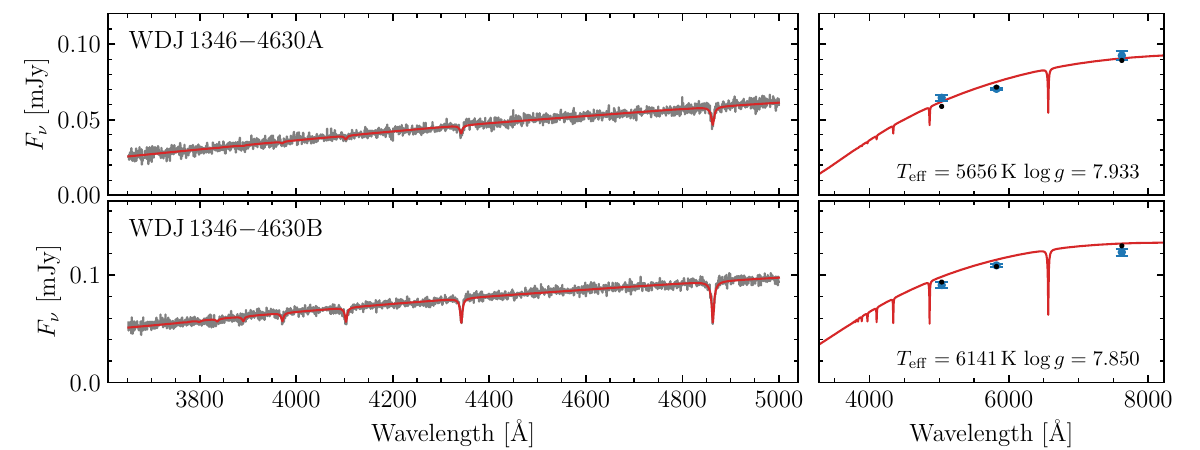}
    \caption{
        Same as Figure~\ref{fig:fits01}, continued.
    }
    \label{fig:fits14}
\end{figure}

\begin{figure}
    \includegraphics[width=\textwidth]{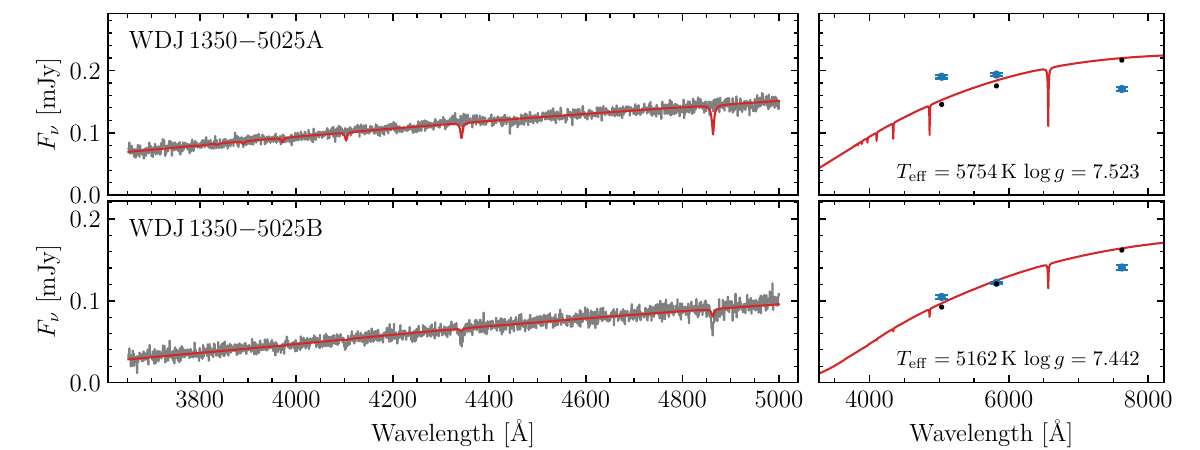}
    \includegraphics[width=\textwidth]{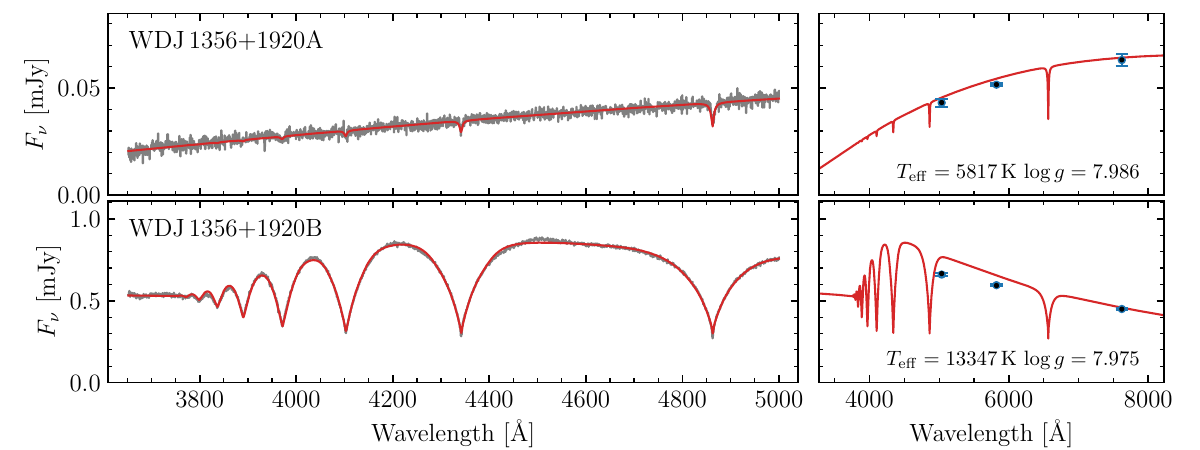}
    \includegraphics[width=\textwidth]{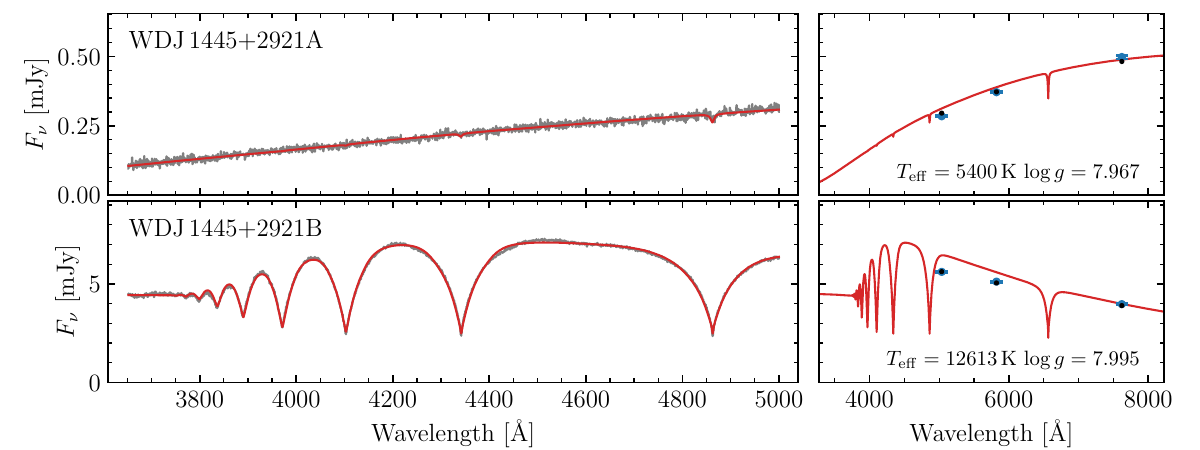}
    \caption{
        Same as Figure~\ref{fig:fits01}, continued.
    }
    \label{fig:fits15}
\end{figure}

\begin{figure}
    \includegraphics[width=\textwidth]{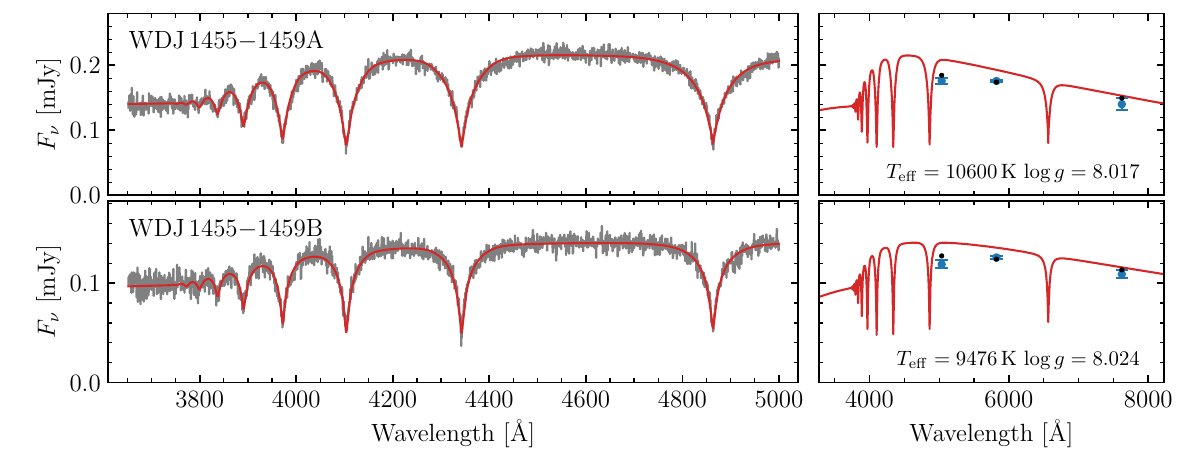}
    \includegraphics[width=\textwidth]{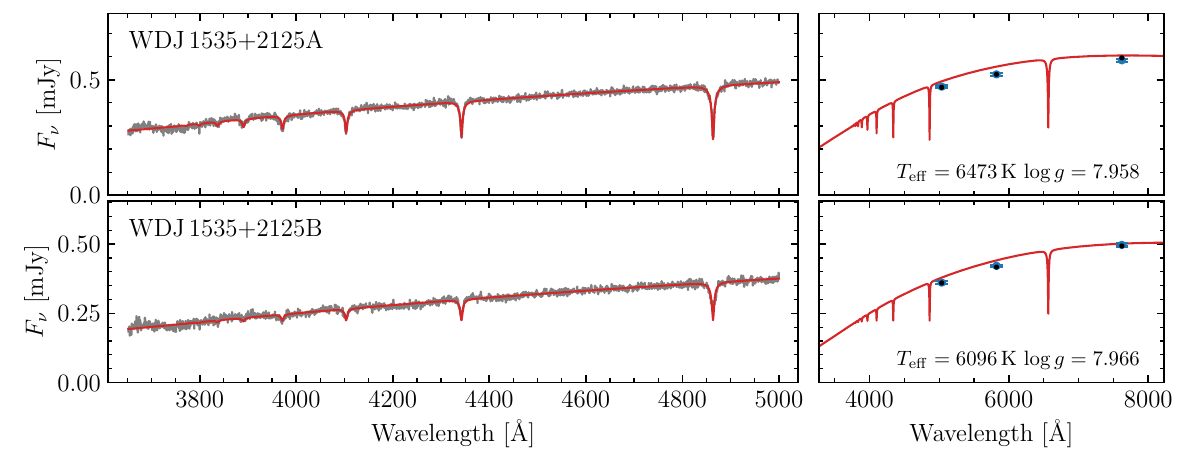}
    \includegraphics[width=\textwidth]{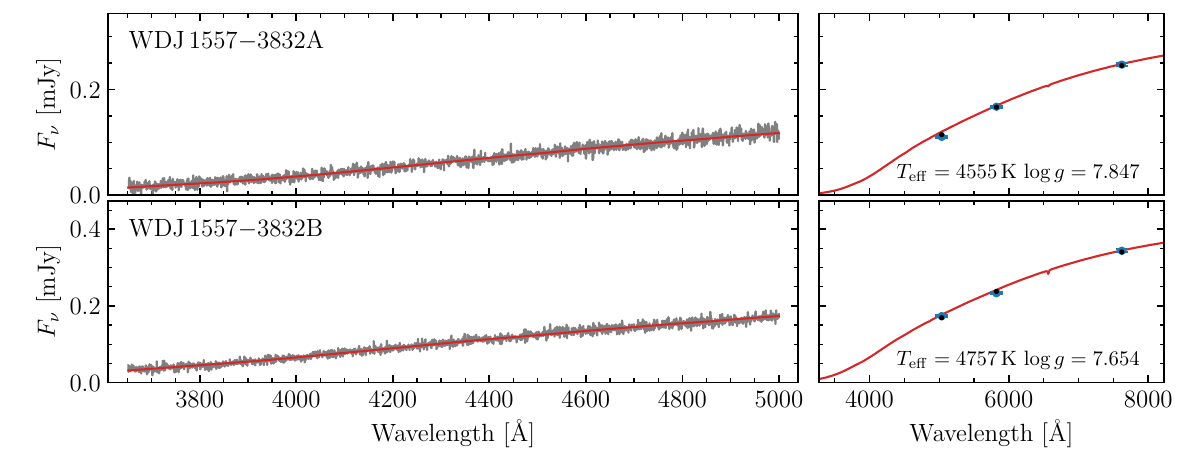}
    \caption{
        Same as Figure~\ref{fig:fits01}, continued.
    }
    \label{fig:fits16}
\end{figure}

\begin{figure}
    \includegraphics[width=\textwidth]{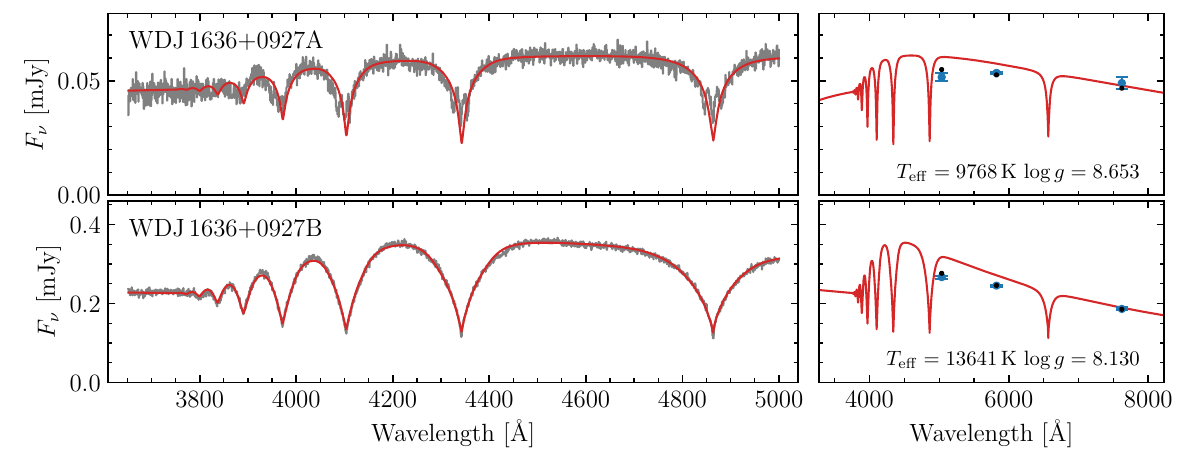}
    \includegraphics[width=\textwidth]{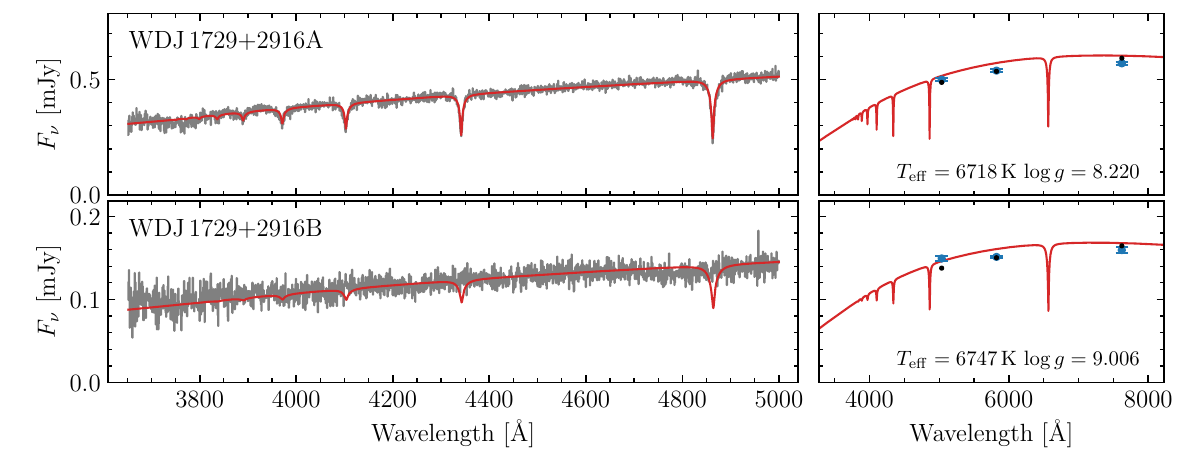}
    \includegraphics[width=\textwidth]{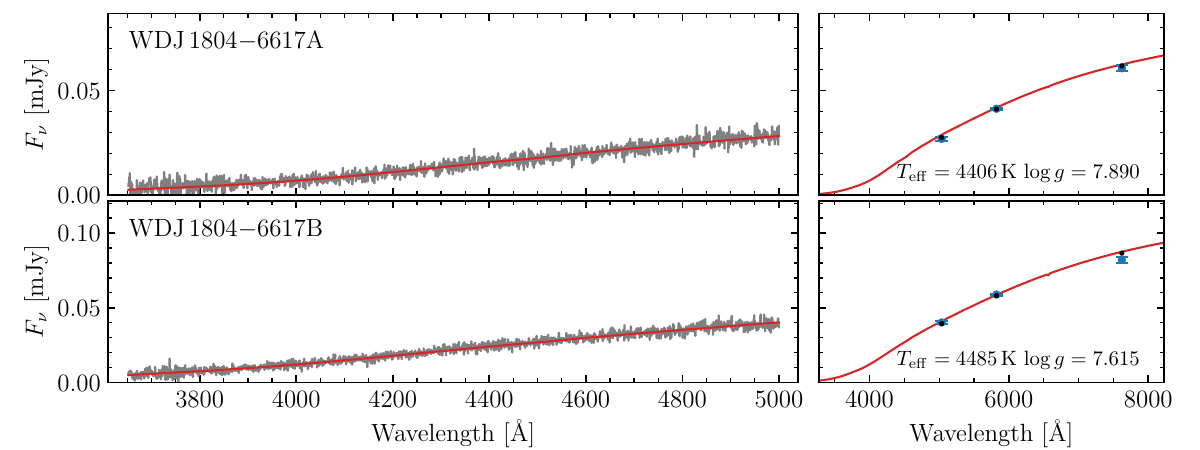}
    \caption{
        Same as Figure~\ref{fig:fits01}, continued.
    }
    \label{fig:fits17}
\end{figure}

\begin{figure}
    \includegraphics[width=\textwidth]{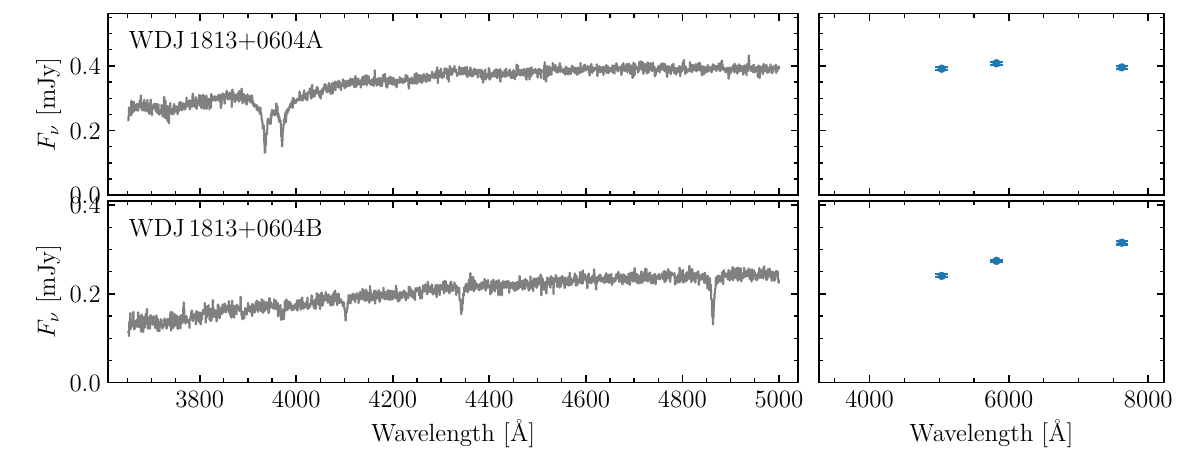}
    \includegraphics[width=\textwidth]{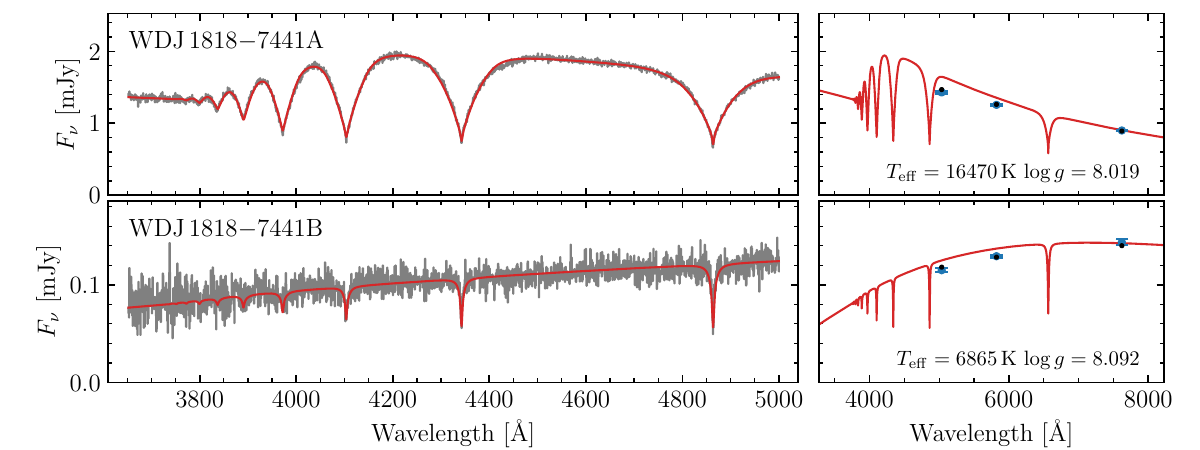}
    \includegraphics[width=\textwidth]{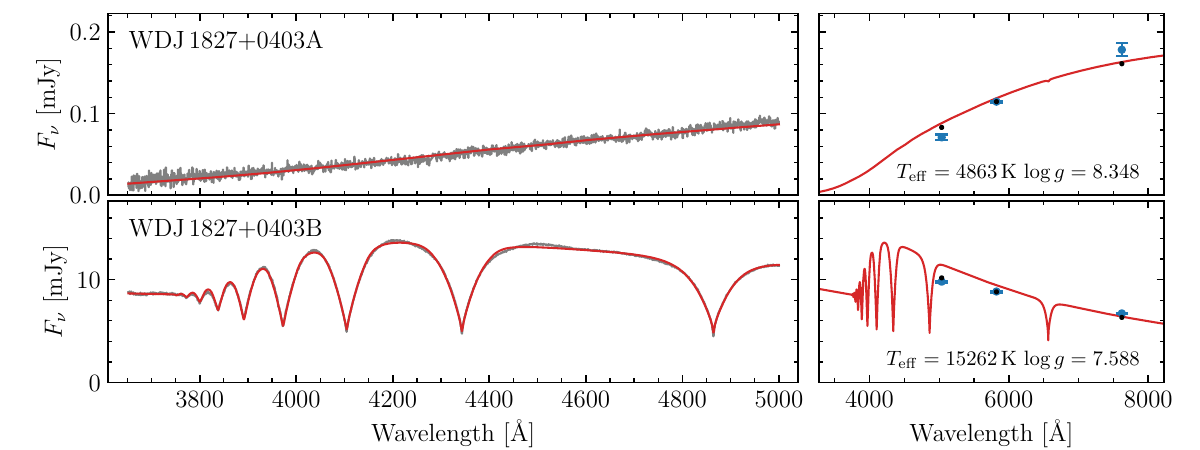}
    \caption{
        Same as Figure~\ref{fig:fits01}, continued.
    }
    \label{fig:fits18}
\end{figure}

\begin{figure}
    \includegraphics[width=\textwidth]{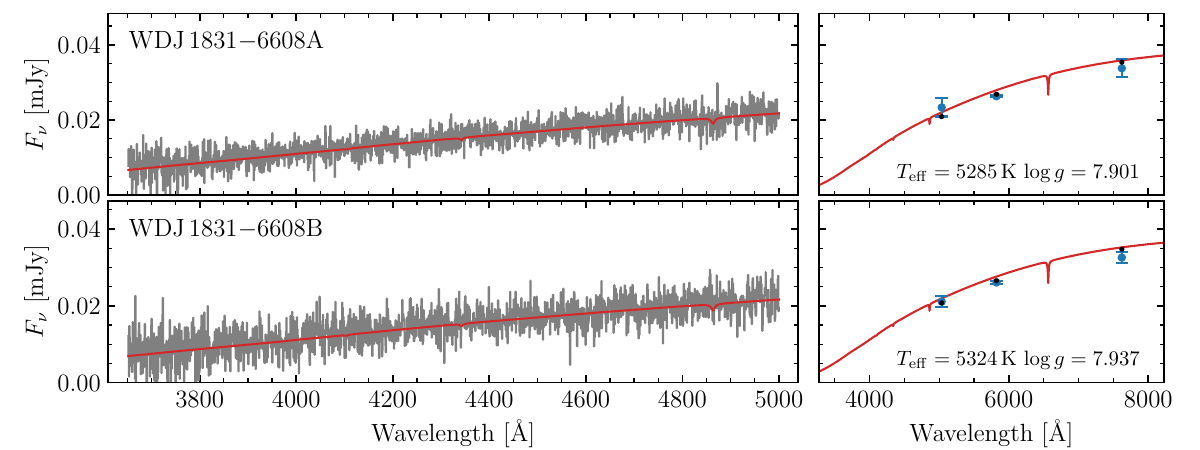}
    \includegraphics[width=\textwidth]{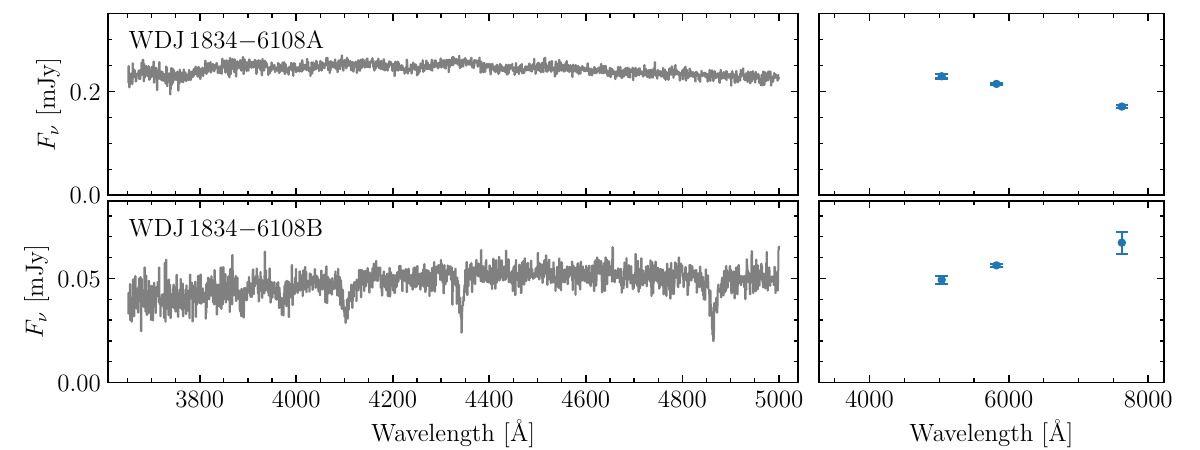}
    \includegraphics[width=\textwidth]{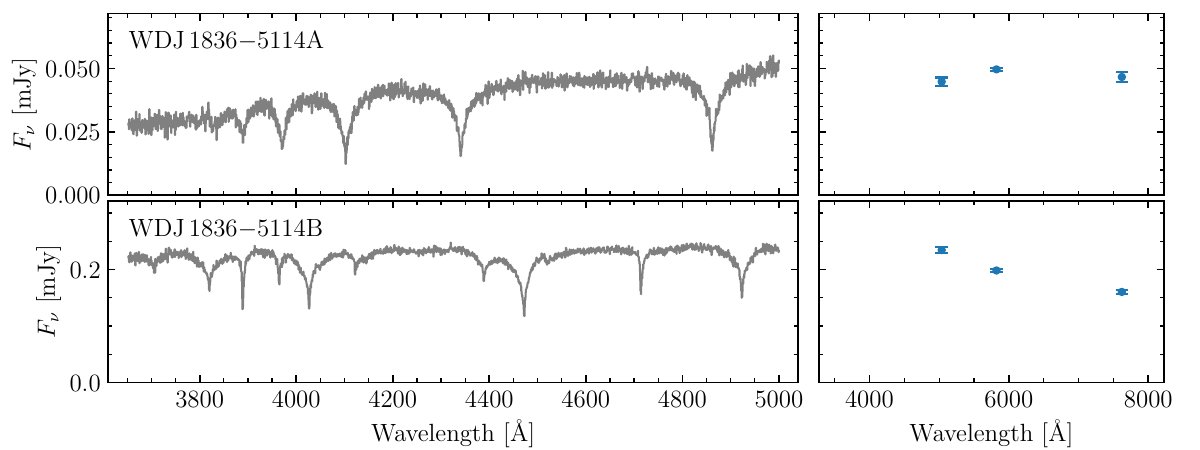}
    \caption{
        Same as Figure~\ref{fig:fits01}, continued.
    }
    \label{fig:fits19}
\end{figure}

\begin{figure}
    \includegraphics[width=\textwidth]{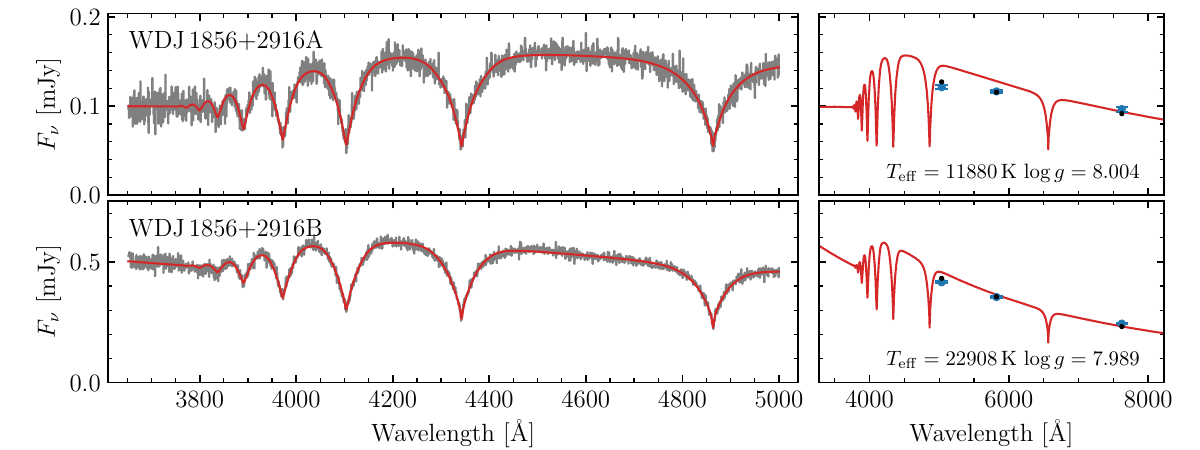}
    \includegraphics[width=\textwidth]{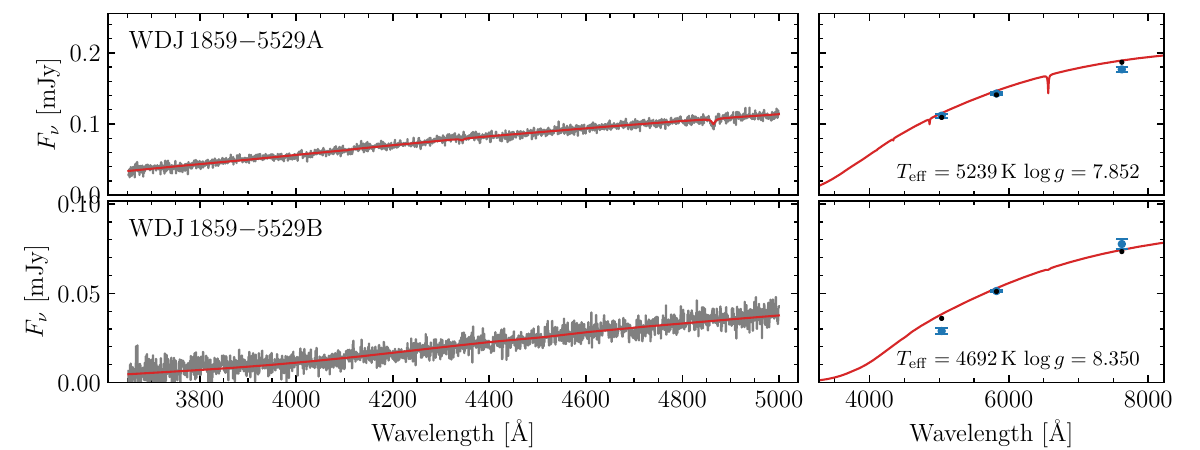}
    \includegraphics[width=\textwidth]{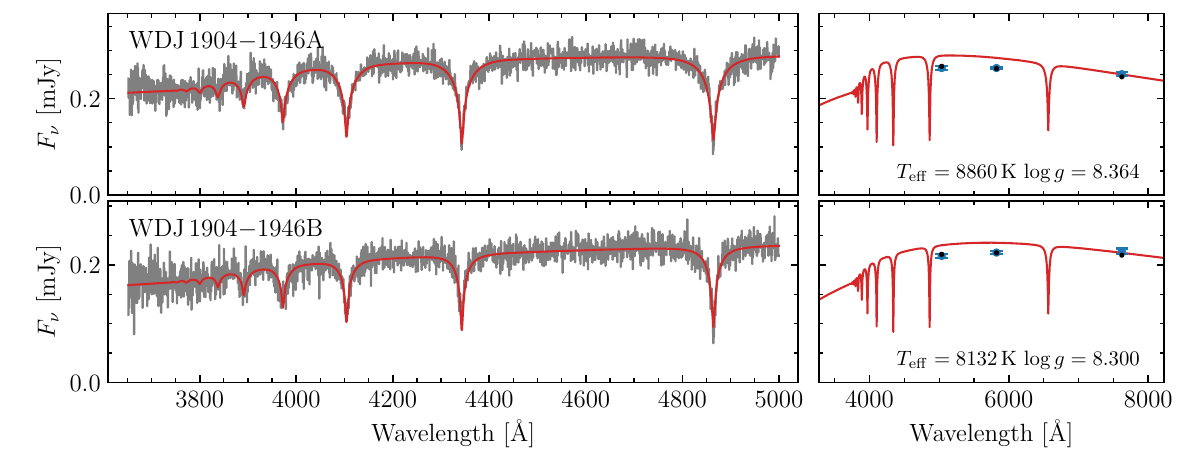}
    \caption{
        Same as Figure~\ref{fig:fits01}, continued.
    }
    \label{fig:fits20}
\end{figure}

\begin{figure}
    \includegraphics[width=\textwidth]{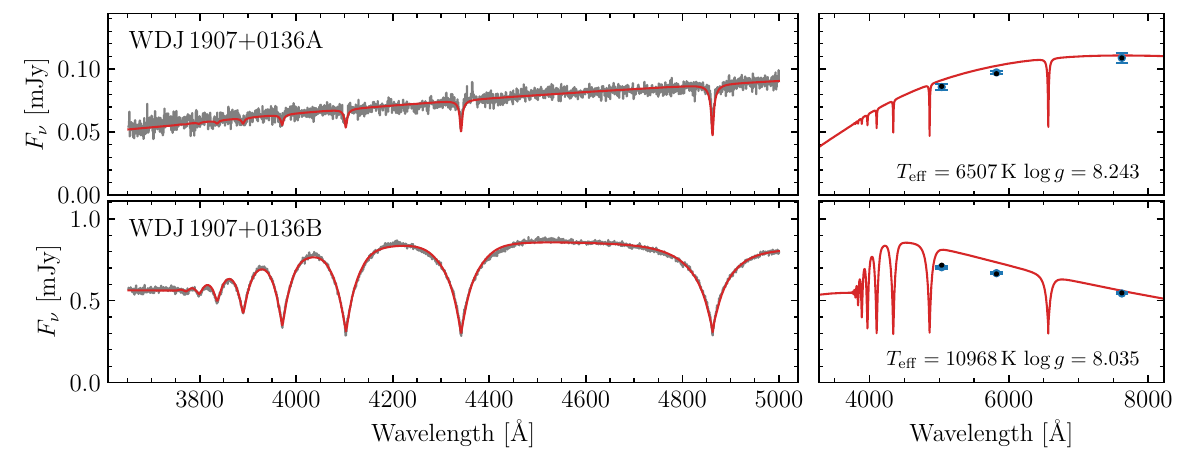}
    \includegraphics[width=\textwidth]{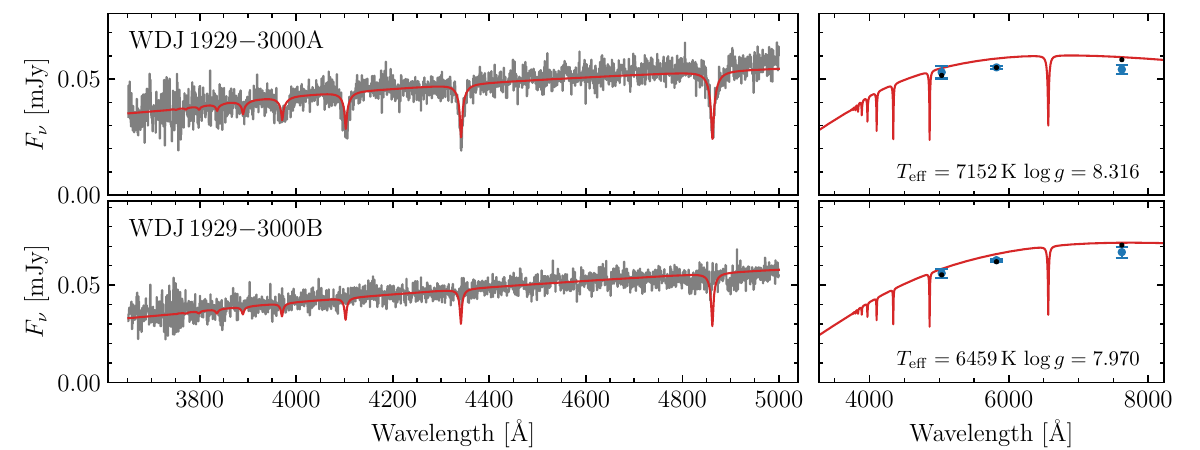}
    \includegraphics[width=\textwidth]{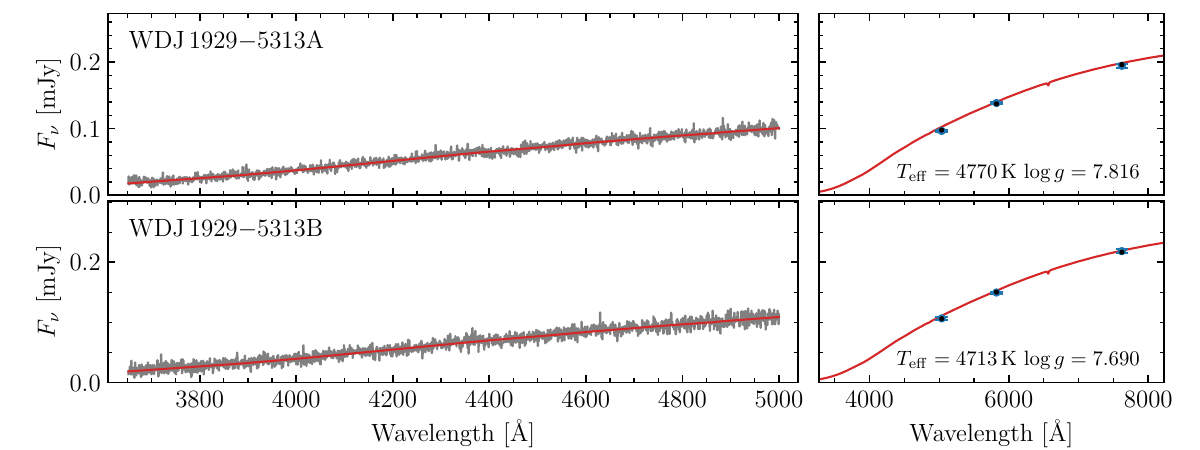}
    \caption{
        Same as Figure~\ref{fig:fits01}, continued.
    }
    \label{fig:fits21}
\end{figure}

\begin{figure}
    \includegraphics[width=\textwidth]{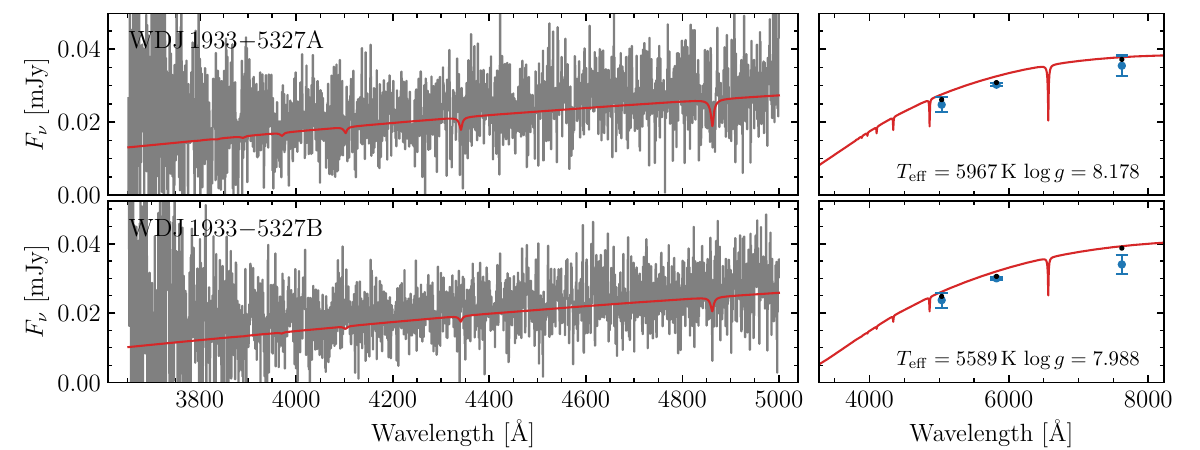}
    \includegraphics[width=\textwidth]{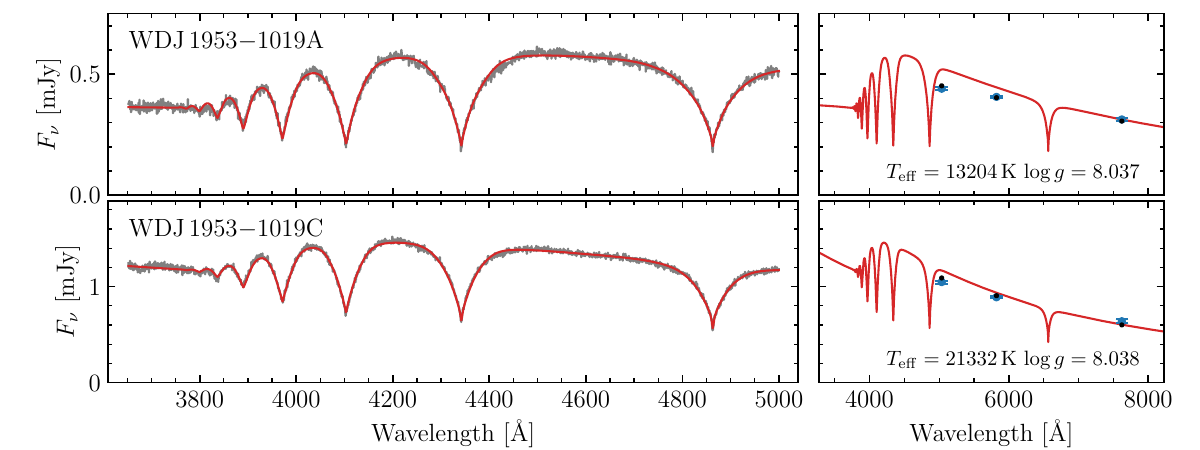}
    \includegraphics[width=\textwidth]{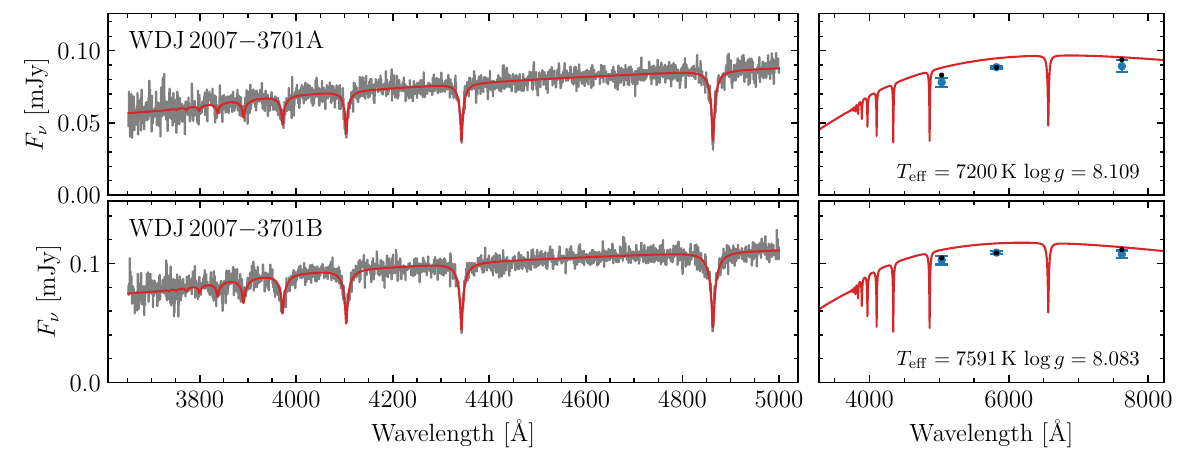}
    \caption{
        Same as Figure~\ref{fig:fits01}, continued.
    }
    \label{fig:fits22}
\end{figure}

\begin{figure}
    \includegraphics[width=\textwidth]{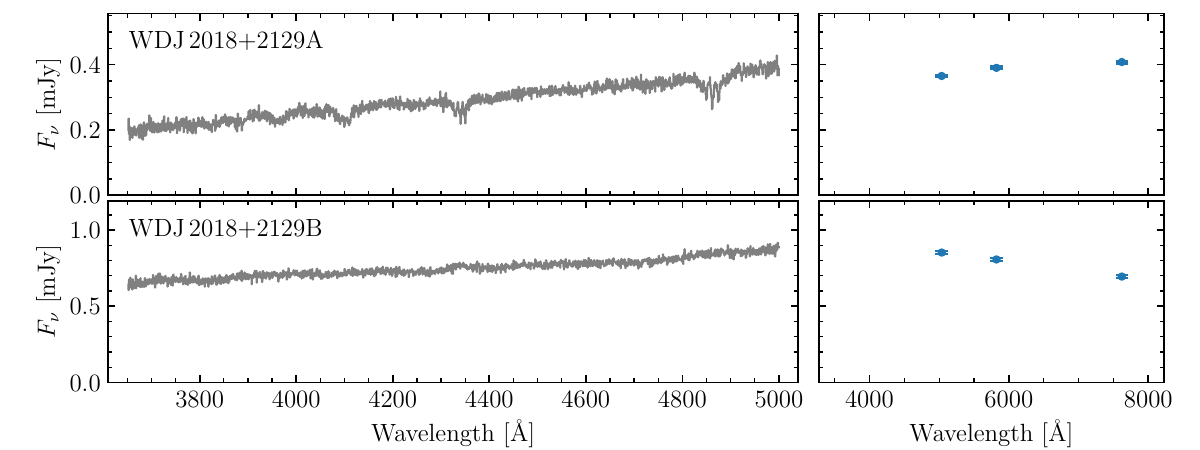}
    \includegraphics[width=\textwidth]{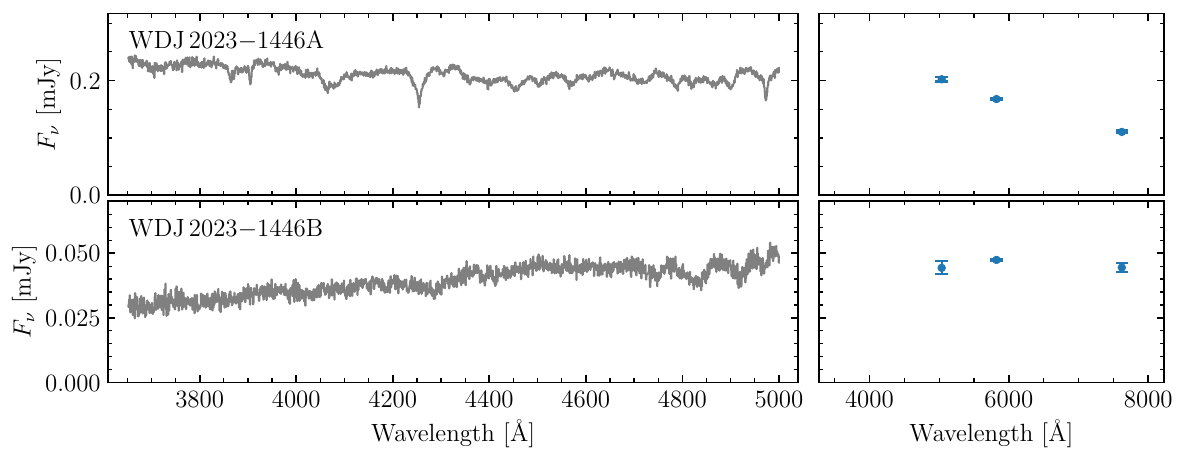}
    \includegraphics[width=\textwidth]{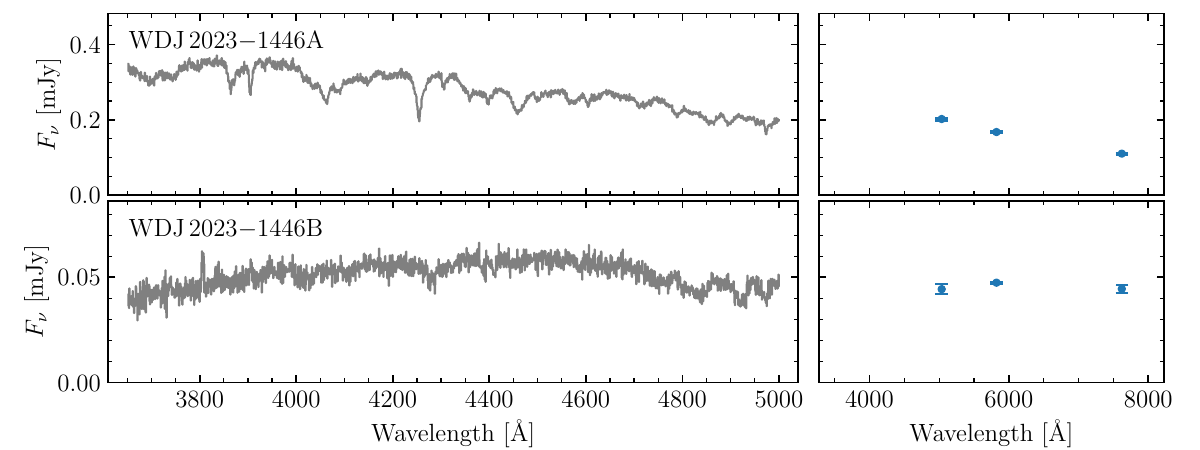}
    \caption{
        Same as Figure~\ref{fig:fits01}, continued.
    }
    \label{fig:fits23}
\end{figure}

\begin{figure}
    \includegraphics[width=\textwidth]{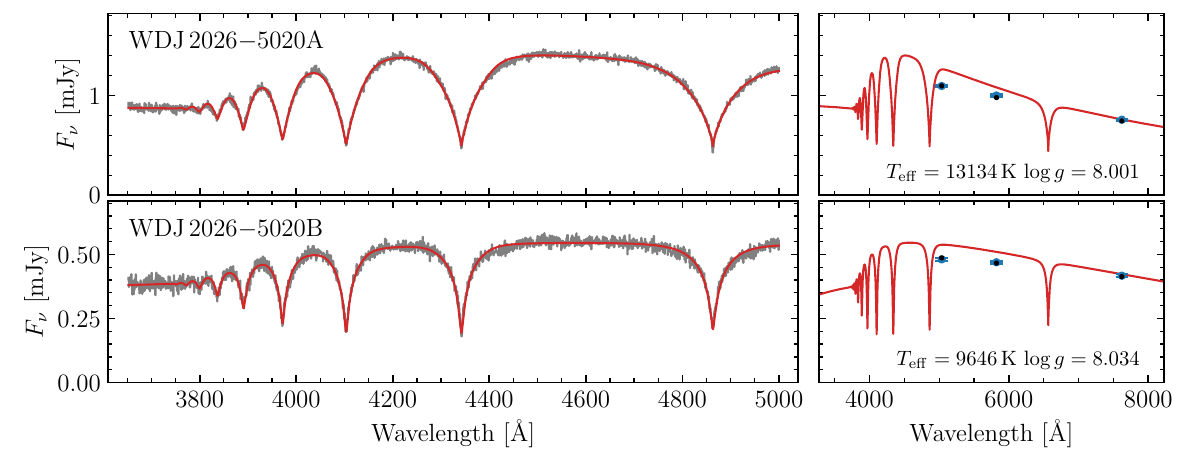}
    \includegraphics[width=\textwidth]{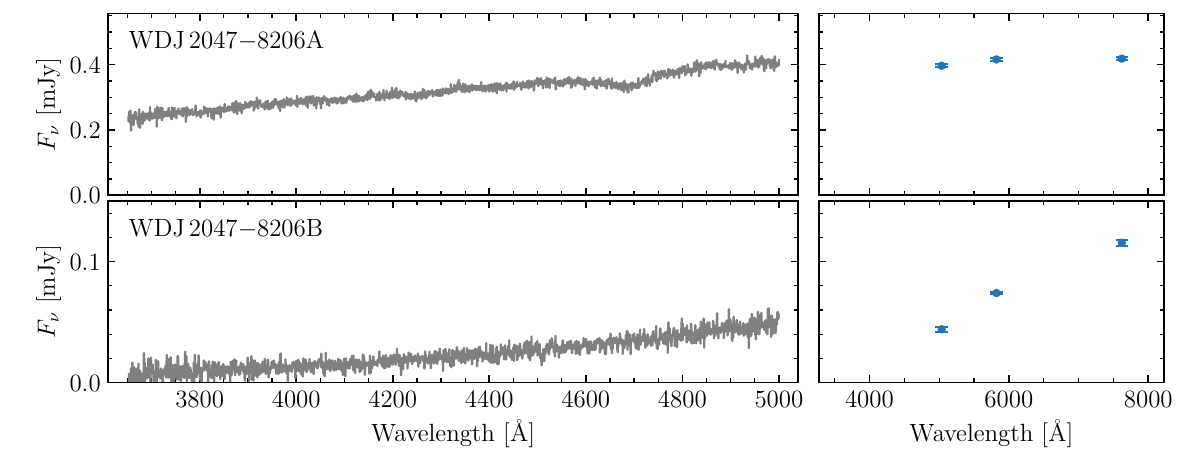}
    \includegraphics[width=\textwidth]{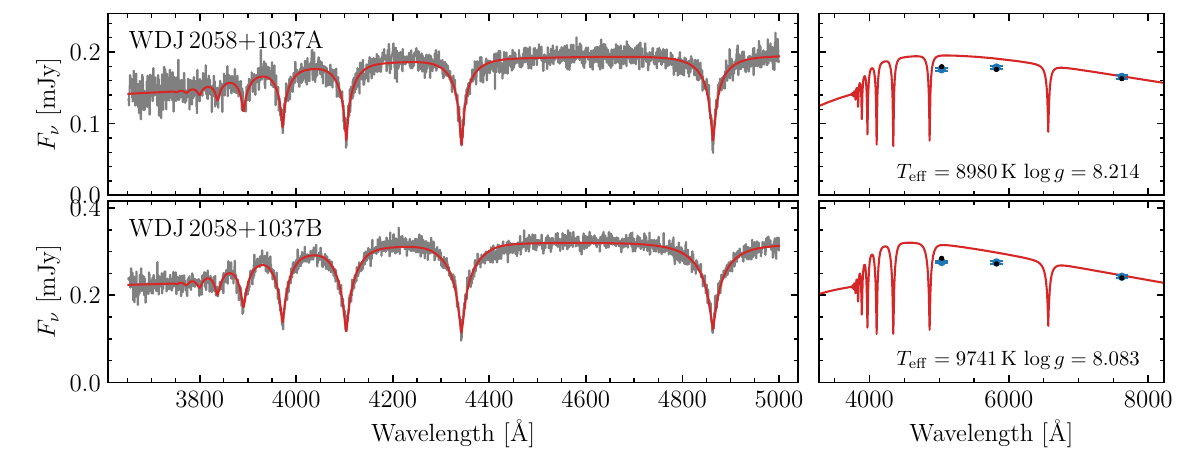}
    \caption{
        Same as Figure~\ref{fig:fits01}, continued.
    }
    \label{fig:fits24}
\end{figure}

\begin{figure}
    \includegraphics[width=\textwidth]{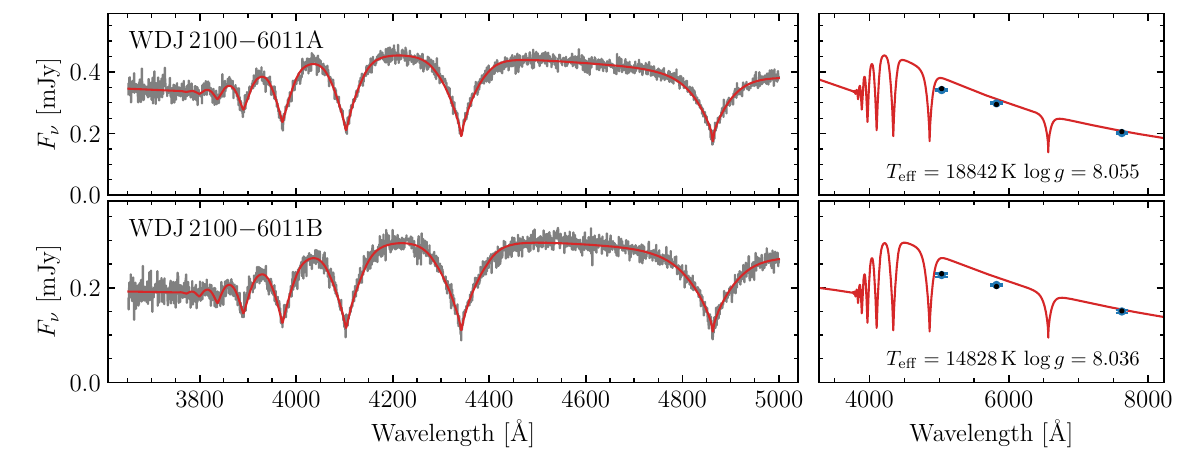}
    \includegraphics[width=\textwidth]{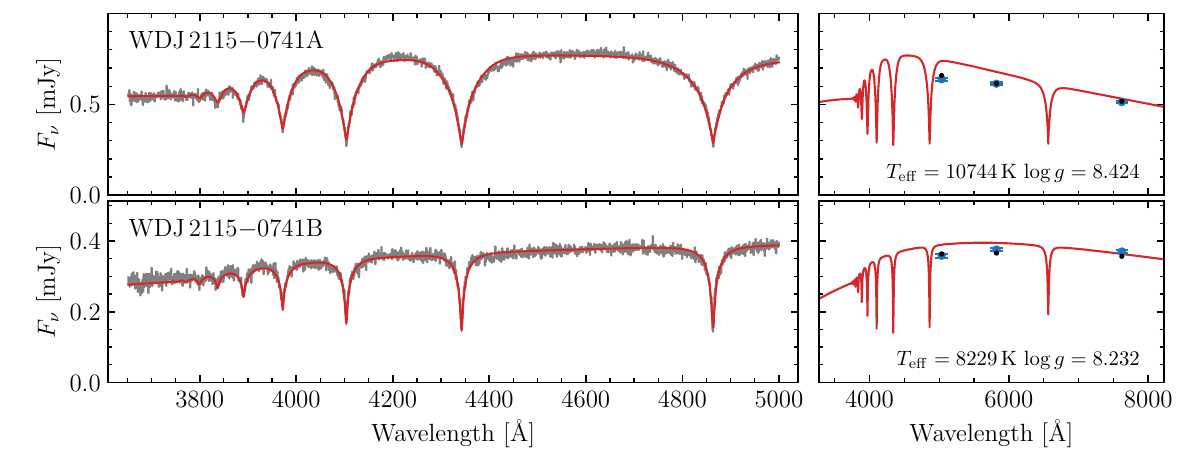}
    \includegraphics[width=\textwidth]{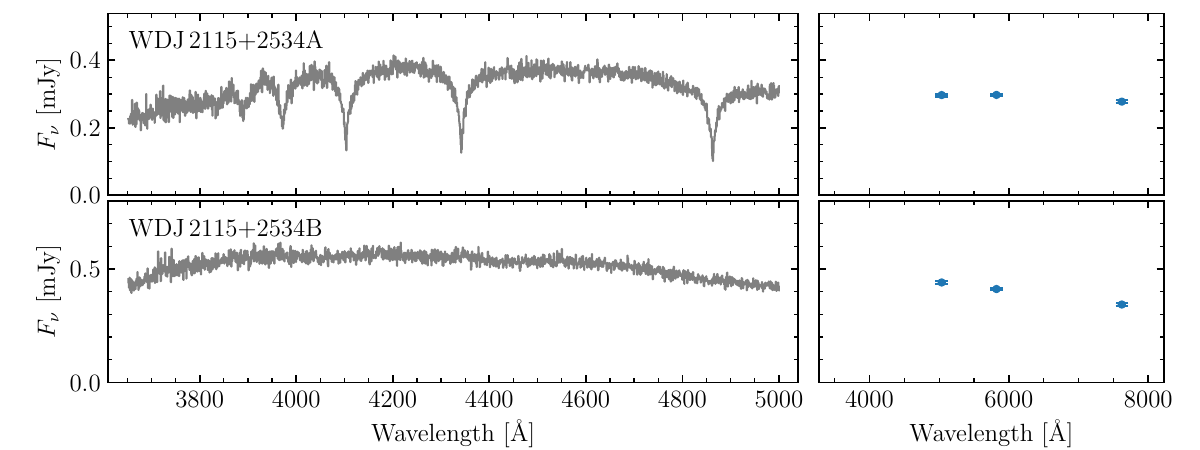}
    \caption{
        Same as Figure~\ref{fig:fits01}, continued.
    }
    \label{fig:fits25}
\end{figure}

\begin{figure}
    \includegraphics[width=\textwidth]{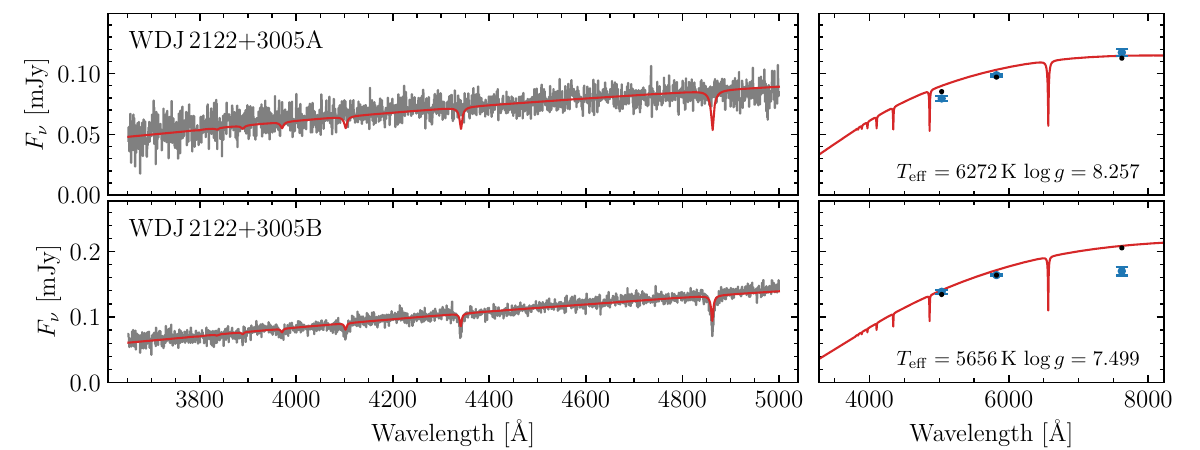}
    \includegraphics[width=\textwidth]{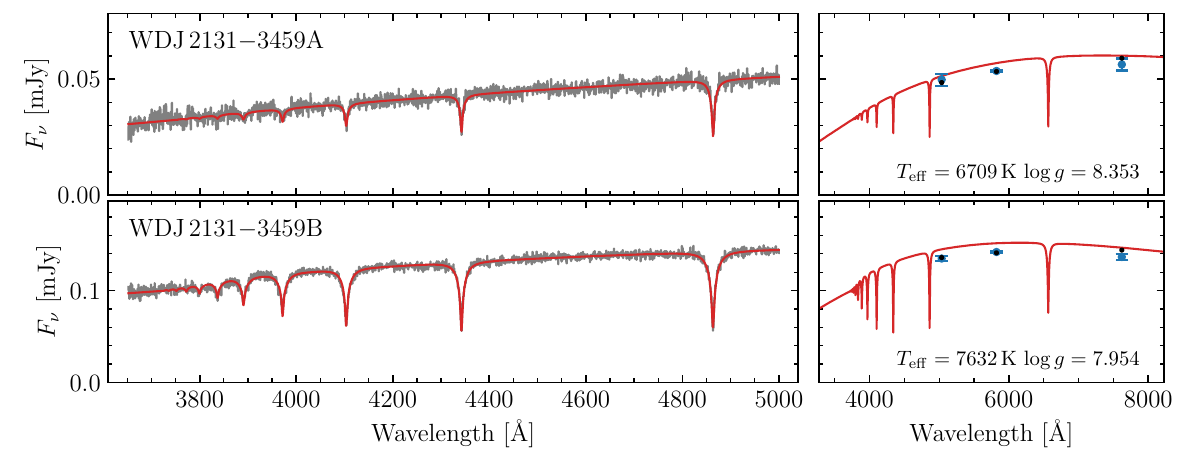}
    \includegraphics[width=\textwidth]{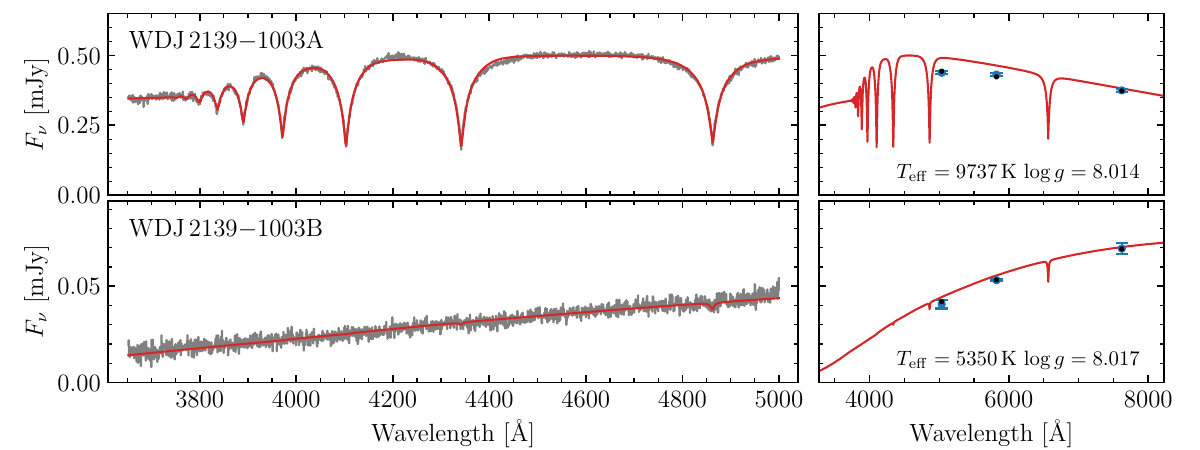}
    \caption{
        Same as Figure~\ref{fig:fits01}, continued.
    }
    \label{fig:fits26}
\end{figure}

\begin{figure}
    \includegraphics[width=\textwidth]{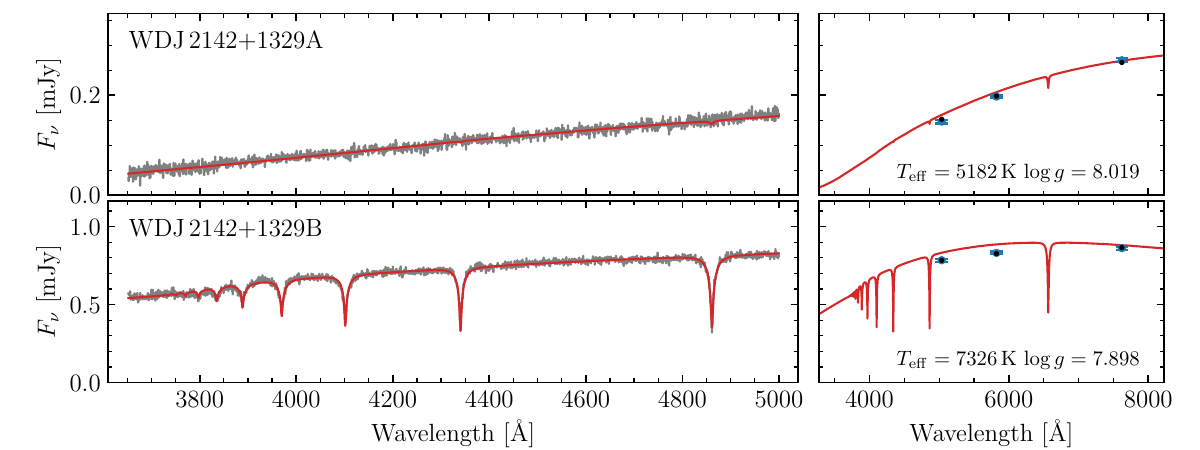}
    \includegraphics[width=\textwidth]{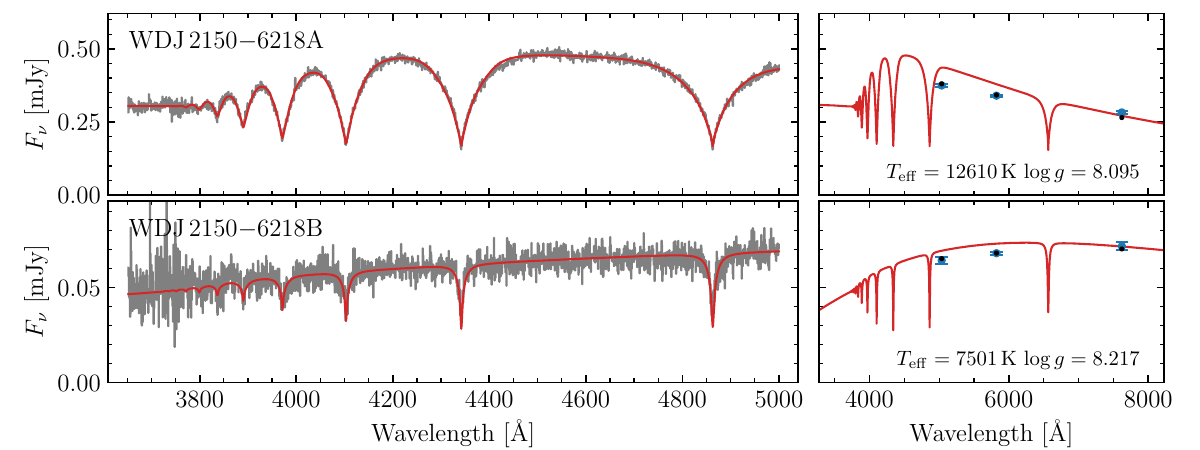}
    \includegraphics[width=\textwidth]{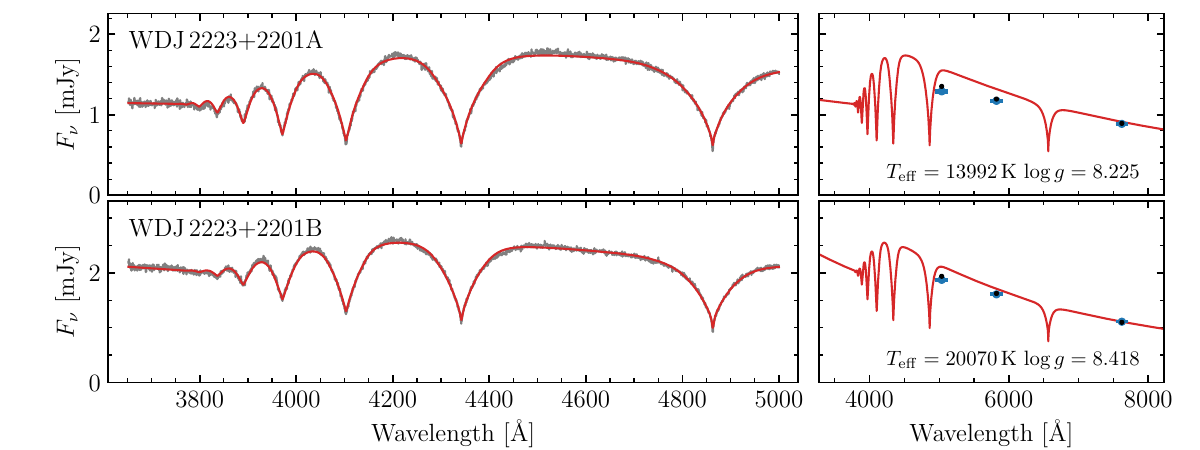}
    \caption{
        Same as Figure~\ref{fig:fits01}, continued.
    }
    \label{fig:fits27}
\end{figure}

\begin{figure}
    \includegraphics[width=\textwidth]{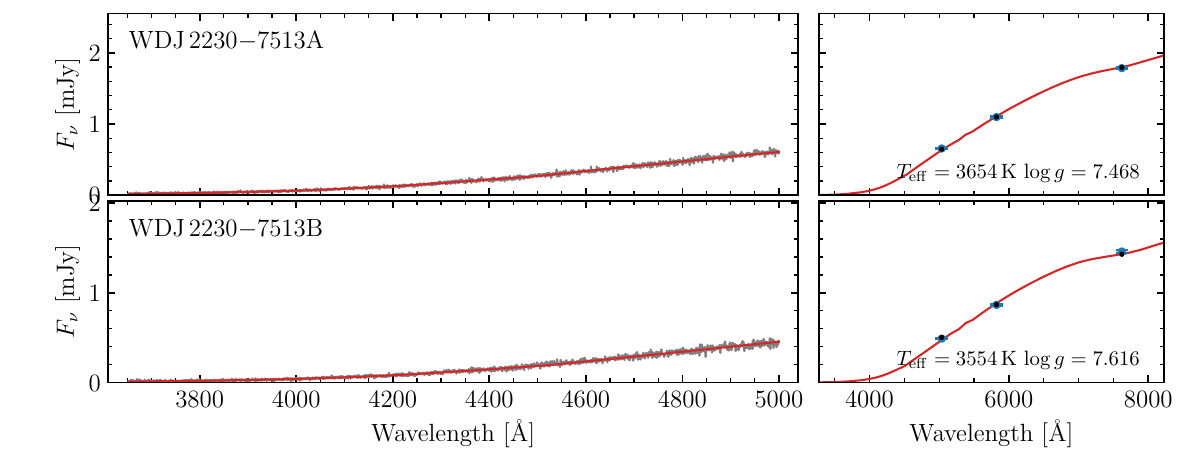}
    \includegraphics[width=\textwidth]{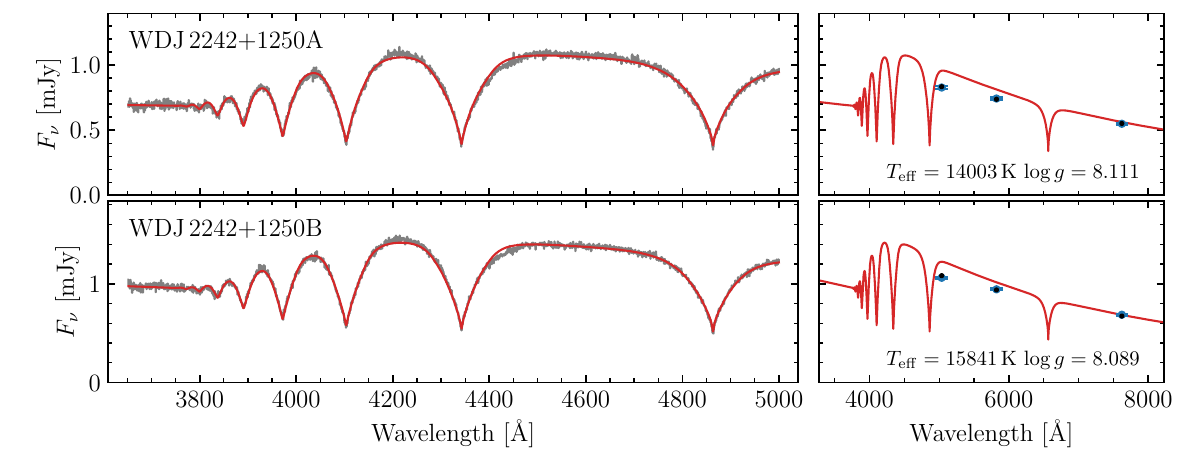}
    \includegraphics[width=\textwidth]{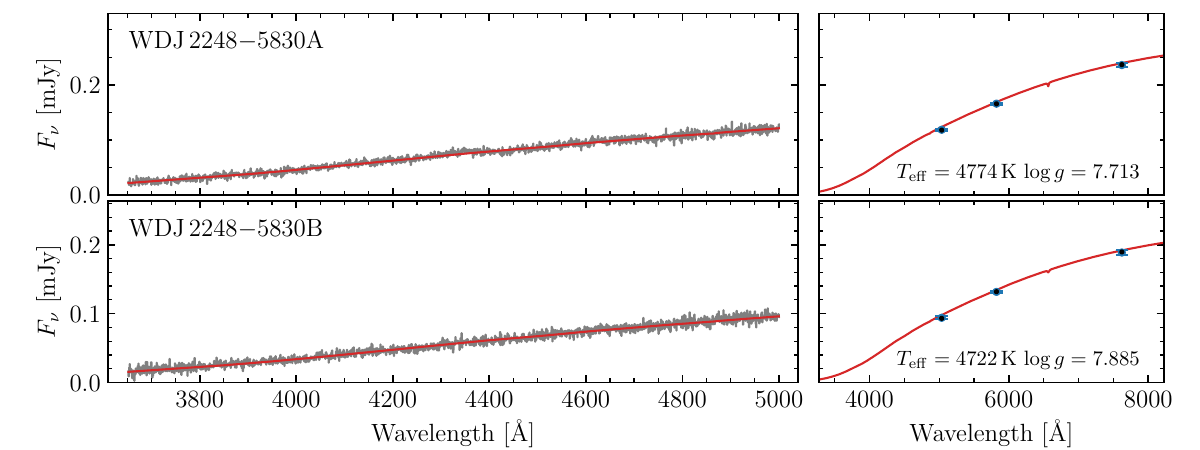}
    \caption{
        Same as Figure~\ref{fig:fits01}, continued.
    }
    \label{fig:fits28}
\end{figure}

\begin{figure}
    \includegraphics[width=\textwidth]{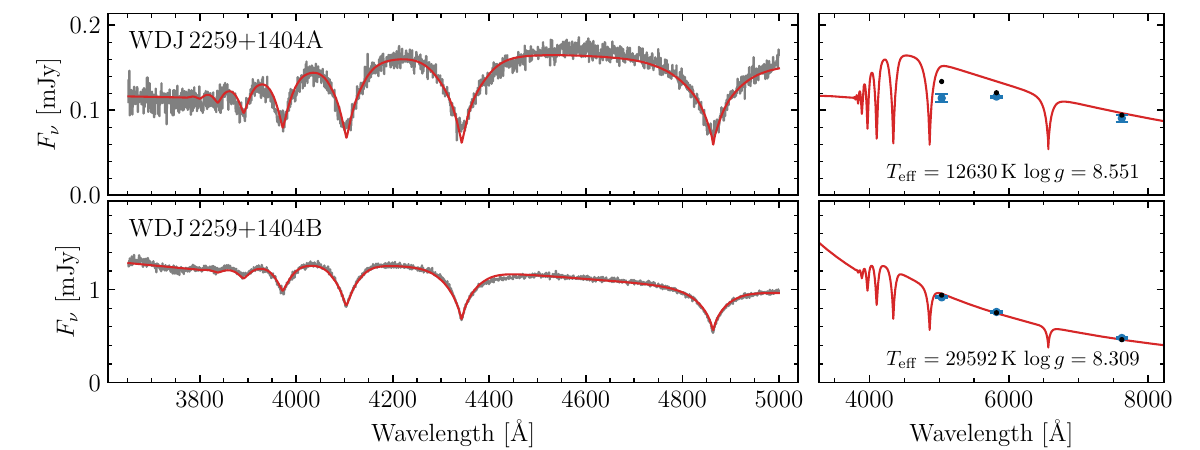}
    \includegraphics[width=\textwidth]{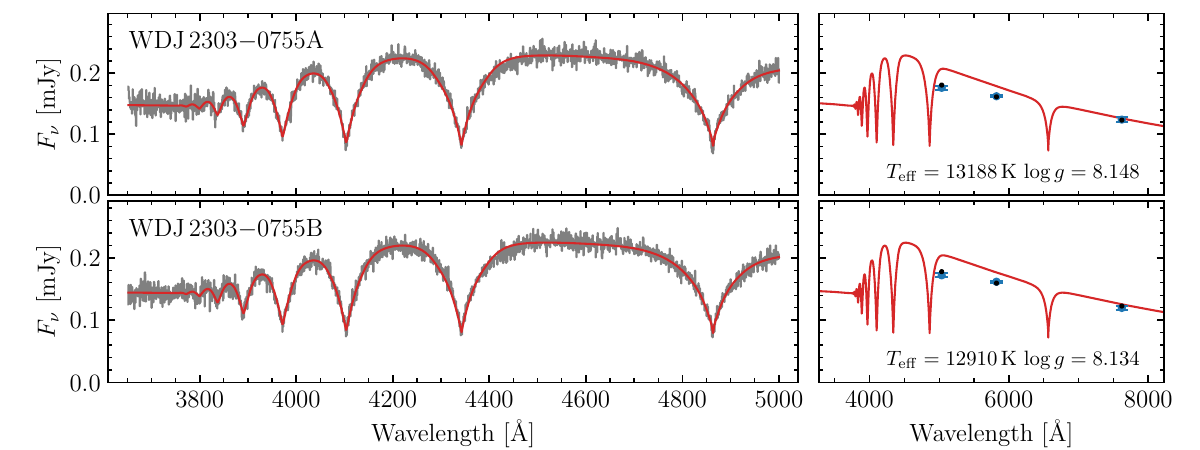}
    \includegraphics[width=\textwidth]{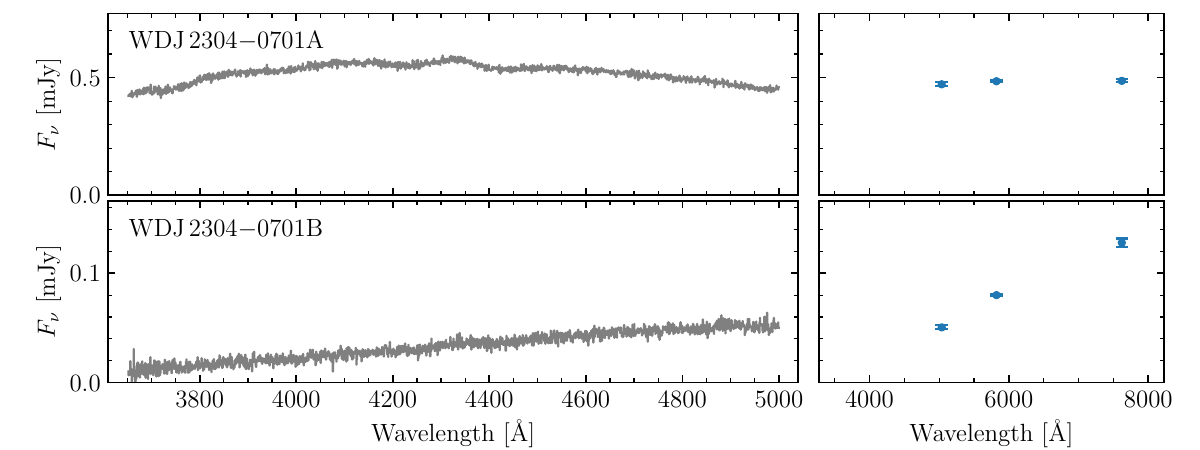}
    \caption{
        Same as Figure~\ref{fig:fits01}, continued.
    }
    \label{fig:fits29}
\end{figure}

\begin{figure}
    \includegraphics[width=\textwidth]{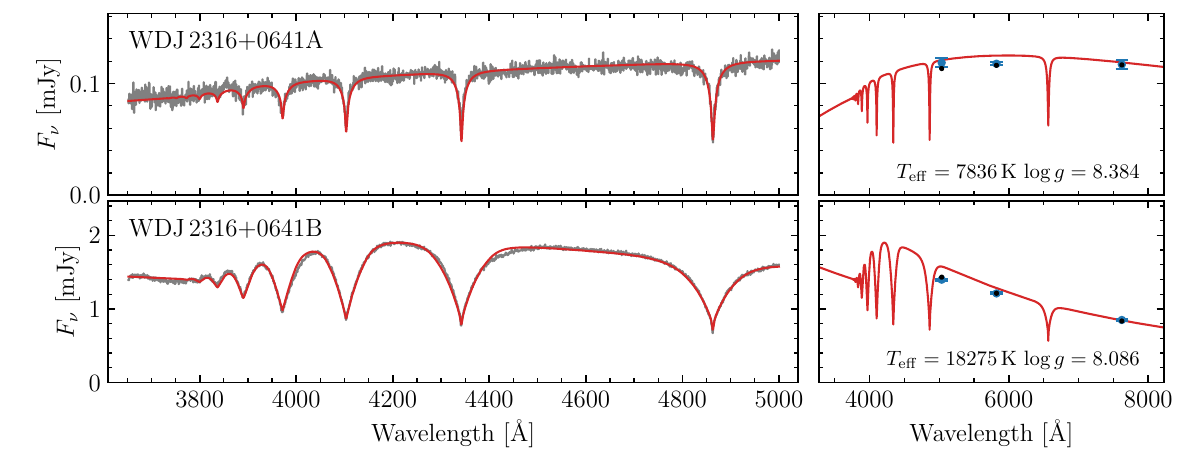}
    \includegraphics[width=\textwidth]{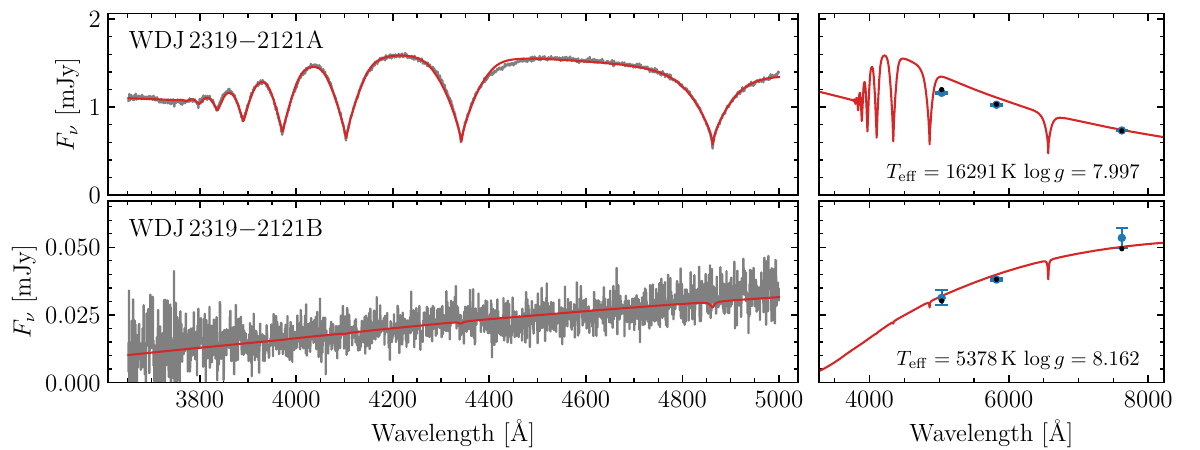}
    \includegraphics[width=\textwidth]{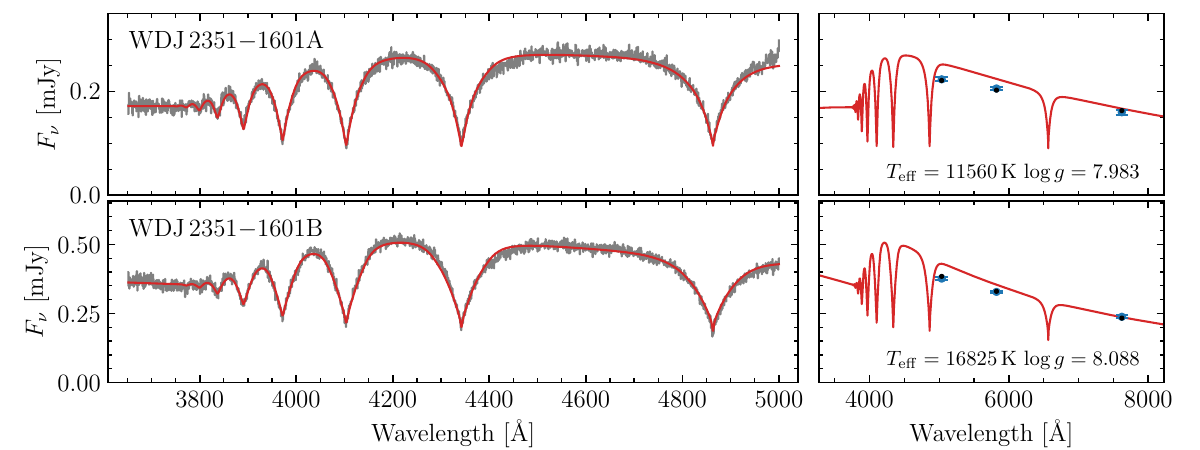}
    \caption{
        Same as Figure~\ref{fig:fits01}, continued.
    }
    \label{fig:fits30}
\end{figure}

\begin{figure}
    \includegraphics[width=\textwidth]{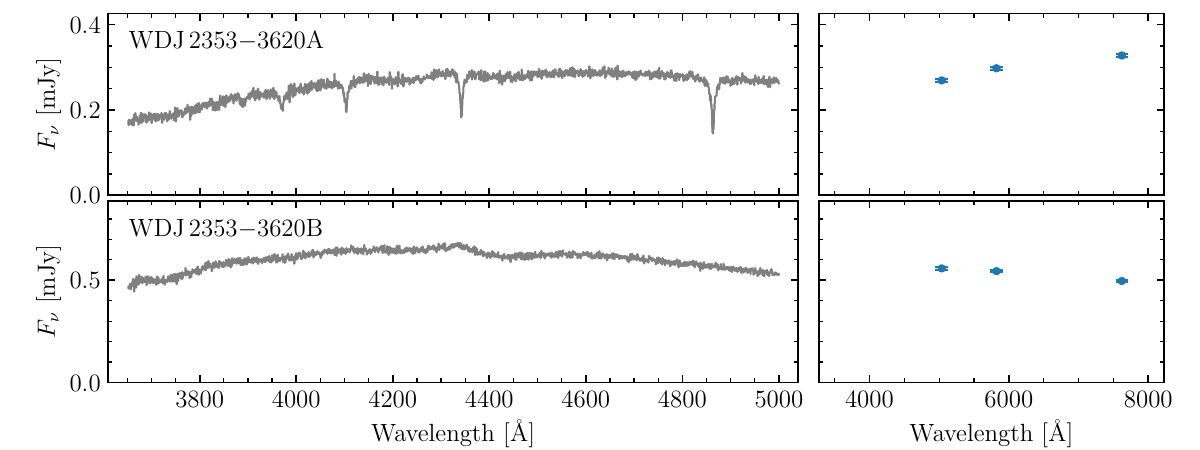}
    \includegraphics[width=\textwidth]{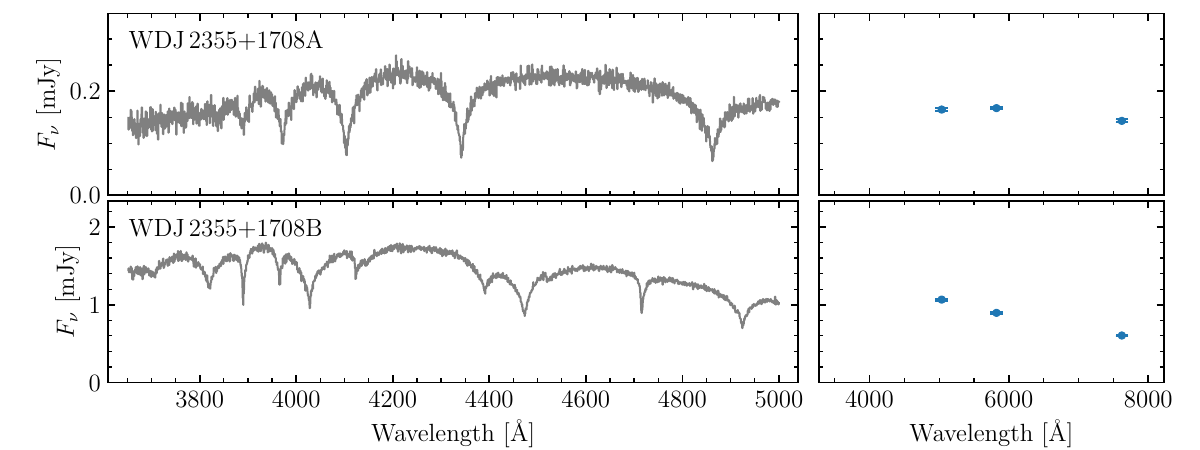}
    \caption{
        Same as Figure~\ref{fig:fits01}, continued.
    }
    \label{fig:fits31}
\end{figure}


\bsp	
\label{lastpage}
\end{document}